\documentclass[twocolumn]{aastex631}

\let\tablenum\relax
\usepackage[version=4]{mhchem}
\usepackage{chemmacros}
\chemsetup{
  modules = {reactions} ,
  formula = mhchem
}

\chemsetup[reactions]{
  before-tag = R ,
  tag-open = ( ,
  tag-close = )
}

\usepackage{graphicx}	
\usepackage{amsmath}	
\usepackage{hyperref}

\begin{document}

\title{Earth-like exoplanets in spin-orbit resonances: climate dynamics, 3D atmospheric chemistry, and observational signatures}

\author[0000-0002-9076-2361]{Marrick Braam}
\email{marrick.braam@unibe.ch}
\affiliation{School of GeoSciences, University of Edinburgh, Edinburgh, EH9 3FF, UK}
\affiliation{Centre for Exoplanet Science, University of Edinburgh, Edinburgh, EH9 3FD, UK}
\affiliation{Institute of Astronomy, KU Leuven, Celestijnenlaan 200D, 3001 Leuven, Belgium}
\affiliation{Center for Space and Habitability, University of Bern, Gesellschaftsstrasse 6, 3012 Bern, Switzerland}

\author[0000-0002-1487-0969]{Paul I. Palmer}
\affiliation{School of GeoSciences, University of Edinburgh, Edinburgh, EH9 3FF, UK}
\affiliation{Centre for Exoplanet Science, University of Edinburgh, Edinburgh, EH9 3FD, UK}

\author[0000-0002-5342-8612]{Leen Decin}
\affiliation{Institute of Astronomy, KU Leuven, Celestijnenlaan 200D, 3001 Leuven, Belgium}

\author[0000-0001-6707-4563]{Nathan J. Mayne}
\affiliation{Department of Physics and Astronomy, Faculty of Environment Science and Economy, University of Exeter, Exeter EX4 4QL, UK}

\author[0000-0003-4402-6811]{James Manners}
\affiliation{Met Office, Fitzroy Road, Exeter EX1 3PB, UK}

\author[0000-0003-1620-7658]{Sarah Rugheimer}
\affiliation{Department of Physics and Astronomy, York University, 4700 Keele St., Toronto, ON M3J 1P3, Canada}



\begin{abstract}
Terrestrial exoplanets around M- and K-type stars are important targets for atmospheric characterisation. Such planets are likely tidally locked with the order of spin-orbit resonances (SORs) depending on eccentricity. We explore the impact of SORs on 3D atmospheric dynamics and chemistry, employing a 3D coupled Climate-Chemistry Model to simulate Proxima Centauri b in 1:1 and 3:2 SOR. For a 1:1 SOR, Proxima Centauri b is in the Rhines rotator circulation regime with dominant zonal gradients (global mean surface temperature 229~K). An eccentric 3:2 SOR warms Proxima Centauri b to 262~K with gradients in the meridional direction. We show how a complex interplay between stellar radiation, orbit, atmospheric circulation, and (photo)chemistry determines the 3D ozone distribution. Spatial variations in ozone column densities align with the temperature distribution and are driven by stratospheric circulation mechanisms. Proxima Centauri b in a 3:2 SOR demonstrates additional atmospheric variability, including daytime-nighttime cycles in water vapour of ${+}$55\% to ${-}$34\% and ozone ($\pm5.2$\%) column densities and periastron-apoastron water vapour cycles of ${+}$17\% to ${-}$10\%. Synthetic emission spectra for the spectral range of the Large Interferometer For Exoplanets fluctuate by up to 36~ppm with orbital phase angle for a 1:1 SOR due to 3D spatial and temporal asymmetries. The homogeneous atmosphere for the 3:2 SOR results in relatively constant emission spectra and provides an observational discriminant from the 1:1 SOR. Our work emphasizes the importance of understanding the 3D nature of exoplanet atmospheres and associated spectral variations to determine habitability and interpret atmospheric spectra.
\end{abstract}


\section{Introduction}
Exoplanet discoveries cover diverse stellar, orbital, and planetary characteristics. Exoplanets are easier to characterise with current observatories around relatively cool stars such as M- and K-type stars, because of observational advantages \citep[][]{charbonneau_dynamics-based_2007}{}{} and the high occurrence rates of such stars and planets orbiting them \citep[e.g.][]{bochanski_luminosity_2010, petigura_prevalence_2013, dressing_occurrence_2015, cloutier_evolution_2020, hsu_occurrence_2020, ment_occurrence_2023}{}{}. Due to their relatively low luminosity, the circumstellar Habitable Zone (HZ) is located close to the star \citep[][]{kasting_habitable_1993, kopparapu_habitable_2013}. Consequently, a planet orbiting in the HZ is likely tidally locked, leading to spin-orbit resonances or SORs \citep[][]{goldreich_spin-orbit_1966, barnes_tidal_2017, pierrehumbert_atmospheric_2019}{}{}.

SORs in the Solar System include the Moon's 1:1 SOR around the Earth and Mercury's 3:2 SOR around the Sun. Although gravitational tides mainly cause 1:1 SORs, processes such as eccentricity tides, orbital scattering, merging events, thermal tides, or secular perturbations can drive higher-order SORs \citep[][]{leconte_asynchronous_2015, barnes_tidal_2017, auclair-desrotour_rotation_2017, auclair-desrotour_final_2019, renaud_tidal_2021, valente_tidal_2024}. Mercury's spin rate slowed down during the planet's evolution as a result of tidal interactions with the Sun. Eventually, Mercury got trapped in a 3:2 SOR due to a non-zero eccentricity that may have exceeded its current value of 0.2056 \citep[][]{goldreich_spin-orbit_1966, correia_mercurys_2004, makarov_conditions_2012, correia_spin-orbit_2019}. Although still debated, previous work has shown that the asynchronous retrograde rotation of Venus can also be explained by a balance between gravitational and thermal tides \citep[][]{gold_atmospheric_1969, ingersoll_venus_1978, correia_four_2001, auclair-desrotour_rotation_2017}. 

For close-in exoplanets around M- and K-type stars, the timescales associated with tidal locking are short compared to planetary lifetimes \citep[][]{goldreich_spin-orbit_1966, pierrehumbert_atmospheric_2019}{}{}. The final state following the process of tidal locking can be expressed in terms of the planet's eccentricity $e$ \citep[][]{goldreich_spin-orbit_1966, dobrovolskis_spin_2007, pierrehumbert_atmospheric_2019}{}{}: a 1:1 SOR or synchronous rotation is the most likely state for planets with $e{\lesssim}0.2$, a 3:2 SOR is most likely for $0.2{\lesssim}e{\lesssim}0.35$. The resulting planetary climates for these two SORs are markedly distinct \citep[e.g.,][]{turbet_habitability_2016, boutle_exploring_2017, del_genio_habitable_2019, colose_effects_2021}{}{}. Higher-order SORs such as 2:1 are possible for planets with $e{\gtrsim}0.4$, but the climatic consequences are more subtle. A 1:1 SOR leads to a permanent dayside and nightside hemisphere, centred at the substellar and antistellar points, respectively. For the 3:2 SOR, the substellar point shifts by 180$^{\circ}$ in longitude for every orbit around the star. Hence, a cycle of daytime and nighttime is completed in two orbits.

To explore the effects of these SORs on habitability, 3D General Circulation Models (GCMs) are employed to simulate the main physical processes in a planetary atmosphere and to predict the resulting climate and habitability. For planets in 1:1 SOR, these show large temperature differences between the dayside and nightside hemisphere, although atmospheric or ocean heat transport can prevent atmospheric collapse \citep[][]{joshi_simulations_1997, hu_role_2014}{}{}. Such exoplanets likely exhibit distinct regimes of atmospheric circulation depending on a planet's rotation rate, radius, and atmospheric composition \citep[e.g.,][]{merlis_atmospheric_2010, edson_atmospheric_2011, carone_connecting_2014, carone_connecting_2015, carone_connecting_2016, noda_circulation_2017, kopparapu_habitable_2017, haqq-misra_demarcating_2018, carone_stratosphere_2018}{}{}. Two distinct circulation regimes are of particular interest for potentially habitable ocean-covered planets in a 1:1 SOR. The `Rhines rotator' regime exhibits weak superrotation with a single equatorial jet, planetary-scale turbulent structures, and mainly shows zonal gradients in atmospheric quantities like temperature. The fast rotator regime exhibits superrotation in a pair of mid-latitude jets and mainly shows meridional gradients in atmospheric quantities like temperature. \cite{sergeev_bistability_2022}{}{} show that a single planet close to a regime transition in terms of its rotation rate can also exhibit different circulation regimes, depending on the initial conditions and model parameterisations. Planet-planet interactions may destabilize planets away from the synchronous 1:1 SOR. \citet{chen_sporadic_2023} show that such spin variations result in libration of the substellar point and thereby affect planetary climates, especially close to the outer edge of the HZ.

A 3:2 SOR results in a changing daytime hemisphere for the planet and exhibits meridional gradients in atmospheric quantities, such as large equator-to-pole differences in temperature that exceed the zonal gradients \citep[][]{turbet_habitability_2016, boutle_exploring_2017, del_genio_habitable_2019}{}{}. Therefore, \cite{del_genio_habitable_2019} suggest that -- at least in the absence of eccentricity -- the circulation is more like that of a slowly rotating Earth \citep[see also][]{genio_comparative_1987}{}{}, with a meridional circulation of one Hadley cell per hemisphere and westward jets over the midlatitudes. Including eccentricity will cause an insolation pattern and associated heating concentrated in two hot spots, one for each hemisphere \citep[][]{dobrovolskis_insolation_2015, boutle_exploring_2017, colose_effects_2021}{}{}. Additionally, an eccentric orbit will produce an increase of the mean flux over one orbit as compared to the circular case \citep[][]{williams_earth-like_2002, dressing_habitable_2010, bolmont_habitability_2016, ji_inner_2023}{}{}. SORs may affect the inner edge of the HZ, as defined by the moist greenhouse limit of the maximum stratospheric water vapour abundances to retain the Earth's oceans \citep[][]{ingersoll_runaway_1969, kasting_runaway_1988}{}{} or the Simpson-Nakajima limit in outgoing longwave radiation \citep[][]{simpson1928some, nakajima_study_1992, goldblatt_low_2013, chaverot_how_2022}{}{}. However, \cite{colose_effects_2021} find that higher-order SORs have little effect on the inner edge since warming of the planet is accompanied by a drier stratosphere.

For both SORs, the interplay between atmospheric dynamics, chemistry, and thermodynamics will control the distribution of atmospheric tracers such as clouds and chemical species. Previous work has shown that the distribution of clouds depends on the rotation rate and circulation regime of planets in a 1:1 SOR \citep[e.g.,][]{komacek_atmospheric_2019, sergeev_bistability_2022}{}{}, affecting the planetary climate \citep[][]{yang_stabilizing_2013, yang_low-order_2014, kopparapu_inner_2016}{}{}. Convective clouds cover the dayside for planets in a 1:1 SOR with zonal gradients and a much thinner nightside cloud deck \citep[see figure 3 of][]{boutle_exploring_2017}{}{}. For a 3:2 SOR, the cloud cover follows the hot spots in temperature and shows banded structures with meridional gradients and a relatively thin high-latitude cloud deck \citep[see figure 10 of][]{boutle_exploring_2017}{}{}. 

The effects of atmospheric dynamics on the distributions of chemical species for planets in a 1:1 SOR are studied using 3D coupled Climate-Chemistry Models or CCMs \citep[][]{proedrou_characterising_2016, chen_habitability_2019, yates_ozone_2020, chen_persistence_2021, braam_lightning-induced_2022, ridgway_3d_2023, luo_coupled_2023, cooke_variability_2023, cooke_lethal_2024}{}{}. A CCM includes a photochemical framework coupled to the GCM to solve for 3D atmospheric chemistry. Even though the production of photochemical species like ozone is limited to the dayside, \citet{braam_stratospheric_2023} demonstrate that a stratospheric dayside-to-nightside circulation can drive the accumulation of such species on the nightside. However, these circulation mechanisms also depend on the distribution of landmasses \citep[][]{lewis_influence_2018, bhongade_asymmetries_2024, martinez_how_2024}. Interactive chemistry for a 3:2 SOR has not yet been investigated but, given the meridional gradients in temperature and more Earth-like circulation discussed above, we might expect a mechanism similar to the Brewer-Dobson circulation, which controls the distributions of species like ozone and water vapour on Earth \citep[e.g.,][]{brewer_evidence_1949, dobson_origin_1956, butchart_brewer-dobson_2014}{}{}. The Brewer-Dobson circulation describes the ascent of chemically enriched air in the tropics, followed by equator-to-pole transport in the stratosphere, and finally descent at higher latitudes. In the case of radiatively active tracers -- like clouds, ozone, or water vapour -- such circulation mechanisms and the resulting spatial distributions in turn affect the thermal structure of a planetary atmosphere, depending on the spectral energy distribution of the host star \citep[see e.g.,][]{godolt_3d_2015}{}{}. In turn, such radiative feedbacks can affect the dynamical state of the atmosphere \citep[][]{hochman_analogous_2023, de_luca_impact_2024}{}{}. Since near-surface ozone is harmful to life on Earth, the spatial distribution of ozone further affects planetary habitability and the upper limits of near-surface ozone levels on M-dwarf planets are explored by \citet{cooke_lethal_2024}.

The complex 3D interplay between stellar radiation, planetary orbit, atmospheric dynamics, and (photo)chemistry determines the planetary climate and habitability, and has the potential to induce several temporally varying effects. These include seasonal variations \citep[][]{cooke_variability_2023}{}{}, internal atmospheric variability \citep[][]{cohen_longitudinally_2022, hochman_greater_2022, cohen_traveling_2023, braam_stratospheric_2023, luo_coupled_2023, de_luca_impact_2024}{}{}, or variability due to external causes such as flares or cosmic rays \citep[][]{segura_effect_2010, scheucher_new_2018, chen_persistence_2021, ridgway_3d_2023}{}{}. Temporally varying abundances of chemical species have been proposed as a biosignature \citep[][]{olson_atmospheric_2018, schwieterman_exoplanet_2018}. Performing 700~years of simulation, \citet{luo_coupled_2023} show how a strong non-varying NO$_\mathrm{x}$ source (such as biological nitrogen fixation) on the surface of an M-dwarf planet can induce oscillations in the ozone abundances, potentially presenting another biosignature. However, such seasonal variations in biosignatures must also be distinguishable from abiotically induced variations \citep[][]{olson_atmospheric_2018}. The passage through periastron and apoastron for a planet in an eccentric 3:2 SOR might be yet another driver of thermal and chemical variations. 

3D spatial and temporal variations will affect spectroscopic observations of terrestrial exoplanet atmospheres \citep[e.g.][]{turbet_habitability_2016, boutle_exploring_2017, chen_persistence_2021, cohen_traveling_2023, braam_stratospheric_2023, cooke_variability_2023, luo_coupled_2023, de_luca_impact_2024}. The quest to understand terrestrial exoplanets and search for biosignatures has motivated the development of the next-generation space observatories. The Habitable Worlds Observatory (HWO) is a planned coronographic mission focusing on direct imaging of terrestrial exoplanets using reflected light at ultraviolet, visible, and near-infrared wavelengths. \citet{cooke_variability_2023} demonstrate how a HWO-like concept can potentially detect spatial and seasonal variations of chemical species for Earth-like exoplanets around G-type stars. Due to limits on the inner working angle of the coronagraph, HWO likely offers little potential to characterise planets around M-dwarfs \citep[][]{vaughan_chasing_2023, harada_setting_2024}. The Large Interferometer For Exoplanets (LIFE) is a European-led concept for a space-based nulling interferometer \citep[][]{quanz_large_2022}, focusing on thermal emission from terrestrial exoplanets at mid-infrared wavelengths. LIFE's capabilities have been demonstrated for Earth as an exoplanet at 10~parsec \citep[][]{konrad_large_2022}, Earth throughout its history \citep[][]{alei_large_2022}, a Venus twin \citep[][]{konrad_large_2023}, capstone biosignatures on planets around MGK stars \citep[][]{angerhausen_large_2024}, and real Earth observational data \citep[][]{mettler_earth_2023, mettler_earth_2024}. For LIFE, Proxima Centauri b \citep[][]{anglada-escude_terrestrial_2016}{}{} is a golden target \citep[][]{angerhausen_large_2024}. A natural follow-up for LIFE is thus to investigate spatial and temporal variations for exoplanet configurations, in particular for Proxima Centauri b. The smaller Mid-InfraRed Exoplanet CLimate Explorer (MIRECLE) mission concept also aims to characterise terrestrial exoplanets at mid-infrared wavelengths \citep[][]{mandell_mirecle_2022} and can pave the way for further characterisation with LIFE. 

In this study, we investigate the 3D interplay between stellar radiation, atmospheric dynamics, and (photo)chemistry for a tidally locked planet around an M-star -- nominally Proxima Centauri b -- simulating both a 1:1 and 3:2 SOR. We also present the effects of the simulated spatial and temporal variations on synthetic mid-infrared emission spectra. In Section~\ref{sec:methods} we describe the CCM framework, the planetary configurations, and the generation of emission spectra. We report our findings on the orbital effects on 3D atmospheric chemistry in Section~\ref{sec:results}. In Section~\ref{sec:results} we show the observational consequences of our study, focusing on the LIFE mission concept. Finally, we will put our results into context and provide the main conclusions in Section~\ref{sec:discconc}.

\section{Methods}\label{sec:methods}
This Section begins with an introduction of the 3D CCM. We then describe the main characteristics of both planet configurations (Section~\ref{subseq:planetconfs}) and the general simulation setups (Section~\ref{subsec:simsetups}). Finally, Section~\ref{subsec:PSG} elaborates on the generation of emission spectra from the 3D CCM data. 

\subsection{Coupled Climate-Chemistry Model}
\label{subsec:3dccm} 
The 3D CCM consists of the Met Office Unified Model (UM) as the GCM coupled with the UK Chemistry and Aerosol framework (UKCA), as described by \citet{braam_lightning-induced_2022}. UM-UKCA is used to describe the atmospheric dynamics and chemistry for the configurations of Proxima Centauri b. We assume an aquaplanet with 1~bar surface pressure \citep[see][and references therein]{braam_lightning-induced_2022} and use a horizontal resolution of $2$ by $2.5^\circ$ in latitude and longitude. Vertically, the atmosphere extends up to an altitude of 85~km \citep[][]{yates_ozone_2020, braam_lightning-induced_2022}.

The UM is used in the Global Atmosphere 7.0 configuration \citep{walters_met_2019}, including the ENDGAME dynamical core to solve the equations of motion \citep[][]{wood_inherently_2014}. The UM is a state-of-the-art model for the prediction of Earth's weather and climate and was in recent years adapted to the study diverse exoplanets, including terrestrial planets \citep[e.g.,][]{mayne_using_2014, boutle_exploring_2017, lewis_influence_2018, yates_ozone_2020, eager_implications_2020} but also Mini-Neptunes \citep[e.g.,][]{drummond_effect_2018} and hot Jupiters \citep[e.g.,][]{mayne_unified_2014, mayne_results_2017}. Furthermore, the UM was part of the Trappist-1 e Habitable Atmosphere Intercomparison (THAI) project \citep[][]{turbet_trappist-1_2022, sergeev_trappist-1_2022, fauchez_trappist-1_2022}. Parametrized sub-grid processes include convection (mass-flux based on \citealt{gregory_mass_1990}), water cloud physics \citep[][]{wilson_pc2_2008}, turbulent mixing \citep[][]{lock_new_2000, brown_upgrades_2008} and the emergence of lightning. The parametrization of lightning flash rates follow scaling relations in terms of the cloud-top height, relating the vertical extent of a convective cloud, large-scale charge separation, and the electrical power of the cloud \citep[][]{vonnegut_facts_1963, williams_large-scale_1985, price_simple_1992, luhar_assessing_2021}. The parametrization was origininally developed to represent lightning on Earth and its first adaptation to exoplanets is described by \citet{braam_lightning-induced_2022}. Radiative transfer through the atmosphere is treated by the Suite of Community Radiative Transfer codes based on Edwards and Slingo (SOCRATES) scheme \citep[][]{edwards_studies_1996}. 

UKCA is used to simulate the global atmospheric chemical composition and its coupled interaction with the climate. It includes gas-phase chemistry and uses the UM for large-scale advection, convective transport and boundary layer mixing of the chemical tracers \citep[][]{morgenstern_evaluation_2009, oconnor_evaluation_2014, archibald_description_2020}. Additionally, the Fast-JX photolysis scheme is implemented within UKCA, calculating photolysis rates of chemical species in the atmosphere \citep[][]{wild_fast-j_2000, bian_fast-j2_2002, neu_global_2007, telford_implementation_2013}. Fast-JX uses multi-scattering eight-stream radiative transfer -- taking into account the varying optical depths of Rayleigh scattering, absorbing gases, clouds, and aerosols -- to provide an interactive treatment of photolysis in calculating the 3D distribution of chemical species in the atmosphere. Fast-JX was adapted to enhance its flexibility in terms of stellar radiation by \citet{yates_ozone_2020} and \citet{braam_lightning-induced_2022}. We distribute the stellar flux over the 18 wavelength bins of Fast-JX (covering 177--850 nm). During the CCM simulations, we rescale these fluxes to the orbital evolution of a planet, interactively calculating photolysis rates over the planetary orbit. The chemistry included is a reduced version of UKCA's Stratospheric-Tropospheric scheme \citep[StratTrop,][]{archibald_description_2020}, including the Chapman mechanism of ozone formation, and the hydrogen oxide (HO$_{\mathrm{x}}$) and nitrogen oxide (NO$_{\mathrm{x}}$) catalytic cycles. This results in 21 chemical species that are connected by 71 reactions, as elaborated in \citet{braam_lightning-induced_2022}. The atmospheres are initialized at an Earth-like atmospheric composition, using pre-industrial values of N$_2$, O$_2$, and CO$_2$. Water vapour abundances are determined interactively following evaporation from the surface ocean. The atmospheric chemistry is driven by photolysis and the atmospheric thermal structure. Additionally, lightning flashes provide a source of nitrogen oxide to drive the NO$_{\mathrm{x}}$ catalytic cycle \citep[][]{braam_lightning-induced_2022}{}{}.

The incoming stellar radiation and its passage through the atmosphere as well as the photolysis rates of chemical species will change over time with a dependence on the orbital parameters. The inclusion of positional astronomy within SOCRATES \citep[][]{edwards_studies_1996, manners_socrates_2021} and UKCA/Fast-JX is based on the description in \citet{smart_text-book_1944} and eccentricity is taken into account following the equation of time as derived by \citet{mueller_equation_1995}. A detailed description of the calculation of incoming stellar radiation is given in Appendix~\ref{subsec:isr}. 

\subsection{Planet configurations}\label{subseq:planetconfs}
Tidal interactions between the planet and its host star or other planets in the system can change a planet's spin state and orbital eccentricity. This is especially significant for the M-dwarf orbiting planets such as Proxima Centauri b \citep[][]{shields_habitability_2016}. We configure CCM simulations for two distinct orbital setups to investigate the coupled 3D evolution of the climate and atmospheric chemistry, as shown in Table~\ref{tab:orbplan_params}. The 1:1 SOR represents synchronous rotation resulting in a permanent dayside and nightside hemisphere for a planet and is the most likely orbital state for $e{\lesssim}0.2$ \citep[][]{goldreich_spin-orbit_1966, dobrovolskis_spin_2007}{}{}. For an eccentricity of ${\sim}$0.3, there is a 55\% chance that the planet orbits in a 3:2 SOR \citep[][]{dobrovolskis_spin_2007}. Therefore, we also simulate Proxima Centauri b in a 3:2 SOR with $e$=0.3.

Proxima Centauri b is the nearest known exoplanet to the Solar System orbiting at 0.0485~AU around its M5.5V host star, and was detected using the radial velocity method \citep[][]{anglada-escude_terrestrial_2016}{}{}. The configurations in Table~\ref{tab:orbplan_params} are based on those of \citet{boutle_exploring_2017} and \citet{braam_lightning-induced_2022}, with the same stellar irradiance and a spectral energy distribution from the v2.2 composite spectrum of the MUSCLES spectral survey \citep[][]{france_muscles_2016, youngblood_muscles_2016, loyd_muscles_2016}. Since the planet is non-transiting, we only have a lower limit on the planet mass of $M\sin(i)$=1.27~M$_\oplus$ \citep[][]{anglada-escude_terrestrial_2016}{}{}. Following \citet{turbet_habitability_2016}, we assume a mass of 1.4~M$_\oplus$ and use Earth's density (5.5 g~cm$^{-3}$) to calculate a radius of 1.1~R$_\oplus$ and surface gravity of 10.9 m~s$^{-2}$. For Proxima Centauri b, we simulate the atmosphere up to an altitude of 85~km \citep[][]{yates_ozone_2020, braam_lightning-induced_2022}, to include stratospheric photochemistry. Using the relation with the scale height $H$ we can translate the altitude into a top-of-the-atmosphere pressure $P_{\mathrm{TOA}}$ of:
\begin{equation}\label{eq:scaleheight}
    P_{\mathrm{TOA}} = P_0 e^{-\frac{z}{H}} = P_0 e^{-\frac{mg}{k_bT}z} = 9.8\times10^{-5} \mathrm{hPa},
\end{equation}
using $P_0$=1,000 hPa, $m{\approx}4.81\times10^{-26}$~kg for an Earth-like atmosphere, $g$=10.9 m~s$^{-2}$, and $T$=200~K or $H$=5,266~m. Proxima Centauri b falls into the Rhines rotator regime for a 1:1 SOR, as described before \citep[e.g.,][]{haqq-misra_demarcating_2018, carone_stratosphere_2018}{}{}.

Radial velocity measurements of Proxima Centauri b indicate upper limits on the eccentricity of 0.35 \citep[][]{anglada-escude_terrestrial_2016} and 0.29 \citep[][]{jenkins_proxima_2019}, with a likely value of 0.25 \citep[][]{brown_eccentricity_2017}. New orbital stability analyses of the Proxima Centauri system by \citet{livesey_orbital_2024} following the discovery of planet d \citep{faria_candidate_2022} find stable configurations for eccentricities up to 0.45 for Proxima Centauri b, but emphasize that values ${<}$0.2 are more likely. For such eccentricities, the planet may orbit in a 3:2 SOR \citep[][]{dobrovolskis_spin_2007}. To study the effects of these higher SORs on climate, habitability, and atmospheric chemistry, we configure the planet with an increased rotation rate that covers $1.5\pi$~radian in one orbital period (see the PCb 3:2 SOR in Table~\ref{tab:orbplan_params}). Due to the eccentricity, the planet's distance from the host star varies between 0.046--0.0485~AU. The impact on the radiative transfer calculations is described in Appendix~\ref{subsec:isr}. Generally, the results apply to similar exoplanets in terms of the host star, orbital configuration, atmospheric composition, and circulation regime.


\begin{table*}
	\centering
	\caption{Orbital and planetary parameters for the setups for Proxima Centauri b (PCb).}
	\label{tab:orbplan_params}
	\begin{tabular}{l|l|l} 
		\hline
		Parameter & PCb 1:1 SOR & PCb 3:2 SOR \\
		\hline
		Semi-major axis (AU) & \multicolumn{2}{c}{0.0485}  \\
		Irradiance at 1 AU (W~m$^{-2}$) &  \multicolumn{2}{c}{2.074}  \\
		Orbital Period (days) &  \multicolumn{2}{c}{11.186} \\
            Radius (R$_\oplus$) &  \multicolumn{2}{c}{1.1}  \\
            Surface gravity (m~s$^{-2}$) &  \multicolumn{2}{c}{10.9} \\
		Rotation rate (rad~s$^{-1}$) & $6.501\times10^{-6}$ & $9.7517\times10^{-6}$ \\
            Eccentricity & 0 & 0.3  \\
		Obliquity & 0 & 0  \\           
            P$_{\mathrm{TOA}}$ (hPa) & 9.0$\times10^{-5}$ & 1.3$\times10^{-4}$ \\
            \hline
	\end{tabular}
\end{table*}

\subsection{Simulation Setups}\label{subsec:simsetups}
We spin up each of the models from the initial atmospheric state as defined in Section~\ref{subsec:3dccm} to a steady state, determined by the time evolution of key atmospheric variables. The stabilisation of the surface temperature and radiative balance at the top of the atmosphere determine the dynamical steady state that takes about 2,000~Earth days (hereafter, just days). Using ozone as a long-lived species \citep[with chemical lifetimes of ${>}$25~years, see][]{braam_stratospheric_2023}{}{}, the chemical steady state was determined by stabilisation of the ozone column density and mole fraction (the amount of ozone compared to total amount of air), and takes about 4,000~days for the 1:1 SOR and 15,000~days for the eccentric 3:2 SOR. After the spin-up and an additional 2,000--3,000 days, we use 600 days of simulation for the 1:1 SOR (corresponding to ${\sim}$50~orbits for Proxima Centauri b) to determine the climatologies and steady state conditions. We use 120 days of daily output for the analysis of the 3:2 SOR. For both setups, we also use the first 1,000 simulation days to diagnose the atmospheric circulation and its effect on 3D distributions of atmospheric tracers. Lastly, we use daily outputs of the steady states for each case to produce emission spectra at different timesteps, as described in the next section.

\subsection{Planetary Spectrum Generator}\label{subsec:PSG}
We use the NASA Planetary Spectrum Generator \citep[PSG:][]{villanueva_planetary_2018}, an online radiative transfer tool, to explore the effect of 3D spatial variations in atmospheric properties on emission spectra. The PSG GlobES\footnote{\href{https://psg.gsfc.nasa.gov/apps/globes.php}{https://psg.gsfc.nasa.gov/apps/globes.php}} 3D mapping tool translates the simulated 3D data into synthetic spectra. We include CCM output for 3D distributions of temperature, pressure, water clouds, ice clouds, and gaseous chemical compounds H$_2$O, O$_3$, NO, NO$_2$, N$_2$O, HNO$_3$. Additionally, we take constant vertical abundance profiles of N$_2$, O$_2$, and CO$_2$ into account, following the CCM assumptions (see Section~\ref{subsec:3dccm}). The application of GlobES follows \citet[][]{fauchez_trappist-1_2022}{}{}, based on publicly available scripts designed for the UM\footnote{\href{https://github.com/nasapsg/globes}{https://github.com/nasapsg/globes}} that were complemented to also process output from the UM-UKCA framework\footnote{\href{https://github.com/marrickb/eccent\_3dchem\_PSJ}{https://github.com/marrickb/eccent\_3dchem\_PSJ}}.
Since emission spectra are affected by the whole observed disk of the planet, PSG weights the projected area of each latitude-longitude bin and uses a layer-by-layer pseudo-spherical ray-tracing algorithm to perform the radiative transfer calculations across the whole disk. The spectra from each latitude-longitude bin are then combined into one observed spectrum, taking the observing geometry into account. The full details of these calculations can be found in \cite{villanueva_planetary_2018}. Molecular absorptions are from the HITRAN database \citep[][]{gordon_hitran2020_2022}{}{}. The radiative transfer also accounts for collision-induced absorption (CIA) by CO$_2$-CO$_2$, H$_2$O-H$_2$O, H$_2$O-N$_2$, O$_2$-O$_2$, O$_2$-N$_2$, and N$_2$-N$_2$ pairs and for aerosol properties following Mie theory. 

From an observer's perspective, the observed hemisphere and thus the 3D distribution of atmospheric properties change as a planet moves through its orbit \citep[see e.g.,][]{olson_atmospheric_2018}{}{}, depending both on the orbital phase and inclination. Therefore, the combined time-dependence of the CCM simulations and the orbital phase angle of the planet is an important factor in generating synthetic observables. For Proxima Centauri b, the orbital inclination is unknown. \citet{kane_orbital_2017} address the dependent mass and inclination of Proxima Centauri b and determine that the planet is $\sim$85\% likely to be terrestrial, corresponding to a relatively high inclination (through the $M~sin(i)$=1.27~M$_\oplus$ dependence). For a mass of about 1.4~M$_\oplus$ we then deduce an inclination of $70^\circ$ for the generation of emission spectra, implying that the orbit is not fully face-on but that we can nevertheless probe a planetary disk throughout the full orbital phase. 

From the CCM simulations, we take the instantaneous output on a daily basis for 12 days to generate emission spectra over a full orbital period. Given Proxima Centauri b's orbital period of 11.186 days, we cover the orbit in phase angle steps of 32.1831$^\circ$ per day. The orbital phase angles corresponding to the 12 simulation days are shown in Table~\ref{tab:phase_angles_pcb}, with periastron corresponding to an arbitrary phase angle of 102.94$^\circ$ following \citet{boutle_exploring_2017}. We use the same phase angles to cover one orbit for the 3:2 SOR, but since the planet no longer rotates synchronously, the substellar longitude also changes as a function of time. For a 3:2 SOR, this corresponds to a shift in substellar longitude of 180$^\circ$ per orbit or 16.09$^\circ$ per day. Therefore, a stellar day technically takes 22.4 days. The first 11.2~days are sufficient to cover the spectral variations, since the second 11.2 days show the same cycle with a 180$^\circ$ phase shift. Therefore, we use 12 days to generate emission spectra with the substellar longitude as shown in Table~\ref{tab:phase_angles_pcb}. Both the phase angles and substellar longitude are given to PSG to calculate the viewing geometry and emission spectra.

\begin{table}
	\centering
	\caption{Orbital phase angles to cover a full orbit for Proxima Centauri b in a 1:1 and 3:2 SOR. For the 3:2 SOR, the substellar point changes as a function of time as specified by the substellar longitude.}
	\label{tab:phase_angles_pcb}
	\begin{tabular}{l|l|l} 
		\hline
		Days & Phase angle ($^\circ$) & Substellar longitude 3:2 ($^\circ$) \\
		\hline
		1    & 102.94     & 0 \\
            2    & 135.12   & 16.09 \\
            3    & 167.31   & 32.18 \\
            4    & 199.49   & 48.27 \\
            5    & 231.67   & 64.36 \\
            6    & 263.85   & 80.45 \\
            7    & 296.04   & 96.54 \\
            8    & 328.22   & 112.63 \\ 
            9    & 0.40     & 128.72 \\ 
            10   & 32.59    & 144.81 \\
            11   & 64.77    & 160.90 \\ 
            12   & 96.95    & 176.99 \\
            \hline
	\end{tabular}
\end{table}

\section{Results}\label{sec:results}
We first compare the planetary climates and the resulting ozone distributions for the simulations in Sections~\ref{sec:plan_climates} and \ref{sec:ozcols}. We identify the dominant chemical processes in the simulations in Section~\ref{sec:chemical3d}, before discussing the temporal evolution of the 3:2 SOR. Each simulation has a distinct interplay with atmospheric circulation that we describe in Section~\ref{sec:dynamical3d}. In Section~\ref{sec:observational}, we present simulated observables based on our 3D simulations.

\subsection{Planetary Climates}\label{sec:plan_climates}
\begin{figure}
\includegraphics[width=\columnwidth]{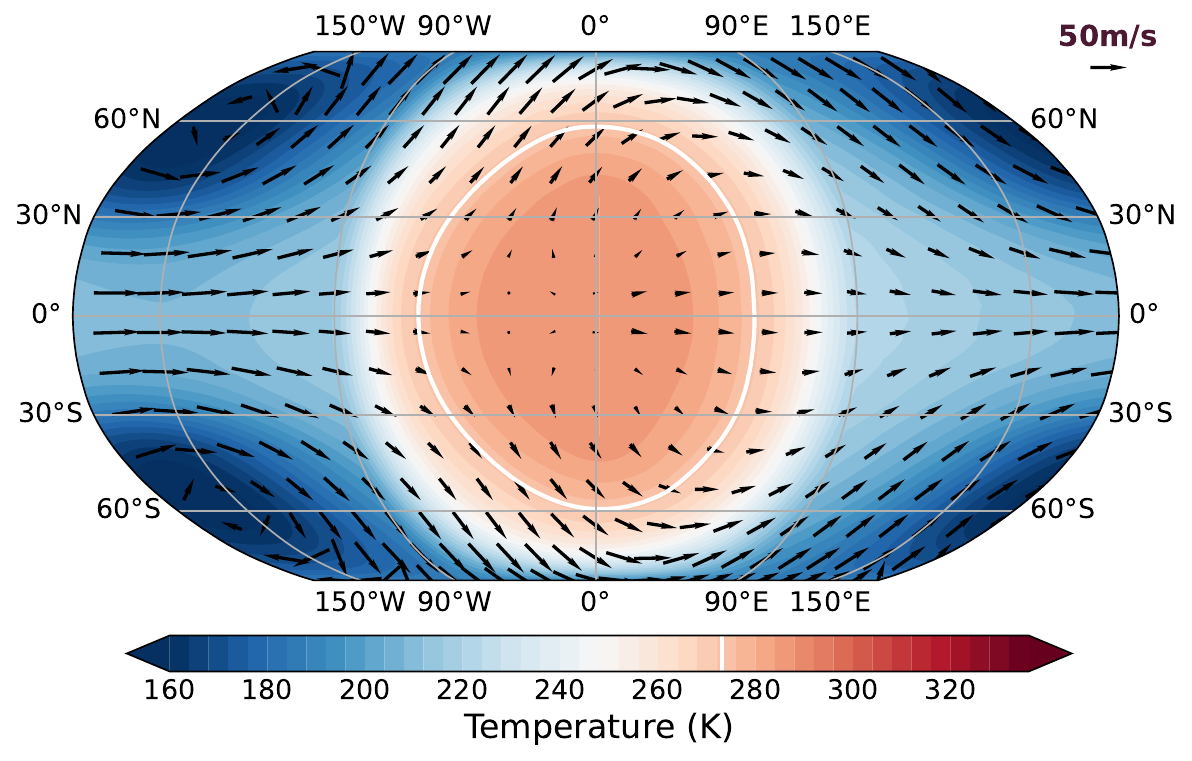}
\caption{Temporal mean surface temperature over 600 days for Proxima Centauri b in a 1:1 SOR. The substellar point (SP) is located at ($\phi,\lambda){=}(0^\circ,0^\circ)$. Overplotted are the horizontal wind vectors at P${\approx}250$~hPa, showing the equatorial jet. The white contour lines indicate surface temperatures of 273.15~K, and therefore the regions that can sustain liquid water.}
\label{fig:temp_11}
\end{figure}

The planetary climate for Proxima Centauri b in a 1:1 SOR with an Earth-like atmospheric composition has been investigated extensively \citep[e.g.,][]{turbet_habitability_2016, boutle_exploring_2017, del_genio_habitable_2019}{}{}, focusing on the dayside-nightside distinctions. For the sake of the discussion and the comparison to a 3:2 SOR, we show the temporal mean distribution of the surface temperature from the CCM simulations in Figure~\ref{fig:temp_11}. The global mean temperature is 229~K and goes from as low as 156~K in the nightside gyres (see also \citealt{braam_stratospheric_2023}) up to a maximum of 288~K. Taking means for the dayside and nightside hemisphere, we find a hemispheric difference of 67~K. The distribution looks similar to \citet{boutle_exploring_2017}, with temperature gradients mainly in the zonal direction, characteristic of the Rhines rotator circulation regime \citep[][]{haqq-misra_demarcating_2018}{}{}. Importantly, the wind vectors indicate the single equatorial jet that characterises the circulation regime of Proxima Centauri b \citep[see e.g.,][]{carone_stratosphere_2018}{}{}. 

\begin{figure}
\includegraphics[width=\columnwidth]{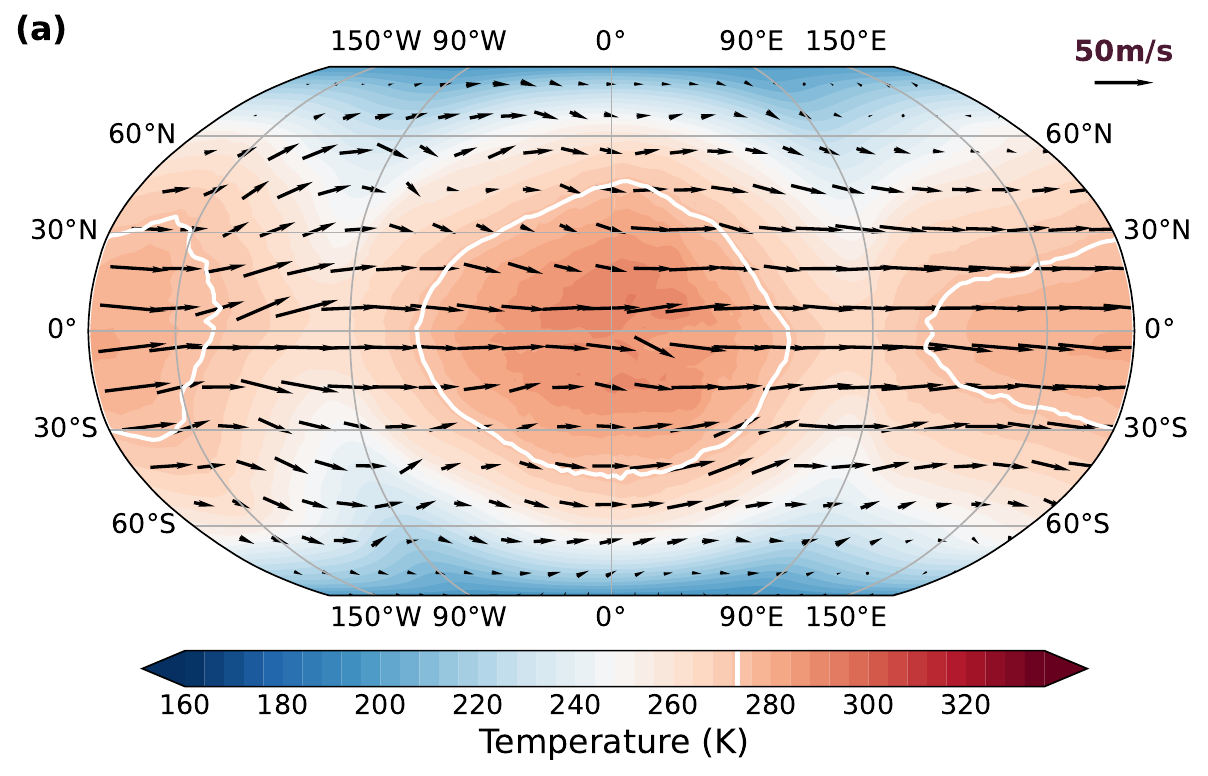}
\includegraphics[width=\columnwidth]{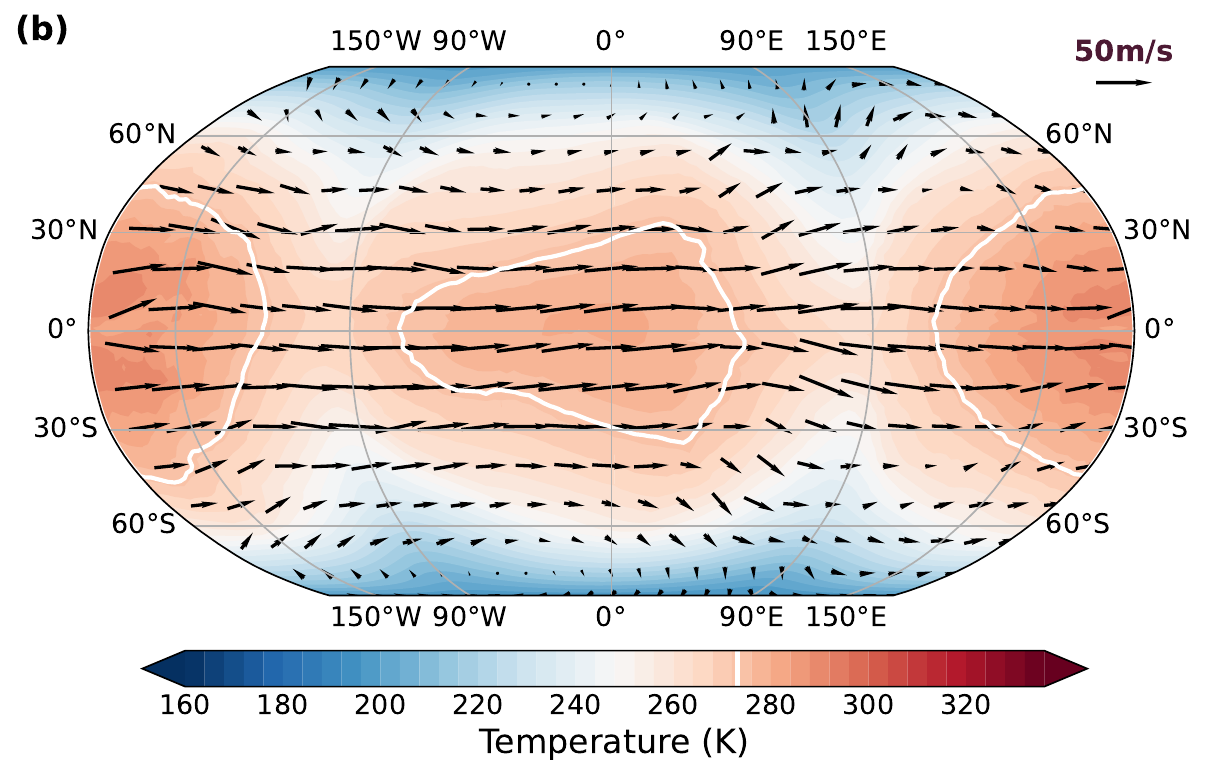}
\caption{Instantaneous surface temperature for Proxima Centauri b in a 3:2 SOR for (a) the substellar point at 0$^{\circ}$ longitude and (b) the substellar point at 180$^{\circ}$ longitude. The horizontal wind vectors are again shown at P${\approx}250$~hPa}
\label{fig:pcb_32_temp}
\end{figure}

When simulating the planet in a 3:2 SOR, the substellar point is no longer fixed. Instead, the substellar point shifts by 180$^\circ$ for every orbit around the star. Figure~\ref{fig:pcb_32_temp} shows how the eccentric 3:2 SOR impacts the surface temperature for Proxima Centauri b. The planet has two warm regions, one centred at 0$^\circ$ and one at 180$^\circ$ longitude. Defining the 0$^\circ$ hemisphere as the region corresponding to longitudes smaller than 90$^\circ$ east and west (see Figure~\ref{fig:pcb_32_temp}) and the 180$^\circ$ hemisphere for longitudes larger than 90$^\circ$ east and west, we find hemispheric differences of only 4.1~K (Figure~\ref{fig:pcb_32_temp}a) and ${-}$4.2~K (Figure~\ref{fig:pcb_32_temp}b) depending on the timestep. These small hemispheric differences illustrate the enhanced homogeneity across the planet as compared to the 1:1 SOR. We show the surface temperature for the substellar point at 0$^\circ$ longitude in Figure~\ref{fig:pcb_32_temp}a and for the substellar point at 180$^\circ$ longitude in Figure~\ref{fig:pcb_32_temp}b. Both timesteps go from a minimum of 196~K to maximum of 292~K, with a global mean surface temperature of 262~K, a 14.4\% increase compared to the 1:1 SOR. The white contour lines show that both hemispheres have regions above the freezing point of water (273.15~K). Temperature gradients are now predominantly visible in the meridional direction, with temperature decreasing as we move to higher latitudes.

\subsection{Ozone column density}\label{sec:ozcols}
\begin{figure}
\includegraphics[width=\columnwidth]{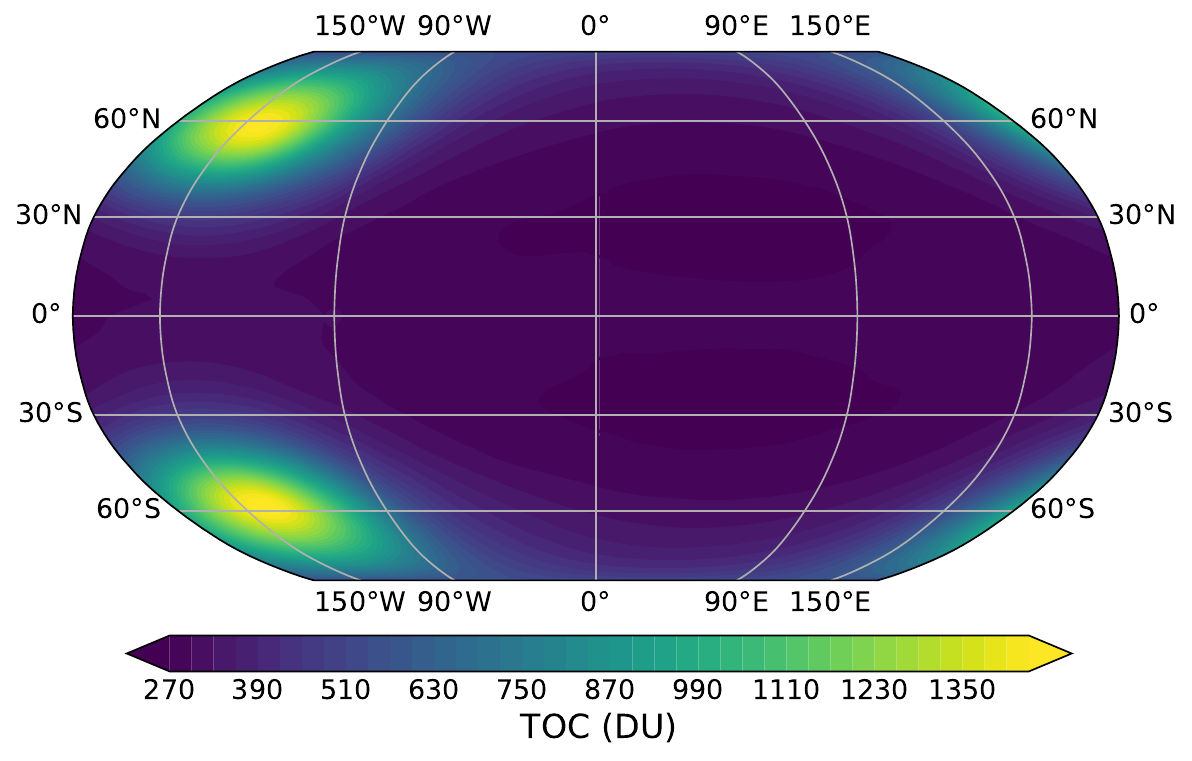}
\caption{Mean over 600 days of the vertically integrated ozone column density for Proxima Centauri b in a 1:1 SOR.}
\label{fig:toc_11}
\end{figure}

In Figure~\ref{fig:toc_11} we show the vertically integrated ozone column density (hereafter ozone column) for Proxima Centauri b in a 1:1 SOR, which shows zonal gradients in line with the surface temperature distribution in Figure~\ref{fig:temp_11}. The global mean ozone column is 387~DU (1~DU${=}2.687\times10^{20}$~molecules~m$^{-2}$), ranging from a minimum dayside value of 262~DU to a maximum of 1466~DU at the location of the nightside gyres. The 3D distribution is driven by a stratospheric dayside-to-nightside circulation, as shown by \citet{braam_stratospheric_2023}.

\begin{figure}
\includegraphics[width=\columnwidth]{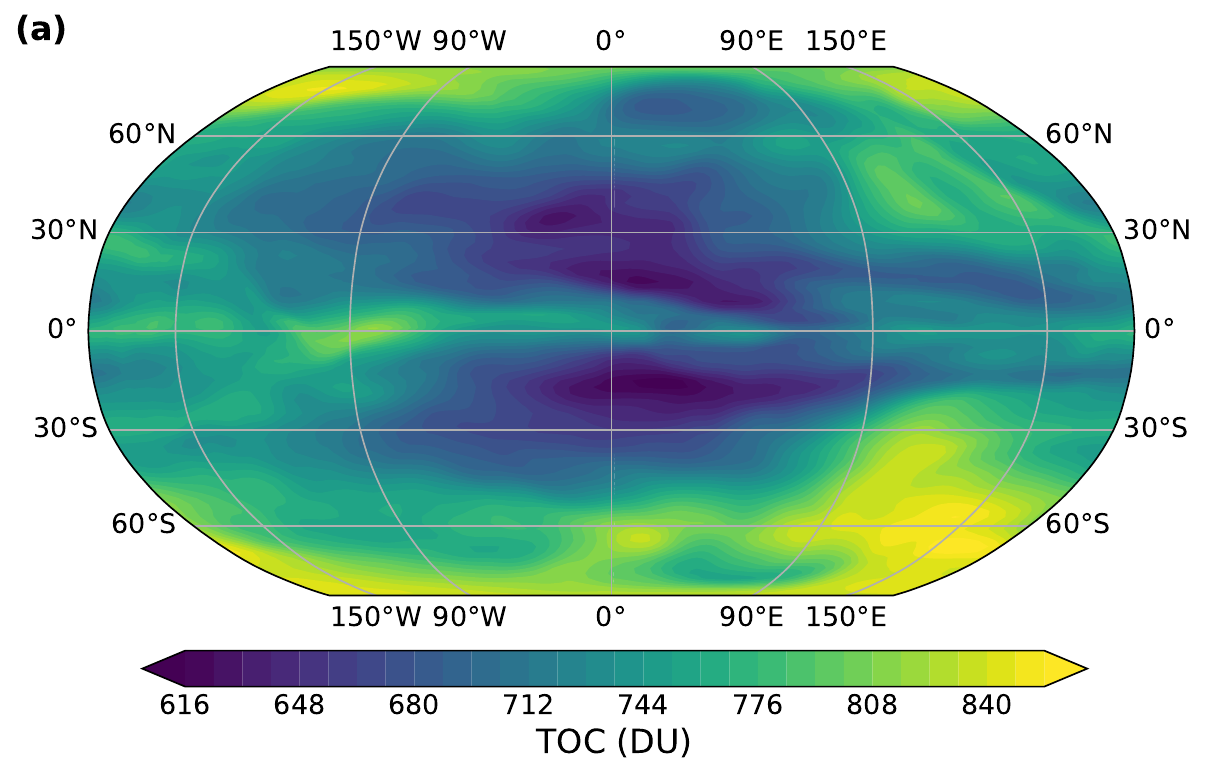}
\includegraphics[width=\columnwidth]{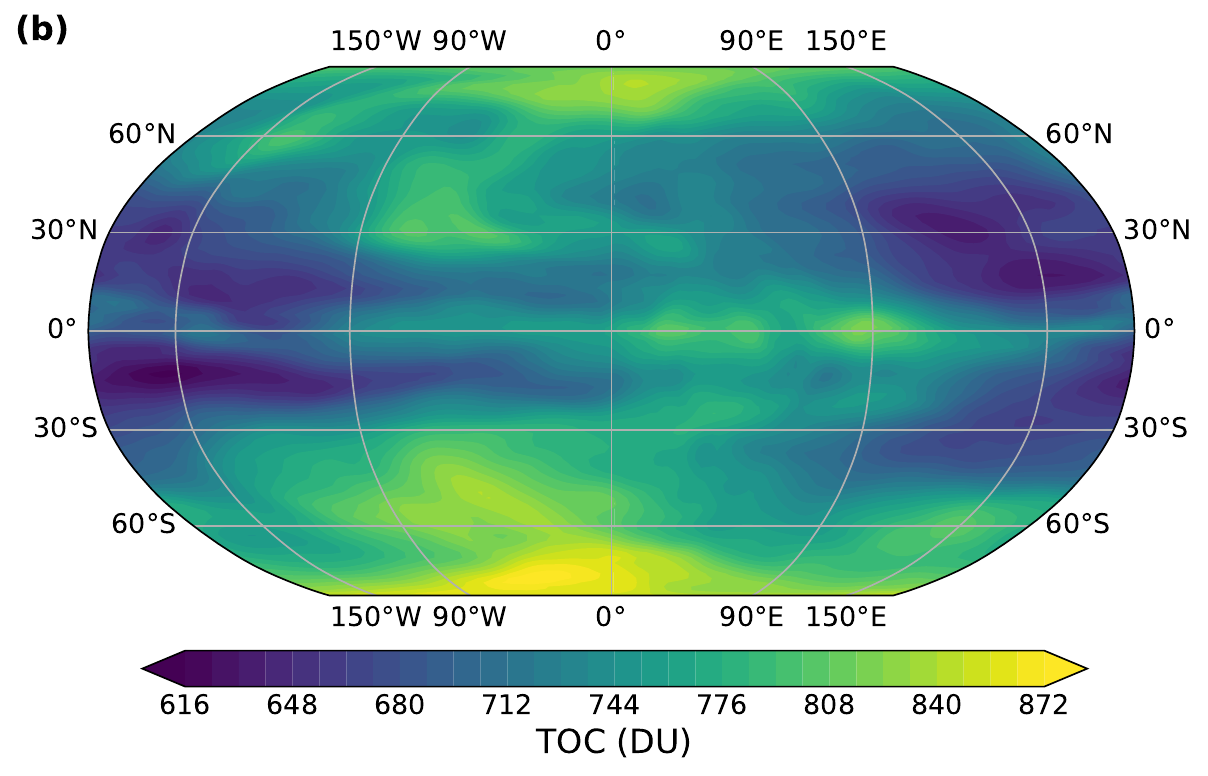}
\caption{Instantaneous vertically integrated ozone column density for Proxima Centauri b in a 3:2 SOR for (a) daytime on the 0$^{\circ}$ hemisphere and (b) daytime on the 180$^{\circ}$ hemisphere.}
\label{fig:pcb_32_toc}
\end{figure}

For Proxima Centauri b in a 3:2 SOR, we instead find meridional gradients in the ozone column as shown in Figure~\ref{fig:pcb_32_toc}, again quantitatively matching the gradients in surface temperature. Comparing Figures~\ref{fig:pcb_32_toc}a and \ref{fig:pcb_32_toc}b, we deduce that the smallest ozone column is found on the daytime hemisphere, which is centred at 0$^\circ$ and 180$^\circ$ in Figures~\ref{fig:pcb_32_toc}a and \ref{fig:pcb_32_toc}b, respectively. Both timesteps have a global mean ozone column of 731~DU, but vary in their minima of 611~DU (a) and 613~DU (b) and maxima of 864~DU (a) and 878~DU (b). Compared to the 1:1 SOR, Proxima Centauri b in a 3:2 SOR shows a much smaller spread in ozone column values, along with an almost doubled global mean. This indicates that there is more homogeneity across the planet for a 3:2 SOR. The spatial distribution of the ozone column is circulation-driven, as we will show in Section~\ref{sec:dynamical3d}. However, the 3:2 SOR induces variability that is visible in the global extrema of both timesteps and, when comparing Figures~\ref{fig:pcb_32_toc}a and \ref{fig:pcb_32_toc}b, on each hemisphere separately. We will further explore this variability in Section~\ref{sec:32res_tevol}.

\subsection{Comparing the dominant chemical processes that determine ozone}\label{sec:chemical3d}
In this section, we analyse the vertical distribution of ozone in the simulations of Proxima Centauri b by comparing the abundances of key chemical species and the most important reactions for the formation and destruction of ozone. Here, we focus on hemispheric means and refer the interested reader to an in-depth analysis of relevant spatial distributions in Appendix~\ref{sec:appendix_chem}. Three chemical mechanisms are of key importance to explain the abundance of ozone molecules in an Earth-like atmospheric composition \citep[for previous work in the context of M-dwarf radiation, see e.g.,][]{segura_biosignatures_2005, grenfell_sensitivity_2014, rugheimer_effect_2015, harman_abiotic_2018, chen_biosignature_2018, yates_ozone_2020, braam_lightning-induced_2022, kozakis_is_2022,ridgway_3d_2023}{}{}. First, the photolysis of O$_2$ drives the Chapman mechanism of ozone formation \citep[][]{chapman_xxxv_1930}{}{}:\\
\\
\ce{O$_2$ + h$\nu$ -> O($^3$P) + O($^3$P)}, \hfill (R1) \\
\ce{O($^3$P) + O$_2$ + M -> O$_3$ + M}, \hfill (R2) \\
\ce{O$_3$ + h$\nu$ -> O$_2$ + O($^3$P)}, \hfill (R3) \\
\ce{O$_3$ + O($^3$P) -> O$_2$ + O$_2$}.  \hfill (R4) \\

Reaction~R1 initiates the production of ozone and Reaction~R4 represents the termination step by destroying ozone. Reactions~R2 and R3 describe the rapid interchange between atomic O($^3$P), O$_2$, and ozone. Once formed, ozone can then be destroyed by a number of catalytic cycles \citep[][]{bates_photochemistry_1950, crutzen_influence_1970, johnston_reduction_1971, grenfell_chemical_2006}{}{}, that are commonly grouped as the HO$_\mathrm{x}$(=H+OH+HO$_2$) and NO$_\mathrm{x}$(=NO+NO$_2$) cycles, presenting the other main chemical mechanisms. In such catalytic cycles, ozone is destroyed without consuming the catalyst species (in this case, HO$_\mathrm{x}$ or NO$_\mathrm{x}$) that remain available for another cycle of ozone destruction.


The HO$_\mathrm{x}$ cycles are initiated by the production of OH from the photolysis or oxidation of water vapour:\\
\\
\ce{H$_2$O + h$\nu$  -> OH + H}, \hfill (R5)\\
\ce{H$_2$O + O($^1$D) -> 2OH}, \hfill (R6)\\
\\
and are terminated when HO$_\mathrm{x}$ species react to destroy themselves:\\
\\
\ce{OH + HO$_2$ -> H$_2$O + O$_2$}. \hfill (R7)\\

NO$_\mathrm{x}$ species are produced by lightning in our simulations and are destroyed by the formation of so-called reservoir species such as HNO$_3$ that are deposited by rainfall \citep[see Appendix~\ref{sec:appendix} and][]{braam_lightning-induced_2022}. 

\begin{figure*}
\includegraphics[width=2\columnwidth]{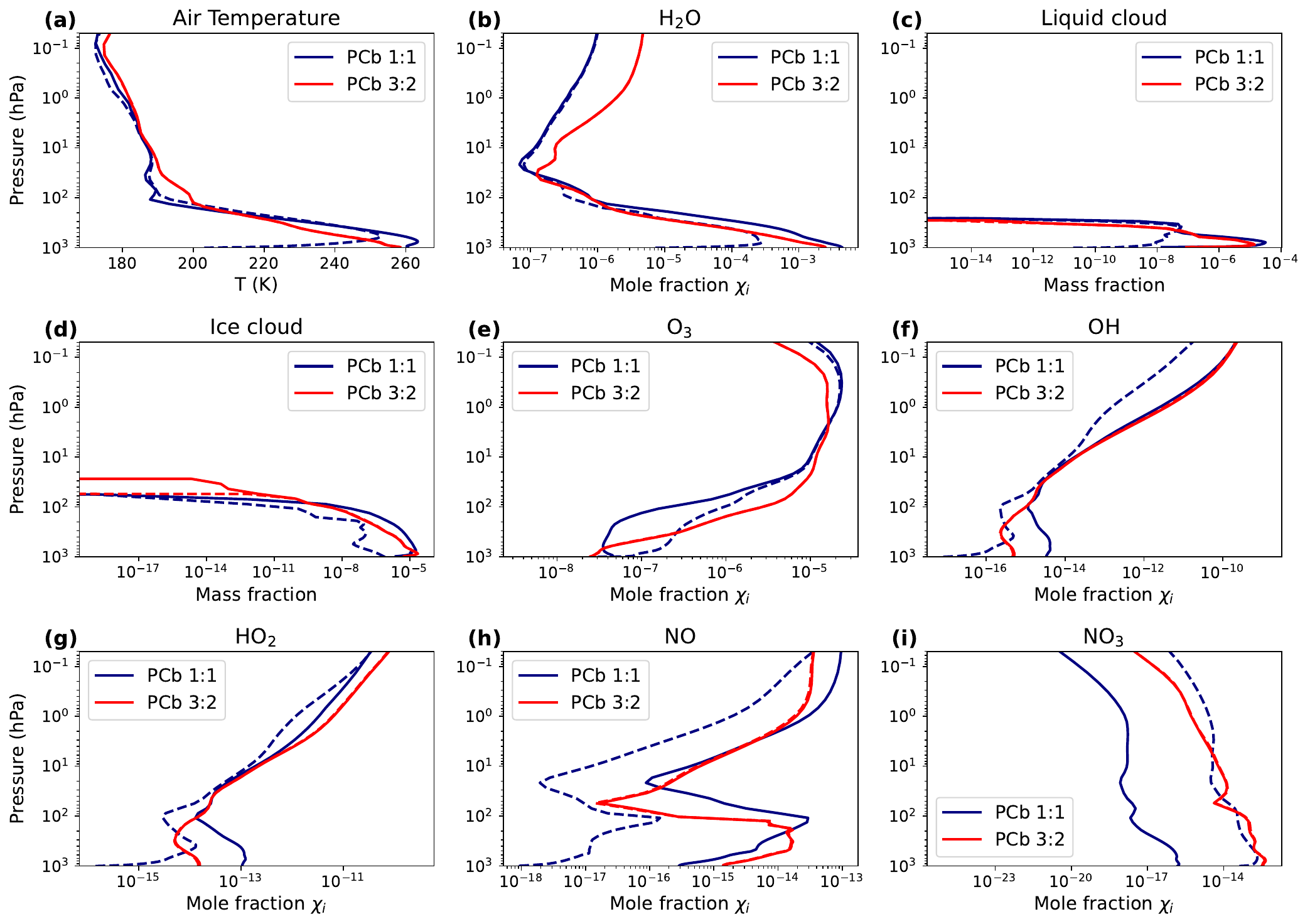}
\caption{Temporal and hemispheric means of the vertical distribution of air temperature, liquid and ice cloud abundances, and mole fractions $\chi_i$ of chemical species i related to ozone formation in both simulations. Each plot has a title corresponding to the variable or chemical species shown. Solid lines indicate the mean over the 0$^\circ$ hemisphere and dashed lines the mean over the 180$^\circ$ hemisphere, representing the dayside and nightside in 1:1 SOR. The vertical distributions illustrate important differences between a 1:1 and 3:2 SOR for Proxima Centauri b (PCb).}
\label{fig:tair_vmrs_intercomp}
\end{figure*}

Figure~\ref{fig:tair_vmrs_intercomp} shows the hemispheric mean vertical distributions of the air temperature, liquid and ice cloud abundances, and the mole fractions $\chi_i$ of chemical species \textit{i} as indicated by the title of each panel. The air temperatures in Figure~\ref{fig:tair_vmrs_intercomp}a illustrate distinctions depending on the SOR. For the 1:1 SOR, we see hemispheric contrasts in temperature for P${>}100$~hPa (or the troposphere), comparing the solid lines (0$^\circ$ hemisphere) and dashed lines (180$^\circ$ hemisphere), in contrast to homogeneity for the 3:2 SOR. We also see that a 3:2 SOR results in a warmer stratosphere for P${<}100$~hPa in addition to the higher global mean surface temperature compared to a 1:1 SOR (see also Section~\ref{sec:plan_climates}). The distribution of water vapour in Figure~\ref{fig:tair_vmrs_intercomp}b illustrates that a 3:2 SOR leads to a much wetter atmosphere than a 1:1 SOR, again especially in the stratosphere. The amount of stratospheric water vapour determines the inner edge of the HZ \citep[][]{simpson1928some, ingersoll_runaway_1969, kasting_runaway_1988, nakajima_study_1992, goldblatt_low_2013}{}{}. Together with the enhanced surface and stratospheric temperatures, the water vapour abundances indicate a sensitivity of the HZ inner edge to a planet's SOR. Whilst the dayside-nightside distinction in troposphere water vapour is clear for the 1:1 SOR, the 3:2 SOR is more homogeneous (see also the spatial distribution in Appendix Figure~\ref{fig:pcb_11_32_h2o}). Similar patterns are visible in the vertical and hemispheric distribution of clouds in Figures~\ref{fig:tair_vmrs_intercomp}c-d.

The vertical distribution of ozone in Figure~\ref{fig:tair_vmrs_intercomp}e shows that the balance between photochemical production and chemical destruction leads to a stratospheric ozone layer on Proxima Centauri b, with a vertical structure depending on the SOR. Ozone peaks at $\chi_{O_\mathit{3}}{\simeq}10$~ppm and, again, the 3:2 SOR shows similar profiles for both hemispheres. Hemispheric differences are visible for the 1:1 SOR, distinctly corresponding to the zonal gradients for Proxima Centauri b (Figure~\ref{fig:toc_11}).

The vertical profiles of OH and HO$_2$ in panels f and g of Figure~\ref{fig:tair_vmrs_intercomp} show that despite similar amounts of daytime water vapour in Proxima Centauri b's troposphere (P${>}$100~hPa) more OH and HO$_2$ is produced on the dayside of the planet in a 1:1 SOR (see Appendix~\ref{sec:appendix_chem} for the spatial distribution). The wetter stratosphere for the 3:2 SOR results in higher OH and HO$_2$ abundances here. From the profiles of NO in Figure~\ref{fig:tair_vmrs_intercomp}h we observe higher NO abundances in the 3:2 SOR for P${>}$200~hPa. Since NO in the lower atmosphere is mainly produced by lightning discharges \citep[][]{braam_lightning-induced_2022}{}{}, this indicates a higher lightning activity and subsequent chemical impact. The slightly taller ice cloud structures for the 3:2 SOR (Figure~\ref{fig:tair_vmrs_intercomp}d) support this idea, as the lightning flash rates in our simulation are parameterised in terms of cloud-top heights \citep[][]{braam_lightning-induced_2022}{}{}, and lead to strong NO emissions. For reference, the associated lightning activities are compared in Appendix~\ref{sec:appendix}. Lightning flashes are mainly found on the dayside of exoplanets in a 1:1 SOR. The dayside production of NO by lightning flashes and the nightside accumulation in long-lived reservoir species such as NO$_3$, HNO$_3$, and N$_2$O$_5$ result in the hemispheric differences in NO as shown in Figure~\ref{fig:tair_vmrs_intercomp}h. Planets in a 3:2 SOR have lightning flashes at all longitudes and a constantly changing daytime hemisphere, thus producing a relatively homogeneous profile in NO. Global NO$_3$ is also enhanced for the 3:2 SOR (Figure~\ref{fig:tair_vmrs_intercomp}i). The dayside-nightside differences in the reservoir species NO$_3$ (see Appendix~\ref{sec:appendix_chem}) for planets in a 1:1 SOR result from its long lifetime on nighttime hemispheres in the absence of stellar radiation. 

\begin{figure*}
\includegraphics[width=2\columnwidth]{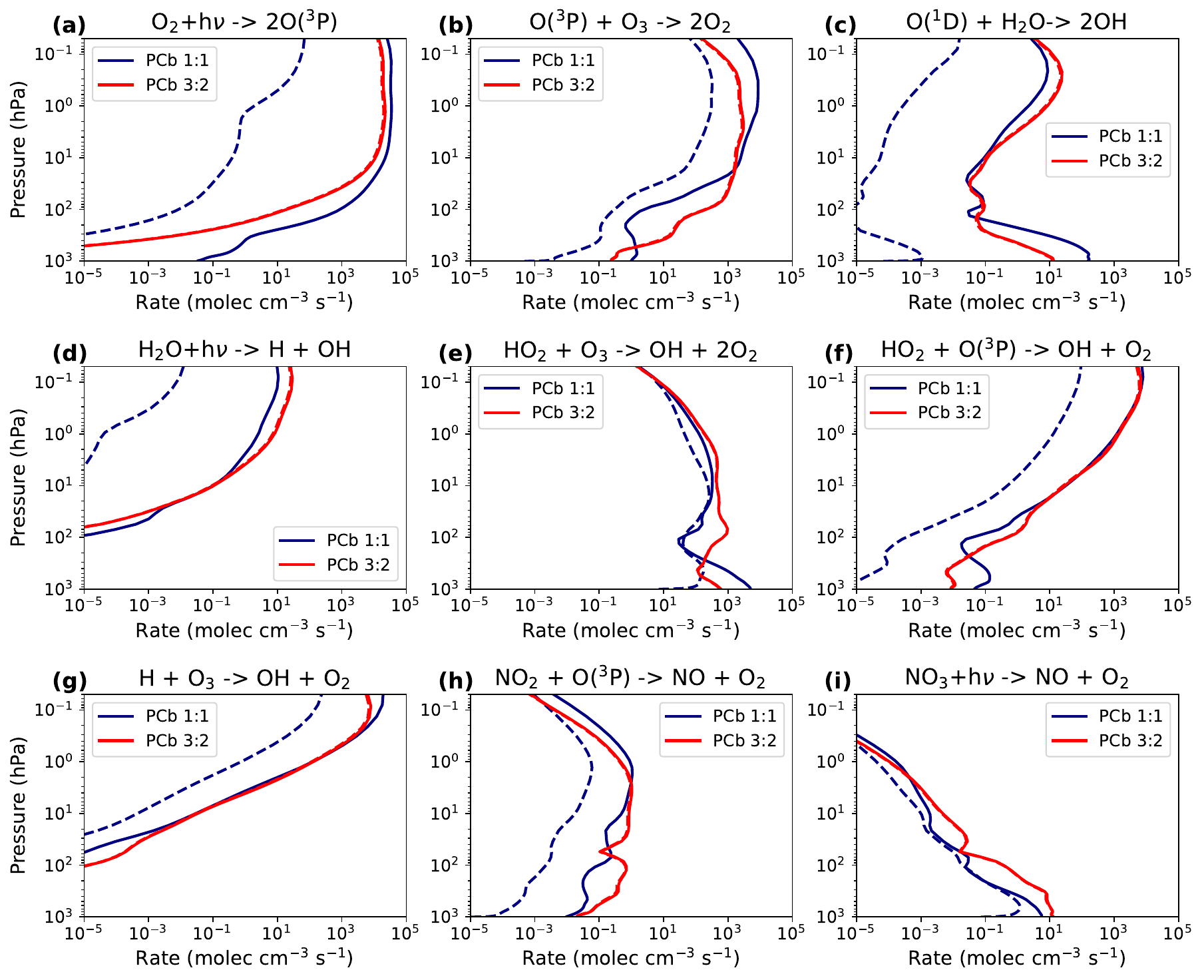}
\caption{Temporal and hemispheric means of the vertical distribution of selected reaction rates for both simulations of Proxima Centauri b (PCb). Each plot title corresponds to the chemical reaction shown. Solid lines indicate the mean over the 0$^\circ$ hemisphere and dashed lines the mean over the 180$^\circ$ hemisphere, representing the dayside and nightside in a 1:1 SOR. The reactions were selected to include the Chapman mechanism of ozone formation (panel a and b), and the rate-limiting steps for HO$_\mathrm{x}$ catalytic cycles (panel c-g) and NO$_\mathrm{x}$ catalytic cycles (panel h and i).}
\label{fig:rratesz_intercomp}
\end{figure*}

To support the interpretation of chemical abundance distributions, we plot the hemispheric means of a selection of reaction rates in Figure~\ref{fig:rratesz_intercomp}. These describe the three main mechanisms for ozone chemistry, including Reaction R1 (Figure~\ref{fig:rratesz_intercomp}a) and R4 (Figure~\ref{fig:rratesz_intercomp}b) of the Chapman mechanism, the initiation reactions (R5 and R6, Figure~\ref{fig:rratesz_intercomp}c and \ref{fig:rratesz_intercomp}d) and rate-limiting steps for three HO$_\mathrm{x}$ catalytic cycles (Figure~\ref{fig:rratesz_intercomp}e--g) and two NO$_\mathrm{x}$ catalytic cycles (Figure~\ref{fig:rratesz_intercomp}h and \ref{fig:rratesz_intercomp}i). 

From Figure~\ref{fig:rratesz_intercomp}a we infer that O$_2$ photolysis rates are highest on the dayside of the 1:1 SOR throughout most of the atmosphere, producing a larger amount of ozone through the Chapman mechanism (Reactions R1--R4). Additionally, we see that nightside photolysis rates for a 1:1 SOR are 3--6 orders of magnitude lower than on the dayside \citep[see also][]{yates_ozone_2020}{}{}, but non-zero due to scattered light from the dayside atmosphere (see Appendix~\ref{sec:appendix_chem}). The Chapman termination reaction in Figure~\ref{fig:rratesz_intercomp}b depends on the vertical profiles of ozone (Figure~\ref{fig:tair_vmrs_intercomp}e) and O($^3$P). The O($^3$P) abundance is driven by photochemistry (Figure~\ref{fig:rratesz_intercomp}a) and therefore the Chapman termination reaction is faster on the dayside for the 1:1 SOR. The termination reaction represents the dominant destruction mechanism between 0.3--40~hPa (1:1 SOR) and 0.4--60~hPa (3:2 SOR) since it exceeds all other ozone destruction reactions at these levels.

The photolysis and oxidation of water vapour (Figures~\ref{fig:rratesz_intercomp}c--d) drive the production of OH, determining the OH profile (Figure~\ref{fig:tair_vmrs_intercomp}f) and initiating the HO$_\mathrm{x}$ catalytic cycles that destroy ozone (or atomic oxygen, preventing ozone from forming). H$_2$O photolysis is stronger for P${<}$10~hPa, whereas H$_2$O oxidation dominates for P${>}$10~hPa. Like \citet{braam_lightning-induced_2022}, we find two HO$_\mathrm{x}$ cycles that are dominant in different altitude ranges. First, for P${>}$40 (1:1) or ${>}$60~hPa (3:2):\\
\\
\ce{OH + O$_3$ -> HO$_2$ + O$_2$}, \hfill (R8)\\
\textbf{\ce{HO$_2$ + O$_3$ -> OH + 2O$_2$}, \hfill (R9)}\\
\hrule width5cm
\noindent Net: \ce{2O$_3$ -> 3O$_2$}, \hfill \\

\noindent where the rate-limiting step is indicated in bold. This rate-limiting step is shown in Figures~\ref{fig:rratesz_intercomp}e. Instead of reacting with O$_3$, HO$_2$ can also react with O($^3$P) to complete another catalytic cycle. However, Figure~\ref{fig:rratesz_intercomp}f shows that -- for P${>}$0.3~hPa -- this cycle has slower reactions than Reaction R9 in Figure~\ref{fig:rratesz_intercomp}e or the Chapman termination reaction in Figure~\ref{fig:rratesz_intercomp}b. In the upper atmosphere (P${<}$0.3 or ${<}$0.4~hPa), another HO$\mathrm{x}$ cycle dominates:\\
\\
\ce{OH + O($^3$P) -> H + O$_2$}, \hfill (R10)\\
\textbf{\ce{H + O$_3$ -> OH + O$_2$}, \hfill (R11)}
\hrule width5cm
\noindent Net: \ce{O$_3$ + O($^3$P) -> O$_2$}. \hfill \\

The rate-limiting steps in these HO$_\mathrm{x}$ cycles (Figures~\ref{fig:rratesz_intercomp}e--g) are generally faster for the 3:2 SOR, which also follows from the higher rates of H$_2$O photolysis and oxidation. However, in the troposphere for P${>}$150~hPa, the 1:1 SOR has stronger water vapour oxidation. The associated HO$_\mathrm{x}$ cycling decreases the tropospheric ozone abundances (Figure~\ref{fig:tair_vmrs_intercomp}e) and thus partly explains the lower global mean ozone column for a 1:1 SOR, due to the dependence on pressure for the vertically integrated column density (see Section~\ref{sec:ozcols}). The amount of tropospheric ozone depends on transport from the stratosphere in the first place, thus indicating a coupled balance between photochemistry, transport, and HO$_\mathrm{x}$ cycling due to the presence of water vapour.

We identify two potential NO$_\mathrm{x}$ catalytic cycles that can affect ozone abundances. The first cycle is:\\
\\
\ce{NO + O$_3$ -> NO$_2$ + O$_2$}, \hfill (R12)\\
\textbf{\ce{NO$_2$ + O($^3$P) -> NO + O$_2$}, \hfill (R13)}
\hrule width5cm
\noindent Net: \ce{O$_3$ + O($^3$P) -> 2O$_2$}. \hfill \\

\noindent The second cycle also involves the formation of NO$_3$:\\
\\
\ce{NO + O$_3$ -> NO$_2$ + O$_2$}, \hfill (R14)\\
\ce{NO$_2$ + O$_3$ -> NO$_3$ + O$_2$}, \hfill (R15)\\
\textbf{\ce{NO$_3$ + h$\nu$ -> NO + O$_2$}, \hfill (R16)}
\hrule width5cm
\noindent Net: \ce{2O$_3$ -> 3O$_2$}. \hfill \\

Figures~\ref{fig:rratesz_intercomp}h--i show that the rate-limiting steps of the NO$_\mathrm{x}$ cycles are faster for the 3:2 SOR, following the enhanced production of NO by lightning discharges (Figure~\ref{fig:tair_vmrs_intercomp}h).
Nightside NO$_3$ photolysis rates (Reaction R16) are only slightly smaller on the nightside of the 1:1 SOR (Figure~\ref{fig:rratesz_intercomp}i), which is due to a combination of scattered light reaching the nightside hemisphere where NO$_3$ is much more abundant than on the dayside (Figure~\ref{fig:tair_vmrs_intercomp}i). The spatial distributions of NO$_3$ and the reaction rate for Reaction R16 in Appendix~\ref{sec:appendix_chem} reinforce these findings. Nevertheless, the HO$_\mathrm{x}$ cycles are at least a hundred times faster at all pressure levels and thus remain the dominant ozone destruction process.

In summary, in terms of decreasing pressure (or increasing altitude) for both a 1:1 and 3:2 SOR, ozone destruction for Proxima Centauri b is controlled by HO$_\mathrm{x}$ cycling (Figure~\ref{fig:rratesz_intercomp}e), the Chapman mechanism (Figure~\ref{fig:rratesz_intercomp}b), and again HO$_\mathrm{x}$ cycling (Figure~\ref{fig:rratesz_intercomp}g).

\subsection{Temporal evolution of the spin-orbit resonances}\label{sec:32res_tevol}
\begin{figure*}
\centering
\includegraphics[width=1.7\columnwidth]{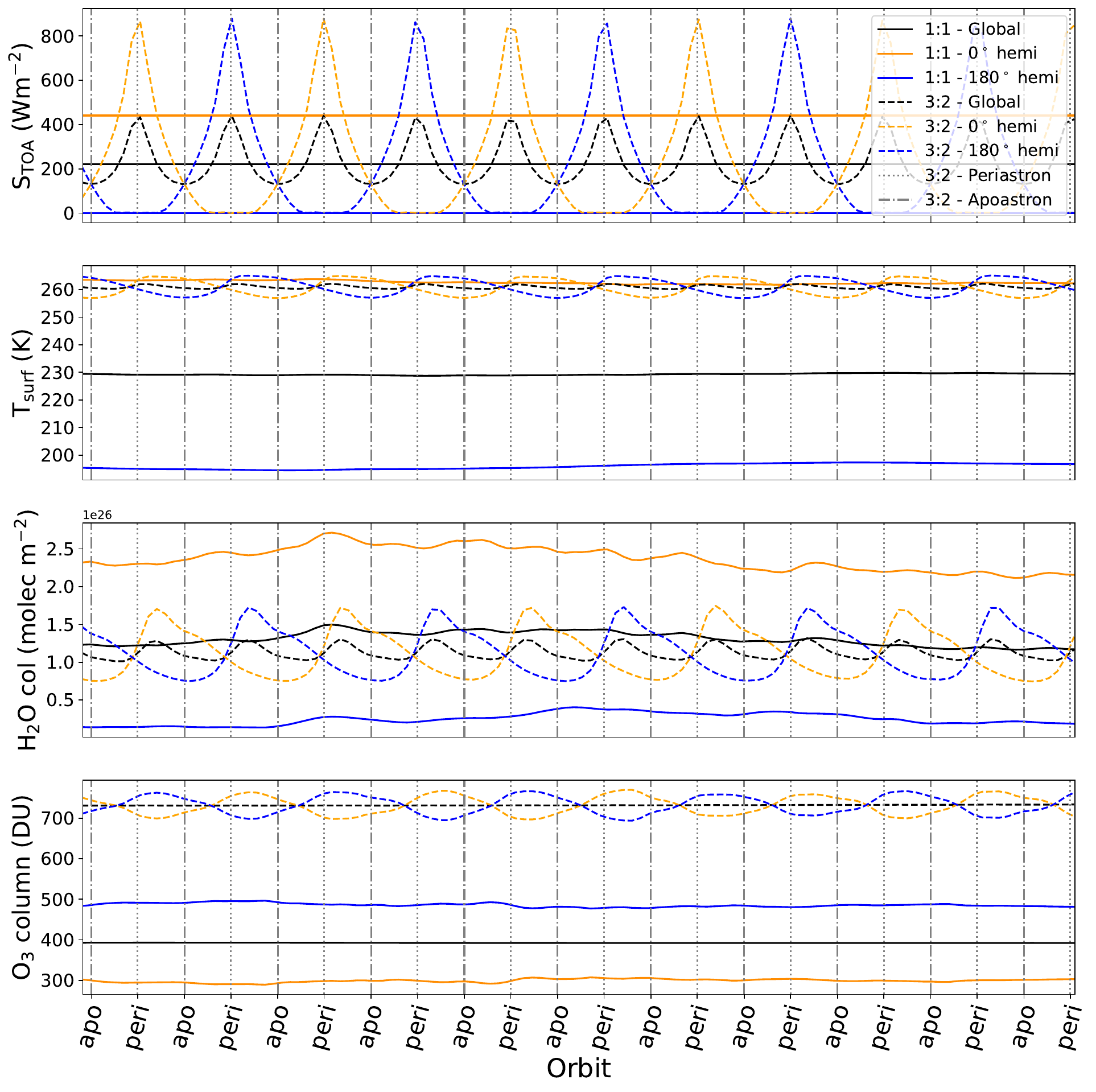}
\caption{Temporal evolution of incoming stellar radiation at the top of the atmosphere ($S_{TOA}$, panel a), surface temperature (b), and vertically integrated water vapour (c) and ozone column densities (d) on Proxima Centauri b in a 1:1 SOR (solid lines) and a 3:2 SOR (dashed lines) over 120~days, corresponding to almost 11~orbits. Orange and blue lines indicate the 0$^{\circ}$ and 180$^{\circ}$ hemispheres, respectively. The global average values are shown as the black line. From the peaks in $S_{TOA}$, we identify the periastron and apoastron passages for the 3:2 SOR, denoted as the dotted and dashdotted grey vertical lines and specified on the $x$-axis. The separation between two successive periastron or apoastron passages represent an orbital period for Proxima Centauri b, independent of the exact SOR.}
\label{fig:pcb_32res_tevol}
\end{figure*}

The calculation for Proxima Centauri in a 3:2 SOR displays variability in the global mean values and spatial distributions of temperature (Figures~\ref{fig:pcb_32_temp}a and \ref{fig:pcb_32_temp}b) and ozone column (Figures~\ref{fig:pcb_32_toc}a and \ref{fig:pcb_32_toc}b). This is supported by the dashed lines in Figure~\ref{fig:pcb_32res_tevol}, showing the temporal evolution of four key properties of the planet: the incoming stellar radiation at the top of the atmosphere ($S_{TOA}$), the surface temperature, the vertically integrated water vapour column density (hereafter H$_2$O(g) column), and the ozone column. For each property, we calculate the mean of the global region (black), the 0$^\circ$ hemisphere centred at 0$^\circ$ latitude and longitude (orange) and the 180$^\circ$ hemisphere centred at 180$^\circ$ latitude and 0$^\circ$ longitude (blue). The temporal evolution of the 1:1 SOR is shown as solid lines, also showing the global and hemispheric means.

$S_{TOA}$ in Figure~\ref{fig:pcb_32res_tevol}a first shows the dayside-nightside contrast for the 1:1 SOR: the maximum $S_{TOA}$ is found on the dayside, whereas the nightside is devoid of incoming radiation. The incoming radiation for the 1:1 SOR is constant. For the 3:2 SOR, Figure~\ref{fig:pcb_32res_tevol}a illustrates the change in the daytime hemisphere: at the first periastron shown (dotted grey lines), $S_{TOA}$ peaks on the 0$^\circ$ hemisphere and at the second periastron, $S_{TOA}$ peaks on the 180$^\circ$ hemisphere. When the planet is at periastron, there is no incoming radiation on the nighttime hemisphere and thus $S_{TOA}{=}0$ there. This daytime-nighttime cycle repeats over the orbital evolution. The passages of apoastron (dashdotted grey lines) represent a minimum in the global mean $S_{TOA}$. The fluctuations in the global mean $S_{TOA}$ exhibit the effect of the orbital eccentricity affecting the star-planet separation.

The varying amount of incoming radiation affects the surface temperature for the 3:2 SOR, as shown in Figure~\ref{fig:pcb_32res_tevol}b. The peaks in surface temperature occur just after periastron passage, whereas the lowest surface temperatures occur after apoastron passage, for the global and the hemispheric means. The time lag between $S_{TOA}$ and temperature extremes indicates the response time of the atmosphere to a change in radiation. We see global temperature variations of up to 1.2~K (0.45\% compared to the global mean) due to the eccentric orbit. Hemispheric variations can reach up to 8~K between daytime peaks and nighttime dips. For 120~days of the 1:1 SOR, we find a non-periodical variability of 0.22\% for the global mean on much longer timescales compared to variations for the 3:2 SOR, along with hemispheric differences in temperature as high as 67~K. Hence, this again shows enhanced hemispheric homogeneity across the planet for a 3:2 SOR.

The H$_2$O(g) column in Figure~\ref{fig:pcb_32res_tevol}c broadly follows the trend of the surface temperature, although the H$_2$O(g) column extremes have a slightly longer timelag after the periastron and apoastron passages. As described in Section~\ref{sec:plan_climates}, increased global temperatures lead to enhanced evaporation of water vapour from the surface ocean, explaining the additional lag in H$_2$O(g) column variations. The maximum H$_2$O(g) column of 1.75${\times}10^{26}$~molecules~cm$^{-2}$ occurs during the daytime, about three days after periastron, and is 55\% higher than the time-averaged global mean. The hemispheric mean H$_2$O(g) column then declines steeply down to its minimum (34\% below the time-averaged global mean) during the nighttime. We see a temporary flattening of this decline in the H$_2$O(g) column around apoastron, when the substellar point is located at 90$^\circ$ or 270$^\circ$ longitude. During apoastron passage, a markedly different atmospheric circulation affects either hemisphere. The temporal evolution of the vertical wind (not shown) indicates a sudden decrease in the efficiency of the overturning circulation, affecting the H$_2$O(g) column. Besides the daytime-nighttime cycle in H$_2$O(g) column we also see a periastron-apoastron cycle in the global mean H$_2$O(g) column (black line) with up to 17\% higher and 10\% lower H$_2$O(g) column levels shortly after periastron and apoastron, respectively. Hence, the relatively small temperature variations significantly affect  H$_2$O(g) columns. For the 1:1 SOR, the mean dayside H$_2$O(g) column equals 2.72${\times}10^{26}$~molecules~cm$^{-2}$ and can be enhanced by up to 107\% compared to the global mean, while the nightside is much drier (up to 89\% depletion compared to the global mean). The global mean H$_2$O(g) column varies by up to 14\% associated with cloud variability on the planet \citep[][]{cohen_traveling_2023}. 

\begin{figure}
\centering
\includegraphics[width=1\columnwidth]{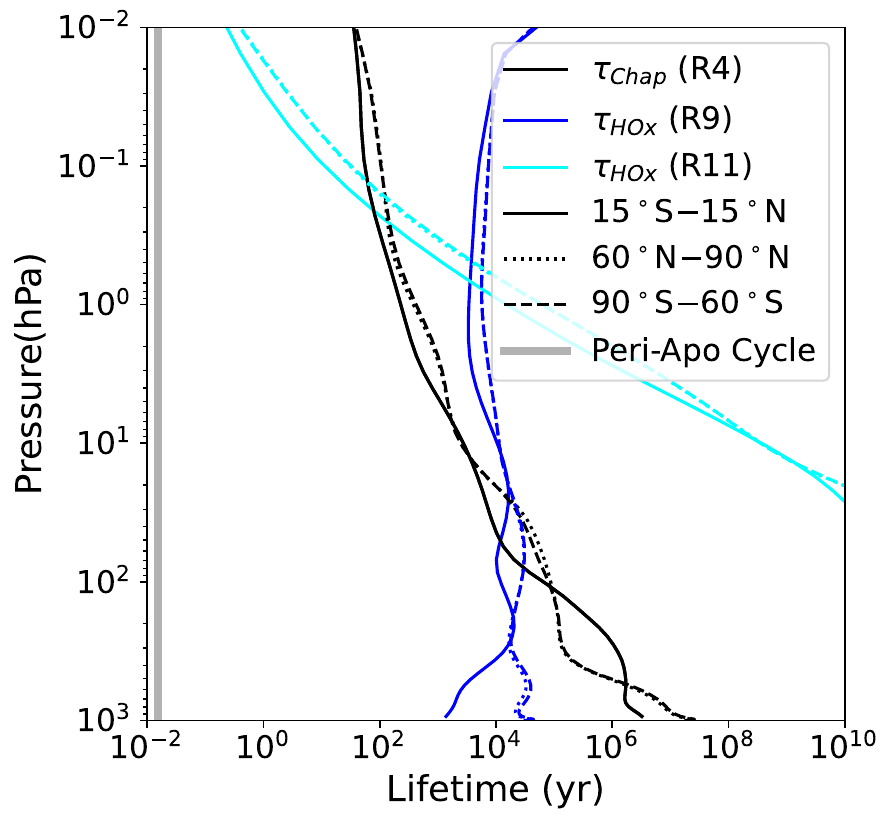}
\caption{Temporally averaged chemical lifetimes for ozone in the planet in a 3:2 SOR. We show the Chapman termination reaction (R4) in black, and the rate limiting steps of the dominant HO$_\mathrm{x}$ cycles in blue (R9) and cyan (R11). The lifetimes are calculated as spatial means for the equatorial region (15${^\circ}$S{-}15${^\circ}$N), the Northern high latitudes (60${^\circ}$N{-}90${^\circ}$N, and the southern high latitudes (90${^\circ}$S{-}60${^\circ}$S), based on the spatial distributions in Figure~\ref{fig:pcb_32_toc}. The grey line shows the timescale between periastron and apoastron passage (5.593~days).} 
\label{fig:pcb_32res_lifetimes}
\end{figure}

The global ozone column (Figure~\ref{fig:pcb_32res_tevol}d) shows no significant fluctuations in the periastron-apoastron cycle for the 3:2 SOR, since the chemical lifetime of ozone far exceeds the periastron-apoastron cycle. In Figure~\ref{fig:pcb_32res_lifetimes}, we show the chemical lifetimes, calculated as:
\begin{equation}
    \tau_{chem} = \frac{n_{\rm{O_3}}}{R_x},
\end{equation}
where $R_x$ represents the loss of ozone in molecules~cm$^{-3}$~s$^{-1}$ due to reaction $x$ and $n_{\rm{O_3}}$ the ozone number density in molecules~cm$^{-3}$. Figure~\ref{fig:pcb_32res_lifetimes} confirms the findings of Section~\ref{sec:chemical3d}, where we found that -- as a function of decreasing pressure -- ozone destruction is dominated by HO$_\mathrm{x}$ cycling through R9, the Chapman termination reaction (R4), and HO$_\mathrm{x}$ cycling through R11. The chemical lifetimes all exceed the periastron-apoastron timescale (the grey line). Comparing the chemical lifetimes of ozone in the equatorial region (solid lines) on one hand and high-latitude regions (dashed and dotted lines) on the other, we see that lifetimes are generally shorter around the equator, leading to faster ozone destruction here. At the same time, ozone production is also enhanced in the equatorial regions. The ozone column density represents an integrated quantity that is most sensitive to the atmospheric layers with the highest air density and thus the lower atmosphere (see e.g., \citealt{braam_stratospheric_2023}). By probing the ozone column at the surface, as in Figure~\ref{fig:pcb_32res_tevol}, we can connect hemispheric variations to dynamically driven chemistry \citep[][]{braam_stratospheric_2023}{}{}. The daytime-nighttime cycles are visible: ozone accumulates in the (changing) nighttime hemisphere with mean ozone columns that are up to 38~DU higher (5.2\%) than the global mean, while daytime ozone is depleted by up to 5.2\% compared to the global mean. Ozone also accumulates on the nightside of the planet in a 1:1 SOR, driven by atmospheric transport processes and related to the long chemical lifetimes here \citep[][]{yates_ozone_2020, ridgway_3d_2023, braam_stratospheric_2023}, with no significant variability in the global mean ozone column. The dayside variability mirrors that on the nightside, indicating that these variations are dynamically driven \citep[][]{braam_stratospheric_2023, de_luca_impact_2024}.

The oxidation and photolysis of H$_2$O(g) drive the HO$_\mathrm{x}$ catalytic cycle, dominating ozone destruction in the troposphere during the daytime on the 3:2 SOR (see Section~\ref{sec:chemical3d} and Figure~\ref{fig:pcb_32res_lifetimes}). The two hemispheres show contrasting trends in H$_2$O(g) and ozone columns as they cycle through daytime and nighttime (Figures~\ref{fig:pcb_32res_tevol}c and \ref{fig:pcb_32res_tevol}d). Because the evaporation rate is higher than the condensation rate, the H$_2$O(g) column peaks during the daytime. Furthermore, incoming daytime radiation is required for HO$_\mathrm{x}$ production (see reactions R5 and R6). Hence, the daytime provides a higher potential for HO$_\mathrm{x}$ catalytic cyling of ozone, evident from the dip in the ozone column. Low nighttime HO$_\mathrm{x}$ abundances result in lower ozone destruction rates. This connection between H$_2$O, HO$_\mathrm{x}$, and ozone is also visible in the spatial distributions in Appendix~\ref{sec:appendix_chem}. The plateauing behaviour at apoastron passage (Figures~\ref{fig:pcb_32res_tevol}c and \ref{fig:pcb_32res_tevol}d) is also seen for both ozone and H$_2$O(g) columns, albeit in the opposite sense. The nighttime ozone exceeds the global mean value at all times and mirrors the temporal evolution of the daytime ozone, which indicates the dynamical connection between the daytime and nighttime hemisphere that we will explore in the next Section.

\subsection{Dynamically-driven 3D distributions}\label{sec:dynamical3d}
\begin{figure*}
\centering
\includegraphics[width=0.666\columnwidth]{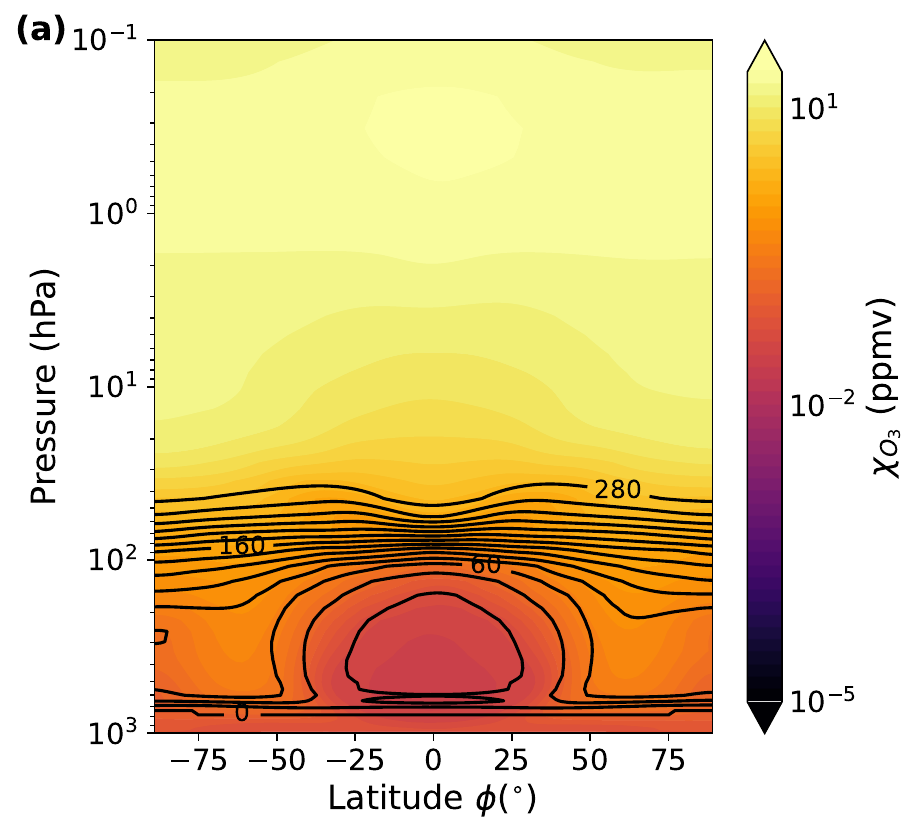}
\includegraphics[width=0.666\columnwidth]{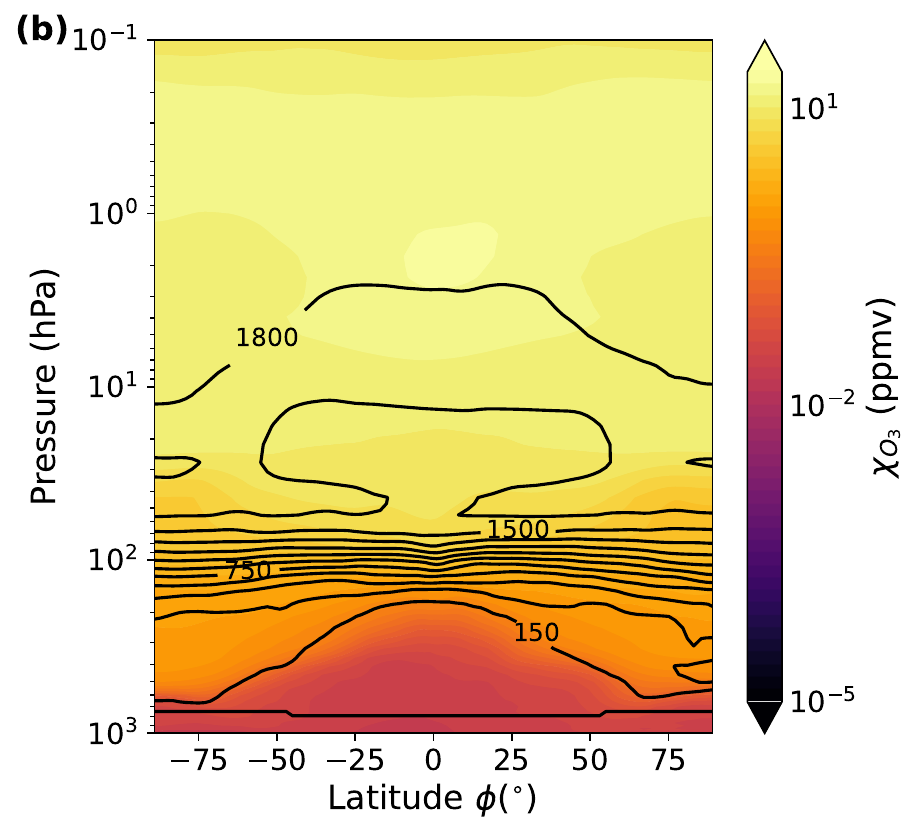}
\caption{Zonal-mean meridional distribution of the mole fraction $\chi_{O_\mathit{3}}$ for (a) Proxima Centauri b in a 1:1 SOR, and (b) Proxima Centauri b in a 3:2 SOR, both temporally averaged. Overplotted black contour lines correspond to the age-of-air tracer in days, to diagnose atmospheric circulation. The age-of-air tracer is is initialized at 0~s everywhere and reset whenever a parcel of air reaches the lowest atmospheric layers (below {$\sim$}2 km or P${>}$700 hPa). In this way, the age-of-air tracer measures the amount of time (in days) since an air parcel was last found in the lowest layers of the atmosphere, probing the main features of the atmospheric circulation.}
\label{fig:o3ageair_all}
\end{figure*}

The zonal-mean meridional distribution of $\chi_{O_\mathit{3}}$ (coloured contours in Figure~\ref{fig:o3ageair_all}) confirms the ozone column distributions as described in Section~\ref{sec:ozcols}. The contour lines in Figure~\ref{fig:o3ageair_all} represent the age-of-air tracer, an inert tracer (no chemical or radiative interaction, purely dynamical) that counts the time (in days) that has passed since an air parcel was last in the lowest layers of the atmosphere (z${<}$2~km or P${>}$700~hPa). In this way, the age-of-air tracer can probe atmospheric dynamics and exchange between the troposphere and stratosphere. Since transport timescales are considerably shorter than chemical lifetimes \citep[see][]{braam_stratospheric_2023}, this allows us to assess the connection between atmospheric dynamics and the ozone distribution.

The main photochemical production regions of ozone are in the stratosphere during the daytime. For Proxima Centauri b, this corresponds to P${<}$90~hPa for the 1:1 SOR and P${<}$20~hPa for the 3:2 SOR, since radiation penetrates deeper into the atmosphere on the permanent dayside for a 1:1 SOR. Figure~\ref{fig:rratesz_intercomp}a shows that the photolysis rates are faster on the dayside for the 1:1 SOR and that equally fast rates occur deeper in the atmosphere. Higher stratospheric H$_2$O abundances (see Figure~\ref{fig:tair_vmrs_intercomp}b) and higher H$_2$O photolysis rates (Figure~\ref{fig:rratesz_intercomp}d) are responsible for the enhanced attenuation of UV radiation for the planet in a 3:2 SOR. 

Following the dayside production close to $\phi$=0$^\circ$ on Proxima Centauri b in a 1:1 SOR, $\chi_{O_\mathit{3}}$ spreads to higher latitudes where it subsides down to ${\sim}$30~hPa (Figure~\ref{fig:o3ageair_all}a). On the nightside, chemical lifetimes of ozone are considerably longer than on the dayside, resulting in regional enhancements in the troposphere (${>}$100~hPa) between $-$80${<}\phi{<}-$40$^\circ$ and 40${<}\phi{<}$80$^\circ$: these represent the gyre accumulation as shown in Figure~\ref{fig:toc_11}. Since we show the zonal mean we lose the zonal variations, but the enhancements are still caused by the nightside accumulation over the gyre regions. The contour lines of the age-of-air tracer in Figure~\ref{fig:o3ageair_all}a demonstrate the flow of air. Since the tracer is reset to 0 s whenever air reaches the lowest model levels (see the 0 days contour line), the larger the number of days, the longer it has been since air was last in the near-surface layers. For low latitudes, we see that relatively young air rises, with the 60-days air found at around 100 hPa. At higher latitudes, the 60-days contour line reaches higher pressures, illustrating the connection between the lower stratosphere and the troposphere here. Therefore, the gyre regions consist of relatively old air compared to the tropospheric surroundings, indicating that the gyres are connected to the older and ozone-rich stratospheric air overhead. \citet{braam_stratospheric_2023} show that a stratospheric dayside-to-nightside circulation drives this connection. 

The high-latitude ozone column is also enhanced for a 3:2 SOR, although now at all longitudes as shown in Figure~\ref{fig:pcb_32_toc}. $\chi_{O_\mathit{3}}$ in Figure~\ref{fig:o3ageair_all}b confirms this spatial distribution, with enhanced $\chi_{O_\mathit{3}}$ at high latitudes for pressures as high as ${\sim}$600~hPa. The age-of-air distribution shows that older, ozone-rich air subsides at these high latitudes, producing the enhanced ozone columns. Hence, following the daytime production of ozone in the equatorial region (around $\phi$=0$^\circ$ for all longitudes), this indicates the existence of a stratospheric equator-to-pole circulation that drives the ozone-rich air from low to high latitudes. The meridional circulation system has similarities to the Brewer-Dobson circulation, which describes the transport of ozone from its equatorial production regions to high latitudes on Earth \citep[e.g.,][]{butchart_brewer-dobson_2014}{}{}. Figure~\ref{fig:pcb_32res_lifetimes} shows that chemical lifetimes are longer in the high-latitude troposphere (dashed and dotted lines) as compared to the equatorial troposphere (solid lines), meaning that any ozone arriving there can survive longer than in the equatorial regions and thus produce the enhanced ozone columns at high-latitudes.

\subsection{Observational prospects}\label{sec:observational}
We translate the 3D data into synthetic observables using the PSG GlobES tool \citep[][]{villanueva_planetary_2018}{}{}, as described in Section~\ref{subsec:PSG}. We use the phase angles in Table~\ref{tab:phase_angles_pcb} to cover the planet in a 1:1 SOR. Additionally, we take the changing substellar longitude for Proxima Centauri b in the 3:2 SOR into account to include a changing daytime hemisphere, also following Table~\ref{tab:phase_angles_pcb}.

\begin{figure}
\centering
\includegraphics[width=1\columnwidth]{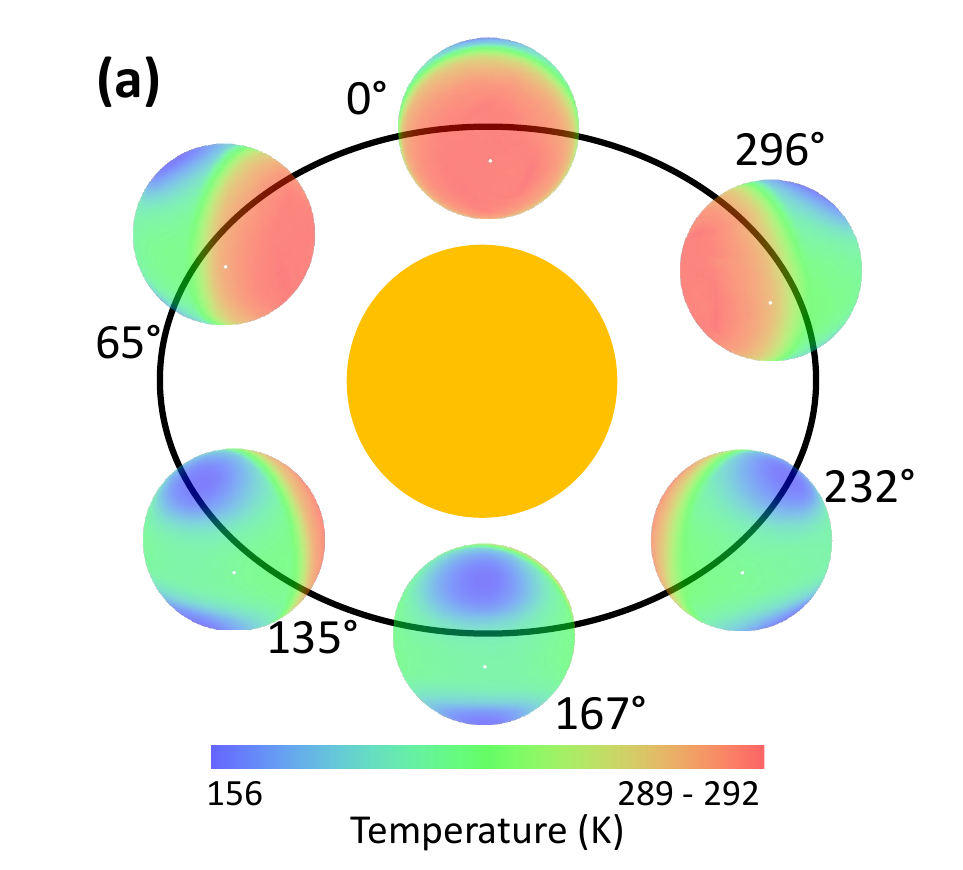}
\includegraphics[width=0.975\columnwidth]{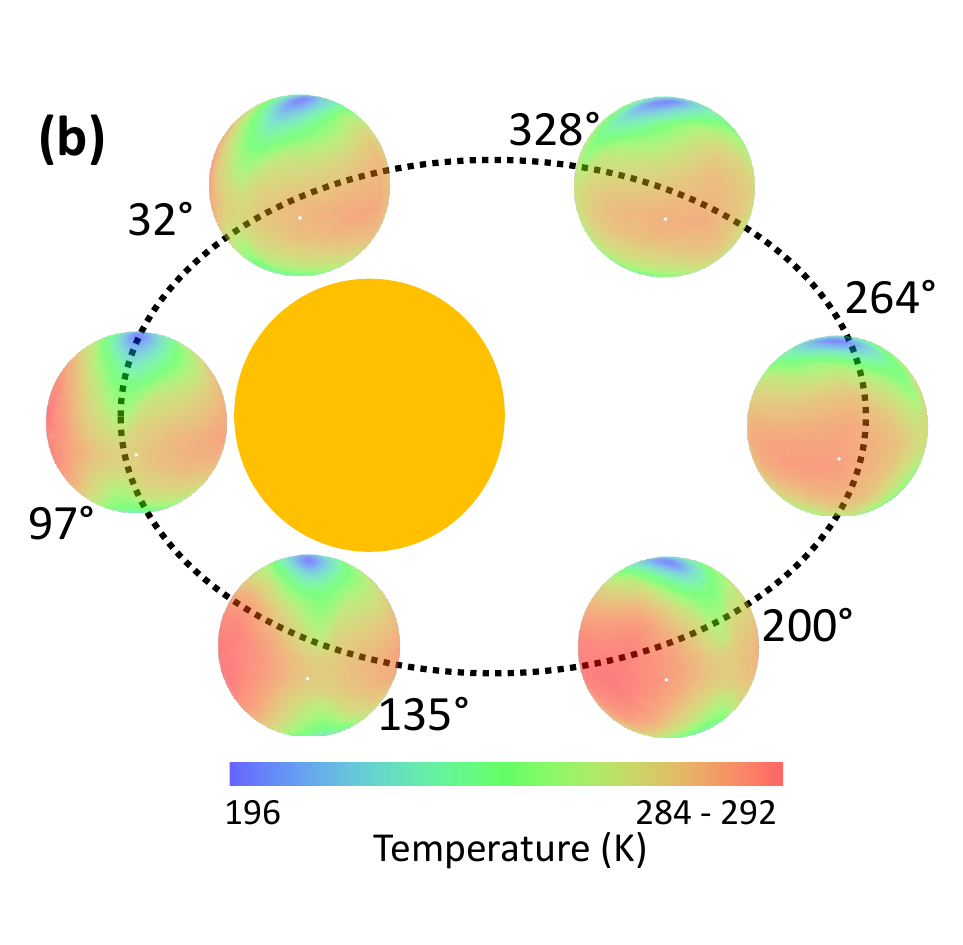}
\caption{Orbital evolution (not to scale) of phase angles and corresponding surface temperature distributions on emission disks as observed by a distant observer. Proxima Centauri b at an orbital inclination of 70$^\circ$ is shown in (a) a 1:1 SOR and (b) a 3:2 SOR. The temperature ranges vary with the timestep, as indicated by the upper limit of the colourmaps.}
\label{fig:orbphase}
\end{figure}

The observed hemisphere and the corresponding distribution of surface temperature as a function of phase angle and substellar longitude are shown in Figure~\ref{fig:orbphase} for Proxima Centauri b in a 1:1 SOR (top) and Proxima Centauri b in a 3:2 SOR (bottom). Due to the orbital inclination, we can cover a full phase for Proxima Centauri b. Figure~\ref{fig:orbphase} shows distinct spatial distributions: there are clear hemispheric differences in surface temperature over phase angles for the planet in a 1:1 SOR, whereas the 3:2 SOR is much more homogeneous. 

\begin{figure*}
\centering
\includegraphics[width=1.75\columnwidth]{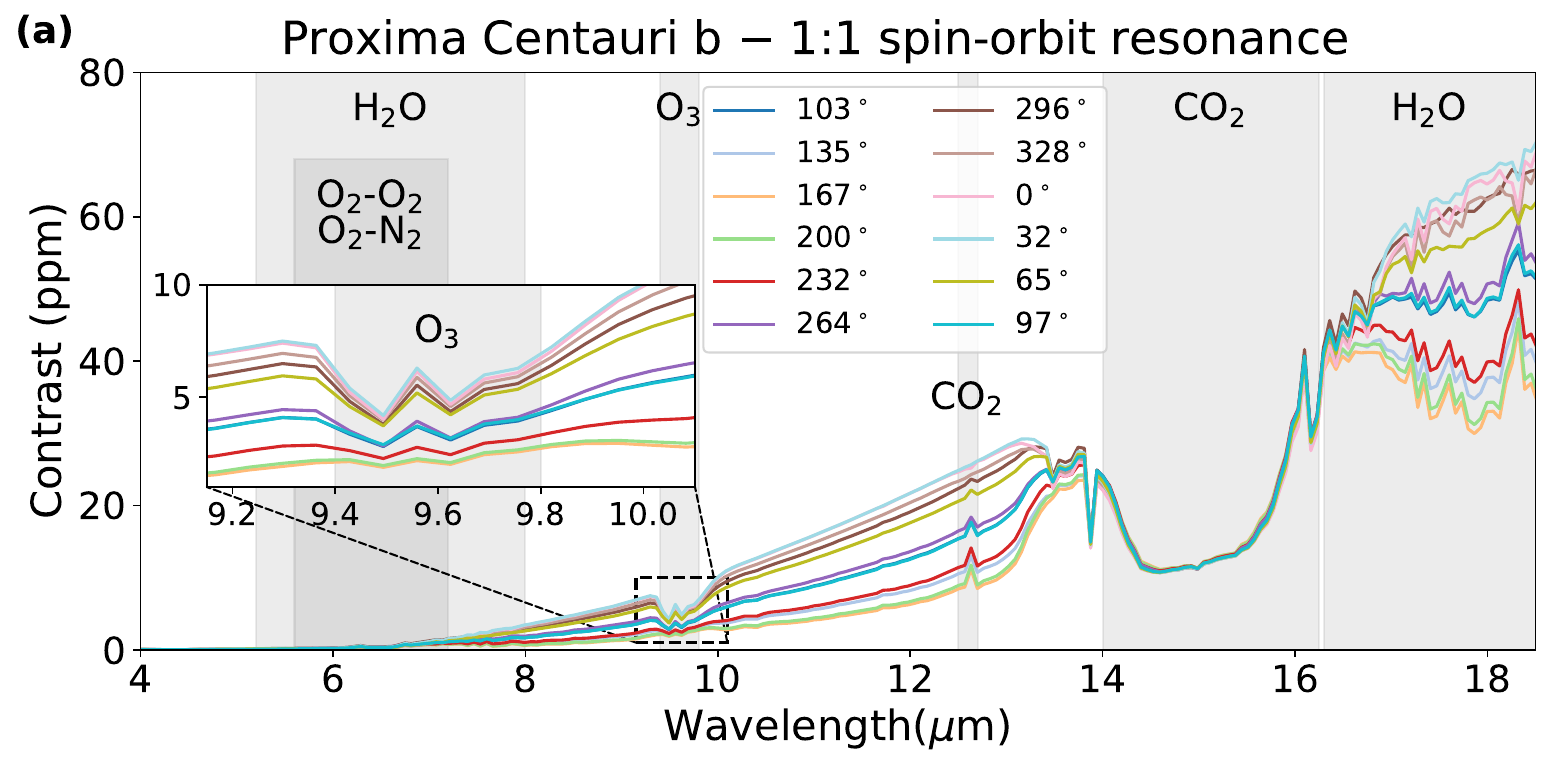}
\includegraphics[width=1.75\columnwidth]{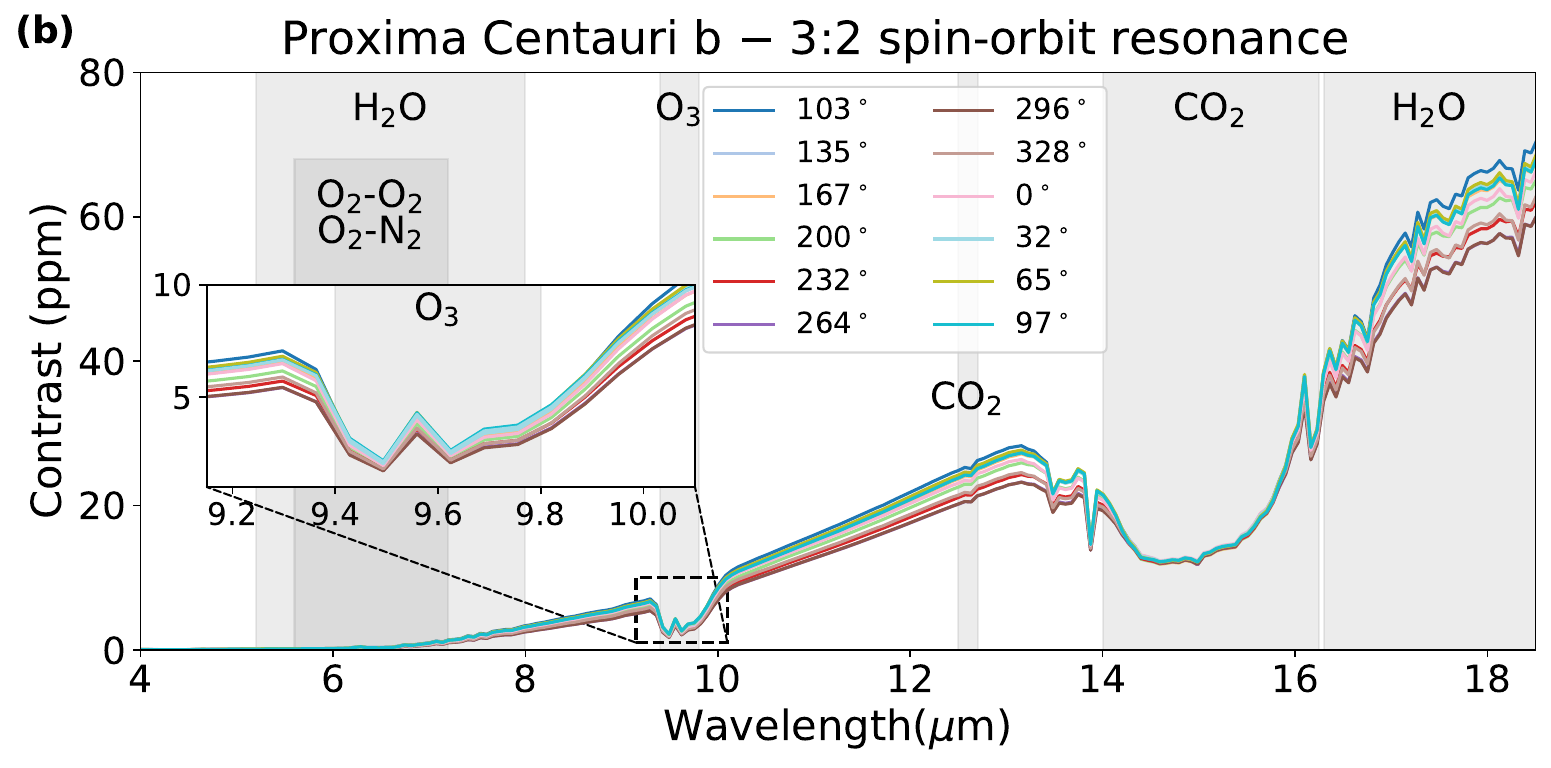}
\caption{Synthetic emission spectra using different orbital phase angles for (a) Proxima Centauri b in a 1:1 SOR, and (b) Proxima Centauri b in a 3:2 SOR. Colours represent the orbital phase and the corresponding timestep of the simulation, using daily output (see Table~\ref{tab:phase_angles_pcb}). We use PSG with the GlobES 3D mapping tool to translate 3D CCM data into synthetic emission spectra.}
\label{fig:emission_spectra}
\end{figure*}

Figure~\ref{fig:emission_spectra} shows emission spectra for the three planet configurations as a function of phase angle and focused on the proposed spectral range for the LIFE mission concept \citep[][]{quanz_large_2022}{}{}. For Proxima Centauri b in a 1:1 SOR, spectral variations are visible in Figure~\ref{fig:emission_spectra}a depending on the observed orbital phase (Figure~\ref{fig:orbphase}a). The most prominent variations are driven by the observed temperature distribution (8--9.5 and 10--13.6~$\mu$m) and water vapour abundances (16--18~$\mu$m). Smaller variations of up to ${\sim}$4~ppm are seen for the ozone feature around 9.6~$\mu$m and water vapour features between 7--8~$\mu$m. The CO$_2$ feature covering 14--16~$\mu$m is constant over time because of its fixed abundance in the simulations and also affects the emission features between 16--18~$\mu$m. The highest contrasts are seen for the 32$^\circ$ and 0$^\circ$ phase angles, for which we observe most of the warm dayside hemisphere. The smallest contrasts correspond to the nightside views at phase angles of 167$^\circ$ and 200$^\circ$. All other phase angles lead to intermediate contrasts. Hence, we see that a synchronously rotating planet provides a cycle of observed contrast levels in an emission spectrum as the planet completes an orbit, varying by up to 20~ppm between 12--13~$\mu$m and up to 36~ppm between 16--18.5~$\mu$m. We find sub-ppm level fluctuations in the 5--8~$\mu$m range. In the wetter dayside atmosphere, water vapour dominates this spectral range but, when the drier nightside hemisphere dominates the view, the spectrum around 6.4~$\mu$m is dominated by the O$_2$-O$_2$ or O$_2$-N$_2$ CIA features \citep[][]{fauchez_sensitive_2020}. 

Moving to Proxima Centauri b in a 3:2 SOR, we see in Figure~\ref{fig:emission_spectra}b that the increased global homogeneity (Figure~\ref{fig:orbphase}b) basically removes all of the phase angle variations in the emission spectrum. Further investigation also reveals that ice clouds mute spectral features particularly between 7--9~$\mu$m and 10--13.6~$\mu$m. The broad structure of the dominant opacity sources (water vapour, ozone, CO$_2$) remains similar to that of the planet in a 1:1 SOR, with contrast levels similar to the highest contrasts in a 1:1 SOR (spectra for 0$^\circ$ and 32$^\circ$ phase angles in Figure~\ref{fig:emission_spectra}a). The ozone features around 9.6~$\mu$m are slightly deeper due to the enhanced global ozone column (see Section~\ref{sec:ozcols}). In this atmosphere, the relatively high amount of water vapour blocks any O$_2$-O$_2$ or O$_2$-N$_2$ CIA features from view. The lack of temporal variability in the 3:2 SOR presents a discriminant to determine whether a planet orbits in a 1:1 or a 3:2 SOR around its host star. Phase curves that focus on the ozone feature see a persistent feature for a 3:2 SOR and will see its disappearance for a 1:1 SOR. The same distinction is seen for any of the wavelengths corresponding to the water vapour features.

Several more subtle distinctions probe the atmospheric pressure-temperature structure when we compare specific spectral bands for CO$_2$ (at 12.6~$\mu$m) and H$_2$O (at 18.3~$\mu$m). Infrared emission features originate in regions of high optical depth and depend on both the abundances and the temperature difference of the absorbing and emitting layers. If the region of high optical depth is in the cooler parts of the atmosphere, this generates a weaker thermal signal than that of the hotter part of the atmosphere. Hence, if temperature decreases with decreasing pressure, we see absorption features. Conversely, in the case of a temperature inversion, we see emission features. Proxima Centauri b in a 1:1 SOR shows absorption features for the CO$_2$ and H$_2$O bands when we mainly probe the dayside hemisphere (e.g., for 0$^\circ$ and 32$^\circ$ in Figure~\ref{fig:emission_spectra}a). However, when we probe most of the nightside hemisphere (e.g., for 167$^\circ$ and 200$^\circ$ in Figure~\ref{fig:emission_spectra}a), both bands become emission features, indicating a different vertical temperature profile between the dayside and nightside hemisphere. Figure~\ref{fig:tair_vmrs_intercomp}a shows that the nightside hemisphere exhibits a strong temperature inversion between pressures 1000~hPa and 500~hPa, with temperatures increasing by over 40~K. On the dayside hemisphere, this inversion is much weaker with only 4~K difference. \citep[][]{guzewich_impact_2020} report a similar switch from absorption to emission in the H$_2$O spectral bands for exoplanets in 1:1 SOR around G-type stars, also due to a tropospheric temperature inversion that is limited to the nightside. Such inversions are reminiscent of the polar troposphere on Earth, although \citet{joshi_earths_2020} show that the inversions are much stronger for simulations of Proxima Centauri b in a 1:1 SOR. The homogeneity of the atmosphere for a 3:2 SOR removes this dayside-nightside distinction in the pressure-temperature profile and thus both the CO$_2$ and H$_2$O features are absorbing over a full orbit.

\section{Discussion and Conclusion}\label{sec:discconc}
Motivated by the range of possible exoplanet environments in terms of the host star, orbital configuration, and planetary parameters, we employ a 3D CCM to investigate the complex interplay between stellar radiation, atmospheric dynamics, and photochemistry for Proxima Centauri b and similar planets around M-dwarfs. These planets are likely tidally locked to their host star \citep[][]{goldreich_spin-orbit_1966, barnes_tidal_2017}{}{} in, for example, a 1:1 or 3:2 SOR depending on eccentricity. In a 1:1 SOR, GCM simulations predict dayside-nightside or zonal gradients and distinct regimes of atmospheric circulation for these planets \citep[e.g.,][]{carone_stratosphere_2018}{}{}, with the existence of an equatorial jet on Proxima Centauri b characteristic of the Rhines rotator regime \citep[][]{haqq-misra_demarcating_2018}{}{}. The changing daytime hemisphere in an eccentric 3:2 SOR produces two hot spots along with meridional gradients in quantities like the temperature that are more like Earth and results in a more homogeneous atmosphere \citep[e.g.,][]{boutle_exploring_2017}{}{}. We compare the main mechanisms controlling the atmospheric chemistry for both a 1:1 and 3:2 SOR, and for the latter setup present the first simulations of interactive (photo)chemistry. The orbital configurations impact the habitability of the planetary surface and observability of chemical species in the atmosphere.

\subsection{Climate}
We find that about half of the dayside surface stays above the freezing point of water for Proxima Centauri b in a 1:1 SOR. Simulating a 3:2 SOR with an eccentricity of 0.3 results in general warming for the planet. The eccentric orbit increases the mean flux over one orbit as compared to the circular case \citep[see also][]{williams_earth-like_2002, dressing_habitable_2010, bolmont_habitability_2016, ji_inner_2023}{}{} and the changing location of the substellar point means that the stabilizing cloud feedback that cools planets in a 1:1 SOR \citep[][]{yang_stabilizing_2013, kopparapu_inner_2016}{}{} is less effective. Nevertheless, Proxima Centauri b in a 3:2 SOR exhibits a stable planetary climate with two habitable regions, one centred at 0$^\circ$ latitude and longitude and one at 0$^\circ$ latitude and 180$^\circ$ longitude \citep[][]{turbet_habitability_2016, boutle_exploring_2017, del_genio_habitable_2019}{}{}. We find a wetter stratosphere for a 3:2 SOR, indicating a dependence of the inner edge of the HZ on the exact SOR. Earlier work by \citet{colose_effects_2021} studied the effects of the 3:2 SOR on the inner edge of the HZ, also reporting enhanced surface temperatures for higher-order SORs. Contrary to our results, \citet{colose_effects_2021} find that their higher-order SORs do not generally result in wetter stratospheres and therefore do not significantly impact the inner edge of the HZ, indicating differences in vertical transport of moisture. \citet{liu_higher_2023} do find a wetter stratosphere for Earth with an eccentricity of 0.4 as compared to a circular orbit. We suggest a model intercomparison project focusing on the impact of SORs on habitability as future work.

The mean hotspot temperatures for the planet in a 3:2 SOR (within the white lines in Figure~\ref{fig:pcb_32_temp}) exceed 279~K, consistent with \citet{boutle_exploring_2017}. They also showed that, for this eccentric orbit, the mean flux approximation does not hold \citep[][]{bolmont_habitability_2016}{}{}. To further unveil the characteristics of the dynamic atmosphere in a 3:2 SOR, we suggest the application of dynamical systems metrics \citep[][]{hochman_greater_2022, hochman_analogous_2023, de_luca_impact_2024}{}{}. Such metrics can be used to quantify the persistence of atmospheric states, allowing for a proper assessment of the stable atmospheric states in a 3:2 SOR and the variability associated with the changing daytime hemisphere and eccentric orbit.

\subsection{Chemistry}
Comparing the 3D distributions of ozone confirms the dependence of (photo)chemistry on an interplay between the stellar radiation, orbital configuration, and atmospheric circulation. The incoming stellar radiation and its wavelength dependence determine the photochemical balance of ozone \citep[see][]{kozakis_is_2022}{}{}. For planets in a 1:1 SOR, a dependence on the circulation state was also suggested by \cite{chen_habitability_2019} for a range of orbital periods and stellar temperatures. On Earth, the Brewer-Dobson circulation transports ozone to the poles and controls the spatial distribution of ozone \citep[][]{brewer_evidence_1949, dobson_origin_1956, butchart_brewer-dobson_2014}{}{}. \citet{braam_stratospheric_2023} show that the spatial distribution of ozone on Proxima Centauri b in a 1:1 SOR is controlled by a stratospheric dayside-to-nightside circulation mechanism, resulting in the zonal asymmetry as shown in Figure~\ref{fig:toc_11}. For the planet in a 3:2 SOR we find meridional gradients in the ozone column. Age-of-air tracer experiments suggest that a stratospheric equator-to-pole circulation mechanism similar to the Brewer-Dobson circulation drives the spatial distribution, with spatially varying chemical lifetimes allowing for ozone buildup on the nightside (1:1) or at high latitudes (3:2).

We explain the differences in vertical ozone profiles (Figure~\ref{fig:tair_vmrs_intercomp}e) by comparing the dominant chemical processes in the simulations. The production of ozone fundamentally depends on the incoming UV flux \citep[as shown in detail by e.g.,][]{selsis2002signature, segura_ozone_2003, grenfell_response_2007, grenfell_sensitivity_2014, rugheimer_effect_2015, kozakis_is_2022}. A low total amount of UV flux produces little ozone with barely discernible absorption features in synthetic spectra \citep[][]{grenfell_response_2007}{}{}. On the other hand, high UV fluxes will be absorbed by ozone and significantly heat the stratosphere, in turn diminishing the strength of the ozone features in emission spectra \citep[][]{kozakis_is_2022}{}{} and thus making it harder to detect the 9.6~$\mu$m features in Figure~\ref{fig:emission_spectra}. Proxima Centauri b seems to receive the right amount of radiation for significant ozone formation without substantial stratospheric heating, like the M7-orbiting exoplanet from \citet{grenfell_sensitivity_2014}. For Proxima Centauri b, HO$_\mathrm{x}$ cycles dominate ozone destruction in the troposphere and upper stratosphere, in line with \citet{yates_ozone_2020} and \citet{braam_lightning-induced_2022}. The production of NO$_\mathrm{x}$ is higher for a planet in a 3:2 SOR due to taller cloud structures (Figure~\ref{fig:tair_vmrs_intercomp}c and d) and associated higher lightning flash rates. Therefore, the extent of NO$_\mathrm{x}$ catalytic cycling and thus the role of lightning in atmospheric chemistry and biosignature interpretation depends on the orbital configuration \citep[see also][]{ardaseva_lightning_2017, braam_lightning-induced_2022}{}{}. Future work should systematically investigate lightning in diverse exoplanet configurations. Catalytic cycling of ozone is also sensitive to the amount of UV flux \citep[e.g.,][]{grenfell_sensitivity_2014}{}{}, which is especially important considering the impact of flares on atmospheric chemistry around M-type stars \citep[e.g.,][]{chen_persistence_2021, ridgway_3d_2023}{}{}. The trade-off between UV radiation driving ozone formation and stratospheric heating and the subsequent impact on atmospheric and climate dynamics \citep[e.g.,][]{de_luca_impact_2024}{}{} needs to be further explored.

Near-surface ozone is harmful to life when exceeding mole fractions of 40~ppb \citep[][]{europe_air_2000}, which \citet{cooke_lethal_2024} use as an upper limit to investigate surface ozone concentrations on exoplanets in a 1:1 SOR. Their analysis includes simulations of Proxima Centauri b, which generally does not exceed the 40~ppb limit \citep[][]{cooke_lethal_2024}. Figure~\ref{fig:tair_vmrs_intercomp}e shows that our simulations of the 1:1 SOR lead to a nightside hemispheric mean below 40~ppb but dayside hemispheric means that exceed the upper limit. Proxima Centauri b in a 3:2 SOR has hemispheric mean ozone levels below the limit. However, the spatial distribution of surface ozone mole fractions (not shown) for the 3:2 SOR also exceeds 40~ppb locally at locations that vary in time. The different levels and distribution of surface ozone from \citet{cooke_lethal_2024} can have a variety of causes including reaction or photolysis rate differences, deposition parameterisations, or distinct features of atmospheric circulation, that warrant model intercomparisons \citep[see e.g.,][]{turbet_trappist-1_2022, sergeev_trappist-1_2022, fauchez_trappist-1_2022}. One clear distinction between our simulations and those of \citet{cooke_lethal_2024} are the inclusion of Earth's continents in the latter case, which affects the atmospheric circulation \citep[][]{lewis_influence_2018, bhongade_asymmetries_2024, martinez_how_2024}. 

On Earth, the presence of reactive halogen gases that include chlorine and bromine can induce additional catalytic cycles that destroy ozone \citep[][]{yung_atmospheric_1980}. These catalytic cycles can proceed both at low latitudes and in the presence of polar stratospheric clouds \citep[][]{tritscher_polar_2021}. Since the low temperatures and dry air over the nightside gyres of exoplanets in 1:1 SOR are reminiscent of the polar stratosphere, these catalytic cycles might end up significantly impacting the ozone distribution on such planets. The presence of CO$_2$ and CH$_4$ in Earth's troposphere catalyzes the production or destruction of ozone, depending on the background NO$_\mathrm{x}$ concentrations \citep[][]{cicerone_biogeochemical_1988}. Since these chemical processes depend on the reaction with OH, they also affect the abundance of OH. Specifically in the context of exoplanets around M-dwarfs, previous 1D photochemical studies have shown that CO$_2$ photochemistry can provide abiotic pathways to high O$_2$ abundance \citep[][]{hu_photochemistry_2012, domagal-goldman_abiotic_2014, tian_high_2014, wordsworth_abiotic_2014, gao_stability_2015, harman_abiotic_2015, luger_extreme_2015}. In our simulations, CO$_2$ is chemically inert and CH$_4$ is absent, due to the focus on 3D ozone distributions in the context of HO$_\mathrm{x}$ and NO$_\mathrm{x}$ chemistry, but future 3D modelling should incorporate interactive halogen, CO$_2$, and CH$_4$ chemistry.

Further analysis of the 3:2 SOR reveals daytime-nighttime cycles as well as periastron-apoastron cycles in temperature, H$_2$O(g) column, and O$_3$ column, ultimately driven by variations in the incoming stellar radiation. We find that extrema in temperature, H$_2$O(g) column, and ozone column follow periastron and periastron passages with a brief time-lag, denoting the response time of the atmosphere to radiation changes. Especially prominent are the cycles in H$_2$O(g) column, with hemispheric changes up to 55\% and eccentricity changes up to 17\% as compared to the time-averaged global mean H$_2$O(g) column. Global mean ozone column fluctuations due to eccentricity are negligible due to the relatively long chemical lifetime of ozone for Proxima Centauri b conditions \citep[see][]{yates_ozone_2020, braam_stratospheric_2023}{}{}. However, the daytime-nighttime cycle with ozone accumulation during the nighttime is clearly visible with hemispheric enhancements and depletion of up to 38~DU (5.2\%) as compared to the time-averaged global mean. The daytime-nighttime cycle illustrates the strong dependence of the ozone distribution on the coupling of photochemistry and atmospheric dynamics \citep[][]{butchart_brewer-dobson_2014, braam_stratospheric_2023}{}{}. For exoplanets in 1:1 SOR around M-dwarfs, \citet{luo_coupled_2023} show that a strong non-varying source of NO$\mathrm{x}$ at the surface can induce oscillations in ozone abundances on multidecadal timescales. The oscillations are driven by a negative feedback loop involving photochemistry, radiative transfer, and atmospheric dynamics. Such large NO$\mathrm{x}$ emissions would require biological nitrogen fixation as a source \citep[][]{luo_coupled_2023}. The ozone layer varies by up to 200~DU during these oscillations, exceeding our reported day-night cycles of up to 38~DU with timescales of about 22~days. Due to the distinct magnitudes and timescales associated with these separate mechanisms, we are able to distinguish between the two. A strong NO$\mathrm{x}$ source can be added to the 3:2 SOR to see if the long-term oscillations reported by \citet{luo_coupled_2023} persist for higher-order SORs. Regardless, abiotic cycles in chemical abundances have to be understood to properly interpret seasonally varying biosignatures \citep[][]{olson_atmospheric_2018, schwieterman_exoplanet_2018} and emphasize the importance of 3D photochemical studies of terrestrial exoplanet atmospheres.

\subsection{Observability}
We report several potential 3D spatial and temporal effects on emission spectra, using the PSG GlobES tool \citep[][]{villanueva_planetary_2018}{}{} to comprehensively generate synthetic spectra from our 3D simulations whilst varying the observed 3D geometry with the orbital phase angle. For Proxima Centauri b in a 1:1 SOR, variations in the pressure-temperature profile as well as horizontal and vertical variations in H$_2$O and ozone abundances cause spectral fluctuations of up to 36~ppm for Proxima Centauri b. The more homogeneous atmosphere of Proxima Centauri b in a 3:2 SOR diminishes spectral variations as a function of orbital phase, as also shown by \citet{turbet_habitability_2016, boutle_exploring_2017}. The presence or absence of these spectral fluctuations then presents a discriminant between a 1:1 or 3:2 SOR. The ozone feature at 9.6~$\mu$m provides an additional window to probe with time-resolved emission spectra or a phase curve, that we can only predict and interpret using 3D CCM simulations. Although favourable predictions of the observability of Proxima Centauri b exist \citep[][]{kreidberg_prospects_2016}, the uncertainties for JWST observations of TRAPPIST-1 b and c \citep[][]{greene_thermal_2023, zieba_no_2023}{}{} indicate that any fluctuations likely disappear under current observational noise levels. Nevertheless, the time-averaged 1-D profiles still depend on the 3D nature of atmospheres.

The spectra in Figure~\ref{fig:emission_spectra} were deliberately generated for the proposed wavelength range of the LIFE mission concept \citep[][]{quanz_large_2022}{}{}, which has the capability to detect P-T profiles on Earth as an exoplanet with constant chemical abundance profiles \citep[][]{konrad_large_2022}{}{}. \citet{alei_large_2022} extend the analysis to the changing atmospheric conditions through Earth's history, finding biases in retrieved P-T profiles. \citet{mettler_earth_2024} find similar biases in the P-T profiles using disk-integrated Earth observations, but suggest these are due to the assumption of constant abundance profiles in the retrievals as compared to non-constant abundance profiles in generating synthetic spectra. As we describe here, the temporal and spatial variations in the P-T profile are a complex interplay between the radiative, thermal, dynamical, and chemical state of the atmosphere that can only be explored using 3D CCM simulations. 

The main spectral features of chemical species such as CO$_2$, H$_2$O, O$_3$, and CH$_4$ can be detected by LIFE to determine the atmospheric composition of terrestrial exoplanets, for the baseline LIFE requirements of spectral resolution $R$=50 and signal-to-noise ratio $S/N$=10, although the identification of O$_3$ and CH$_4$ benefits from enhanced $S/N$=20 \citep[][]{konrad_large_2022, alei_large_2022, mettler_earth_2024}{}{}. \citet{fauchez_sensitive_2020} have shown that O$_2$-O$_2$ or O$_2$-N$_2$ CIA features have strong mid-infrared features in simulated transmission spectra, with a potential additional impact from O$_2$-H$_2$O or O$_2$-CO$_2$. Future work should be directed towards the detectability of these O$_2$-X CIA features in emission spectra \citep[][]{fauchez_sensitive_2020}. Using real disk and time-averaged Earth observations, \citet{mettler_earth_2023} find that the mid-infrared spectrum of Earth as an exoplanet will vary as a function of season and viewing geometry. Variations in temperature or planetary albedo should be detectable with LIFE for the baseline $R$ and $S/N$ requirements \citep{mettler_earth_2024}, but varying abundances of chemical species with season or viewing geometry are currently muted by retrieval uncertainties \citep[][]{mettler_earth_2024}{}{}. However, as \citet{mettler_earth_2024} also note, these retrieval results are biased by simplifying assumptions such as cloud-free atmospheres or vertically constant abundance profiles, with work underway to mitigate these. Variations with season or viewing geometry are planet-specific and, as we show here, zonal asymmetries in a 1:1 SOR result in more dramatic spectral variations as compared to the meridional asymmetries on a 3:2 SOR exoplanet or on Earth. Any of these variations must be understood to distinguish properly interpret seasonally varying biosignatures \citep[][]{olson_atmospheric_2018}. Therefore, an in-depth investigation of the observability of the spectral variations in Figure~\ref{fig:emission_spectra} with LIFE is currently being conducted for various combinations of $S/N$ and $R$.

\section*{acknowledgments}
MB kindly thanks Paul Rimmer, Trent Dupuy, and Denis Defrère for interesting discussions that helped to consolidate the results. MB also thanks Thaddeus Komacek for discussions on the spectral features of oxygen. We are grateful to the two anonymous reviewers whose comments helped to significantly improve the article.

MB, PIP, and LD are part of the CHAMELEON MC ITN EJD which received funding from the European Union’s Horizon 2020 research and innovation programme under the Marie Sklodowska-Curie grant agreement no. 860470. MB appreciates support from a CSH Postdoctoral Fellowship. PIP acknowledges funding from the STFC consolidator grant \#ST/V000594/1. LD acknowledges support from the KU Leuven IDN grant IDN/19/028 and from the FWO research grant G086217N. NJM is supported by a Science and Technology Facilities Council Consolidated Grant [ST/R000395/1], a UKRI Future Leaders Fellowship [grant number MR/T040866/1] and a Leverhulme Trust Research Project Grant [RPG-2020-82]. SR acknowledges NSERC grant RGPIN-2022-04588.

We gratefully acknowledge the use of the MONSooN2 system, a collaborative facility supplied under the Joint Weather and Climate Research Programme, a strategic partnership between the Met Office and the Natural Environment Research Council. Our research was performed as part of the project space ‘Using UKCA to investigate atmospheric composition on extra-solar planets (ExoChem)'. For the purpose of open access, the authors have applied a Creative Commons Attribution (CC BY) licence to any Author Accepted Manuscript version arising from this submission.

%


\software{The \textsc{python} packages \textsc{iris} \citep[][]{met_office_iris_2022} and \textsc{aeolus} \citep[][]{sergeev_aeolus_2022} were used for the post-processing of model output. Scripts to process and visualize the data are available on GitHub: \href{https://github.com/marrickb/eccent\_3dchem\_PSJ}{https://github.com/marrickb/eccent\_3dchem\_PSJ}. The CCM simulations were performed using the Met Office Unified Model and UK Chemistry and Aerosol model (\href{https://www.ukca.ac.uk/}{https://www.ukca.ac.uk/}), both are available for use under licence; see \href{http://www.metoffice.gov.uk/research/modelling-systems/unified-model}{http://www.metoffice.gov.uk/research/modelling-systems/unified-model}. The emission spectra were produced using the Planetary Spectrum Generator \citep[][]{villanueva_planetary_2018}, see \href{https://psg.gsfc.nasa.gov/index.php}{https://psg.gsfc.nasa.gov/index.php}. The data underlying this article will be shared on reasonable request to the corresponding author, mainly motivated by the size of the data.}



\appendix
\section{Incoming stellar radiation}\label{subsec:isr}
The incoming stellar radiation and its passage through the atmosphere as well as the photolysis rates of chemical species will change over time with a dependence on the orbital parameters. The inclusion of positional astronomy within SOCRATES \citep[][]{edwards_studies_1996, manners_socrates_2021} is based on the description in \citet{smart_text-book_1944}. For an eccentric Keplerian orbit, the true anomaly $\nu(t)$ represents the angle (in radians) between the direction of periastron from the barycenter and the position of a planet at time $t$. The mean anomaly $M(t)$ (in radians) is the fictitious angle from periastron for a planet on a circular orbit at the same time $t$, assuming the same semi-major axis as for the true elliptical orbit. Hence, it concerns the mean motion $2\pi/P$ where $P$ is the orbital period in seconds. The mean anomaly is given by:
\begin{equation}
    M(t) = \frac{2\pi(t-t_p)}{P}.
\end{equation}
Here, $t$ indicates the current time, $t_p$ the time when the planet is at periastron, and $P$ the length of a year or orbital period (all in seconds). For a Keplerian orbit of eccentricity $e$, $M(t)$ can be used to calculate the true anomaly $\nu(t)$, or the angular distance of the planet from periastron. Within SOCRATES, $\nu(t)$ is represented as the third-order approximation of a series expansion known as the equation of the center, following the derivation by \citet{smart_text-book_1944}:
\begin{equation}
    \nu(t) = M + \left( 2e-\frac{e^3}{4} \right)\sin(M(t)) + \frac{5}{4}e^2\sin(2M(t)) + \frac{13}{12}e^3\sin(3M(t)).
\end{equation}
Using Kepler's second law, $\nu$ is then used to calculate the normally incoming stellar radiation at the top of the atmosphere ($S_{TOA}$) in W~m$^{-2}$:
\begin{equation}
    S_{TOA}(t) = S_0 \left( \frac{1+e\cos(\nu(t))}{1-e^2} \right)^2,
\end{equation}
where $\mathrm{S}_0$ is the stellar constant, the normally incoming stellar radiation at the mean star-planet distance $a$ in  W~m$^{-2}$.

The incoming radiation will, however, depend on the local stellar zenith angle ($\zeta$). To determine $\zeta$, we use the orbital parameters to calculate the stellar declination $\delta$, denoting the latitude at which the the star is vertically overhead:
\begin{equation}
    \sin(\delta) = \sin(\epsilon)\sin(\theta).
\end{equation}
The declination depends on the obliquity $\epsilon$ (0 in our case) and angular distance of the planet relative to vernal equinox $\theta$. 

Additionally, we need to know the hour angle $\Omega$ of the star, denoting the angle (in radians) through which a planet has rotated since stellar noon or the angle of the star from the local meridian at time t:
\begin{equation}
    \Omega = \lambda + \pi \left( \frac{2t}{T_D} - 1 \right),
\end{equation}
where $T_D$ is the length of the day and for longitude $\lambda$. If we consider the astronomical measurement of time on Earth based on the Earth's rotation, the solar day represents the length of time between two successive returns of the Sun to some meridian. The time system based on the solar day is the apparent solar time. However, the apparent solar day has a variable length with seasonal deviations of up to 16~minutes, due to: 1) the Earth's eccentricity, which causes variations in the Earth's orbital velocity as it speeds up close to periastron and slows down close to apoastron, and 2) the Earth's obliquity, meaning that the the Sun's annual motion lies in the ecliptic which is tilted to the Earth's celestial equator. The mean solar time follows a theoretical motion of the mean sun moving at a uniform rate along the equator. The mean solar day then represents the average of the apparent solar days in a year. The discrepancy between the apparent and mean solar time is described by the equation of time. Two formulations of the equation of time exist in the UM, based on \citet{smart_text-book_1944} and \citet{mueller_equation_1995}. The equation of time derived by \citet{mueller_equation_1995} contains higher-order terms and therefore is more appropriate for the relatively high eccentricity considered in this study. The equation of time then corrects for the variable length of day due to the eccentric orbit of Proxima Centauri b. 

Using $\Omega$ as a function of time allows us to calculate $\zeta$ at latitude $\phi$ and longitude $\lambda$:
\begin{equation}
    \cos(\zeta) = \cos(\phi)\cos(\delta)\cos(\Omega) + \sin(\phi)\sin(\delta).
\end{equation}
The incoming stellar radiation $S$ can then be calculated at any location as:
\begin{equation}
    S(t) =  S_{TOA}(t)\times\cos(\zeta).
\end{equation}

\section{Lightning-induced chemistry}\label{sec:appendix}
\begin{figure}
\centering
\includegraphics[width=0.75\columnwidth]{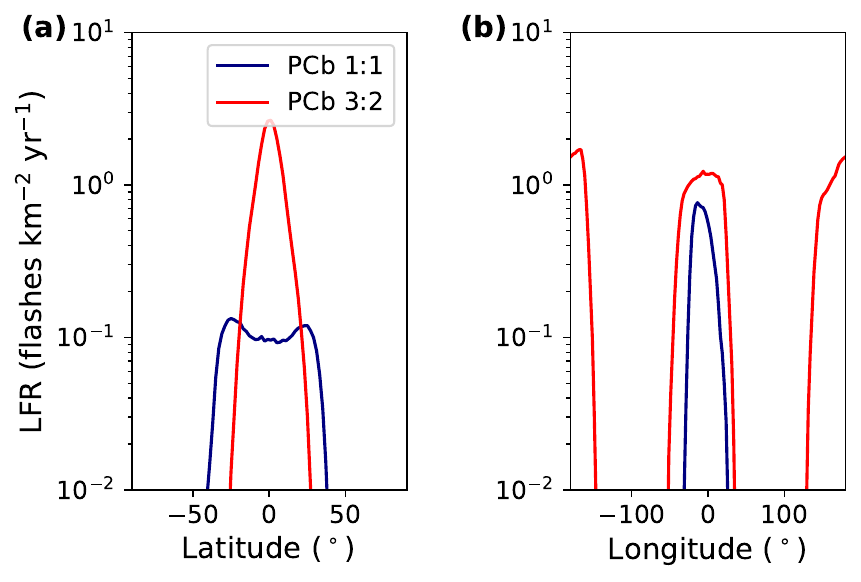}
\caption{Simulated lightning flash rates for Proxima Centauri in a 1:1 SOR (PCb 1:1) and a 3:2 SOR (PCb 3:2). By showing both the zonal-mean meridional distribution in panel a and meridional-mean zonal distribution in panel b, we show that lightning flashes are contained to the dayside hemisphere for planets in 1:1 SOR and are found on both hemispheres for planets in 3:2 SOR.}
\label{fig:lfr_comp}
\end{figure}
The atmospheric chemistry on any planet depends on energetic processes that can break molecular bonds, such as photolysis, flares, or lightning. In this study, we include the effects of photolysis and lightning, as described in Section~\ref{sec:methods}. The CCM interactively calculates the formation of lightning flashes based on scaling relations between the size of a thundercloud and the electrical power output \citep[][]{vonnegut_facts_1963}, so that the lightning flash rates (LFR) can be described in terms of the cloud-top height $H$ \citep{price_simple_1992, luhar_assessing_2021}. The application to exoplanet environments has been described in detail in \citet{braam_lightning-induced_2022}. Parameterisations are derived for continental and oceanic LFRs. Since we assume an aquaplanet, we only use the oceanic parameterisation:
\begin{equation}\label{eq:lfr_o_l20}
    LFR_O=2.0 \times10^{-5}H^{4.38}.
\end{equation}
Each lightning flash is empirically scaled to produce 216~moles of NO per flash to match the observed lightning-NO$_\mathrm{x}$ production on Earth. Figure~\ref{fig:lfr_comp} compares the resulting LFRs for both Proxima Centauri b simulations. For a 1:1 SOR, peaks in LFR are found close to the substellar point (0$^\circ$ latitude and longitude). For the 3:2 SOR, we find that LFR peaks are found on both the 0$^\circ$ and 180$^\circ$ longitude hemispheres, corresponding to the temperature hotspots in Figure~\ref{fig:pcb_32_temp}. LFRs for the 3:2 SOR are up to two orders of magnitude larger for a 1:1 SOR, explaining a more prominent role for lightning-induced NO$_\mathrm{x}$ chemistry in a 3:2 SOR.

\section{Spatial distributions of atmospheric chemistry}\label{sec:appendix_chem}
To expand on the analysis of the dominant chemical processes in Section~\ref{sec:chemical3d}, this Section presents the spatial distribution of selected chemical species and reaction rates for the simulations of both the 1:1 and 3:2 SOR. For the distributions of chemical species, we present the vertically integrated column density similar to the ozone column densities in Figures~\ref{fig:toc_11} and \ref{fig:pcb_32_toc}, by integrating the number densities over the vertical extent of the atmosphere. For the distributions of the reaction rates, we show the distribution at a particular pressure level of interest based on the vertical distributions in Figure~\ref{fig:rratesz_intercomp}. The chosen pressure level is indicated in the title of each plot. For the 1:1 SOR, we take the temporal mean over 600~days. For the 3:2 SOR, we take the output on a single day for daytime on each hemisphere (corresponding to the first two periastron passages in Figure~\ref{fig:pcb_32res_tevol}).

\begin{figure}
\includegraphics[width=0.33\columnwidth]{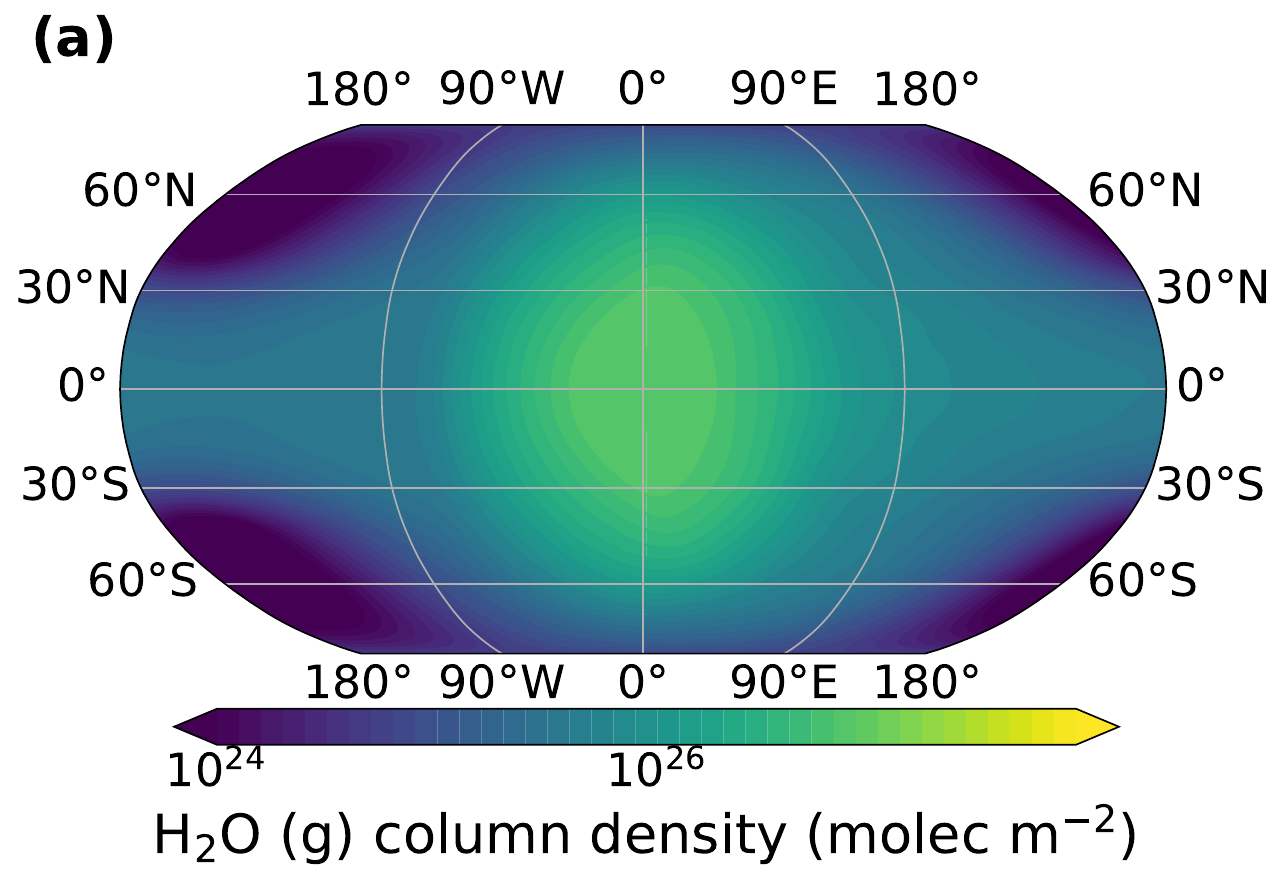}
\includegraphics[width=0.33\columnwidth]{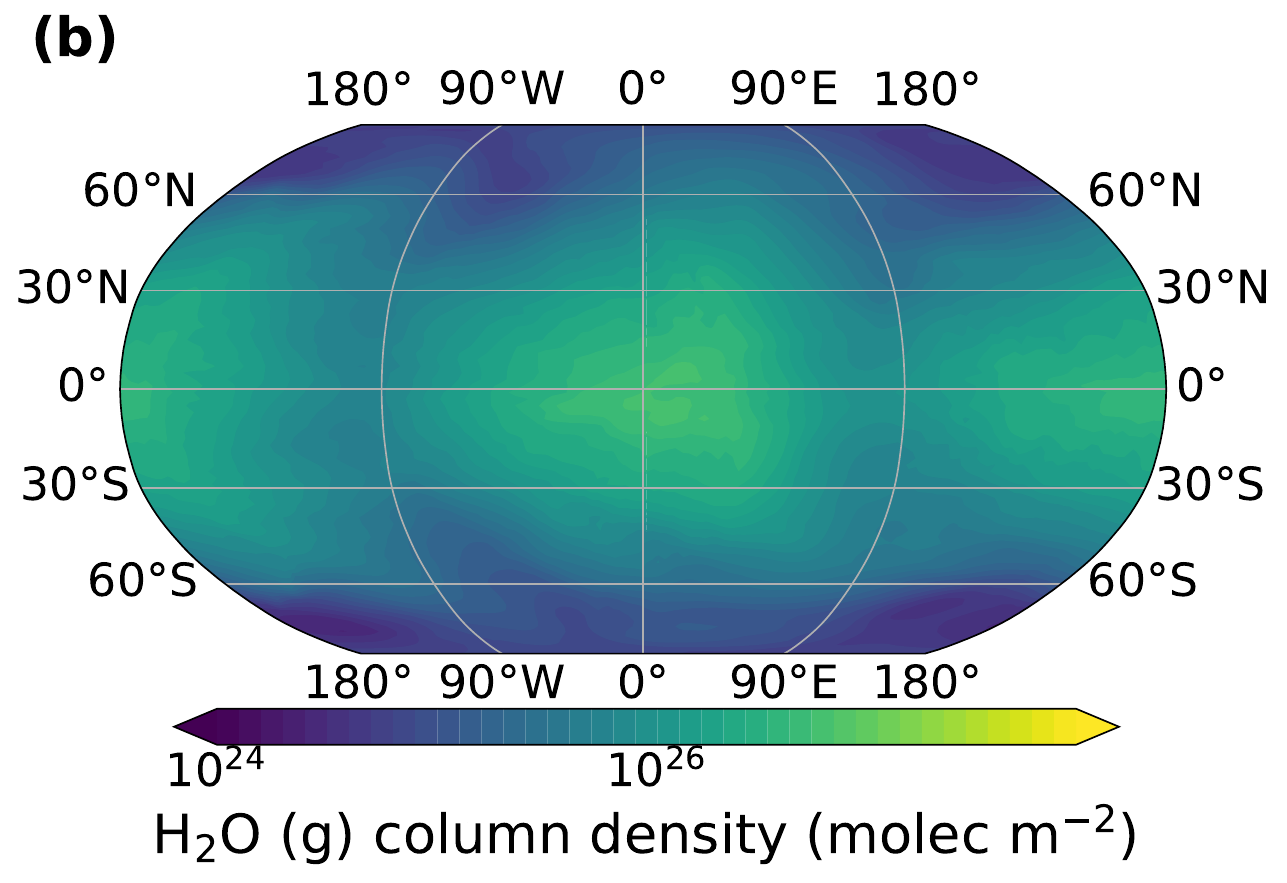}
\includegraphics[width=0.33\columnwidth]{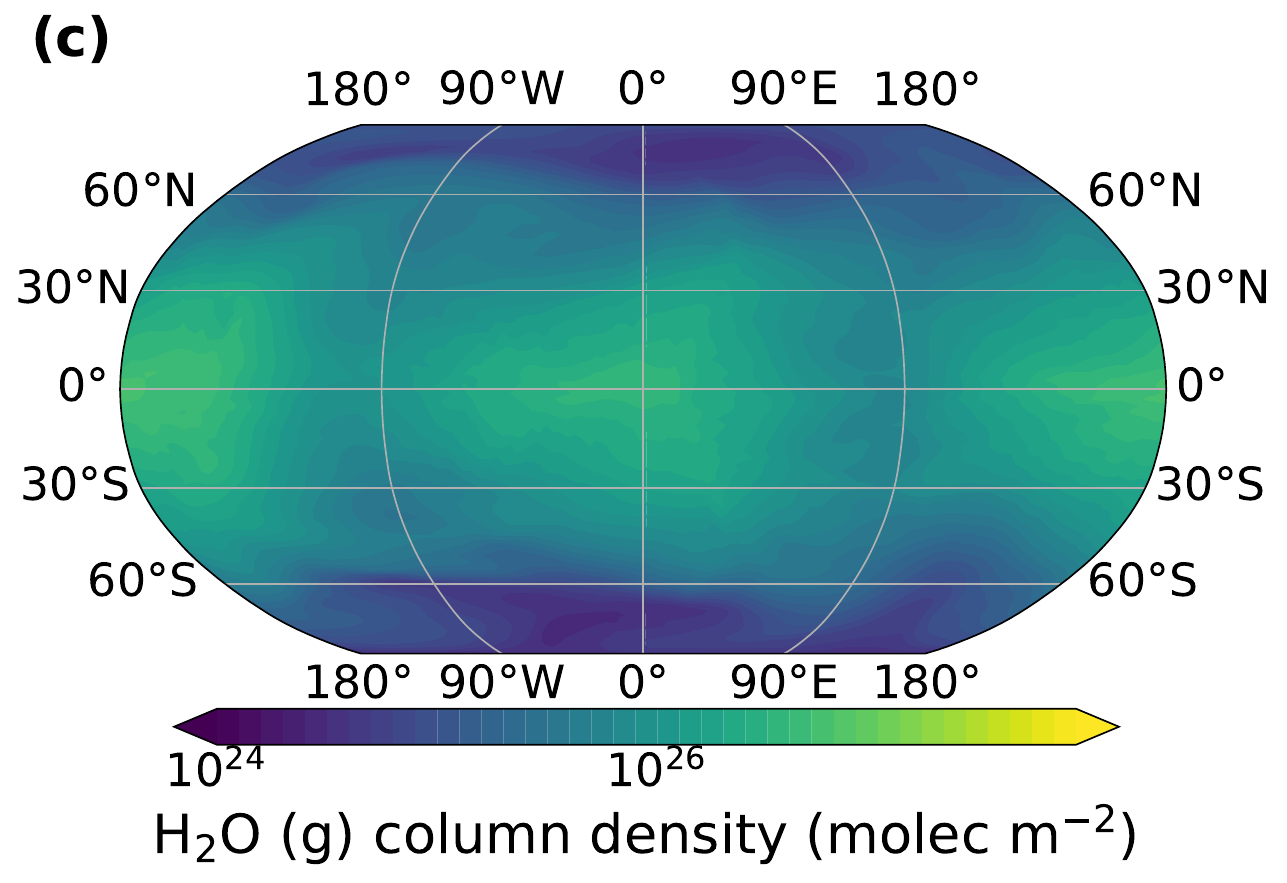}
\caption{Mean vertically integrated H$_2$O(g) column density for Proxima Centauri b in (a) a 1:1 SOR, (b) a 3:2 SOR with daytime on the 0$^{\circ}$ longitude hemisphere, and (c) a 3:2 SOR with daytime on the 180$^{\circ}$ longitude hemisphere.}
\label{fig:pcb_11_32_h2o}
\end{figure}

\begin{figure}
\includegraphics[width=0.33\columnwidth]{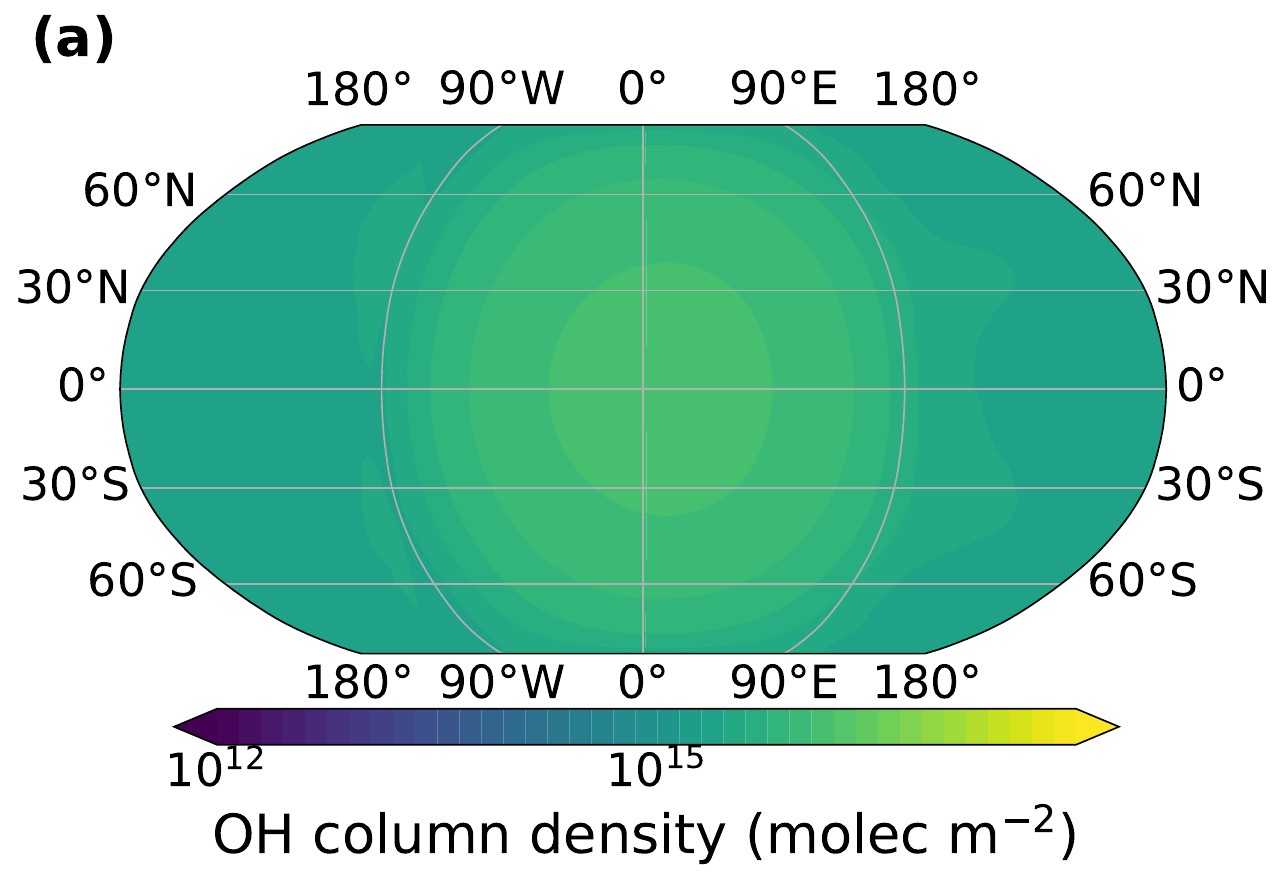}
\includegraphics[width=0.33\columnwidth]{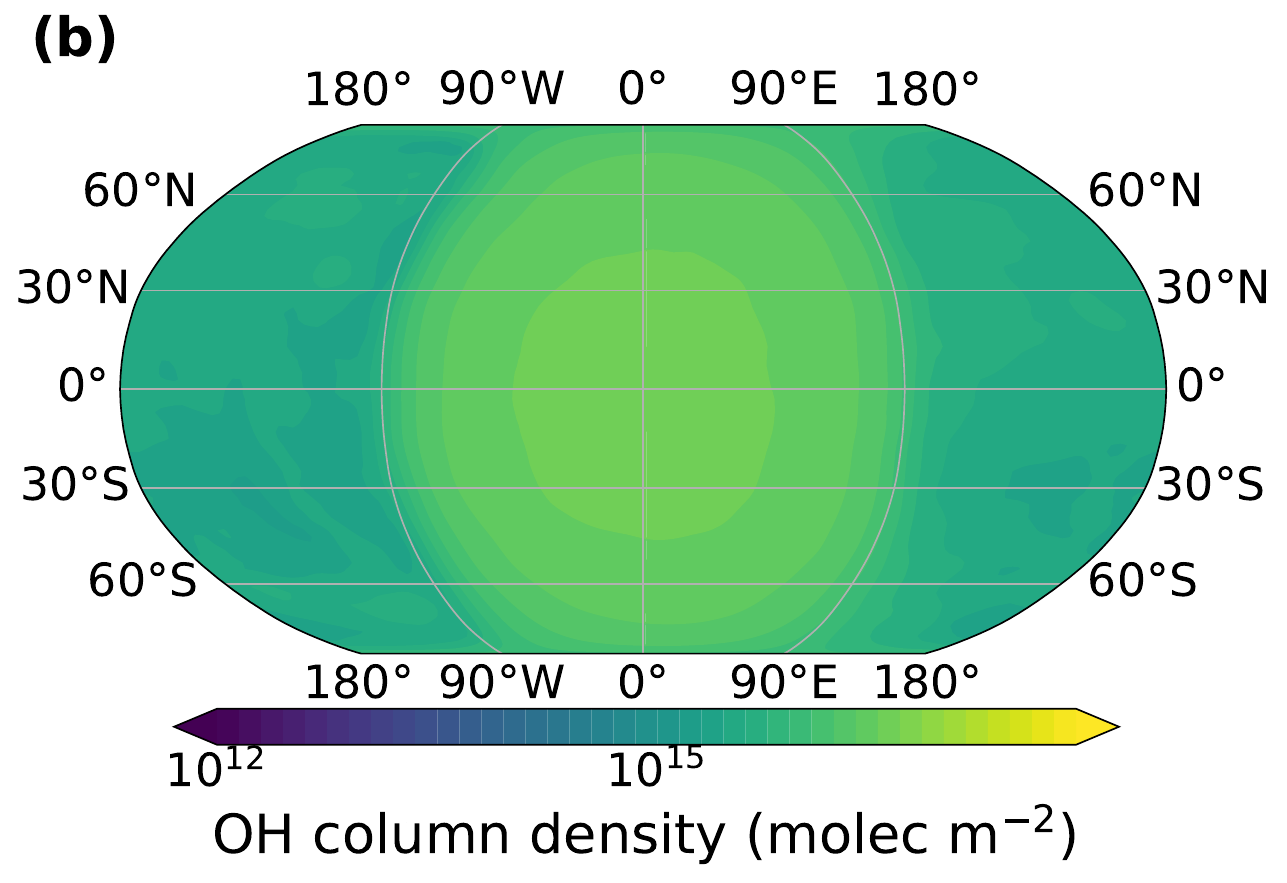}
\includegraphics[width=0.33\columnwidth]{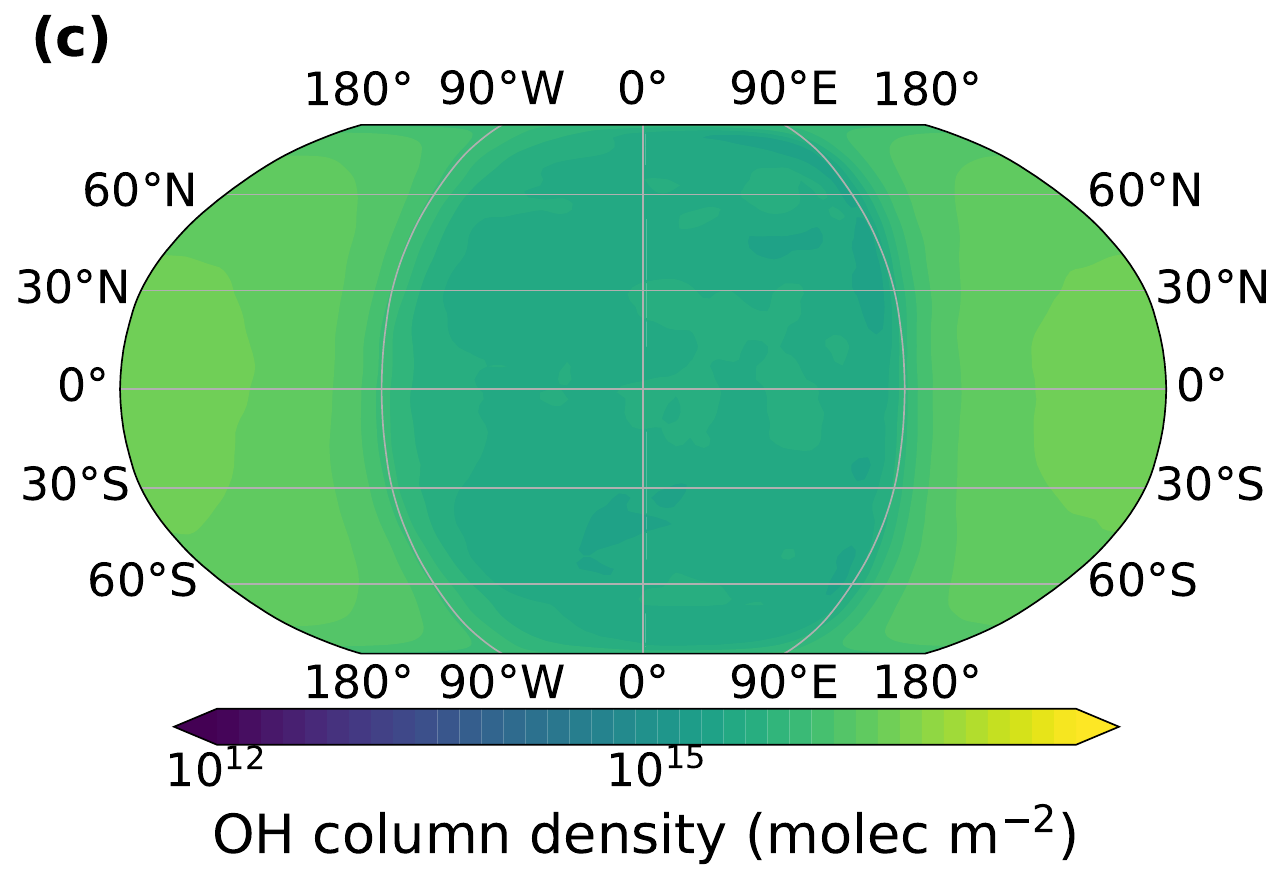}
\caption{Mean vertically integrated OH column density for Proxima Centauri b in (a) a 1:1 SOR, (b) a 3:2 SOR with daytime on the 0$^{\circ}$ longitude hemisphere, and (c) a 3:2 SOR with daytime on the 180$^{\circ}$ longitude hemisphere.}
\label{fig:pcb_11_32_oh}
\end{figure}

Figure~\ref{fig:pcb_11_32_h2o} shows the distribution of the H$_2$O(g) columns for both SORs. Water vapour in the simulated atmosphere results from evaporation from the surface oceans and its abundance is strongly dependent on the balance between evaporation and condensation \citep[see][]{boutle_exploring_2017}. SORs affect this balance by their distinct warming effects (Figures~\ref{fig:temp_11} and \ref{fig:pcb_32_temp}), even more so in the case of interactive sea-ice formation \citep[see][]{turbet_habitability_2016, del_genio_habitable_2019}. In turn, the distribution of water vapour strongly affects the formation of clouds \citep[e.g.,][]{yang_stabilizing_2013, sergeev_trappist-1_2022}. In Figure~\ref{fig:pcb_11_32_h2o}a we see that the H$_2$O(g) column is thickest for the dayside hemisphere in the 1:1 SOR, with a relatively dry nightside hemisphere especially over the gyre regions. The only source of water vapour on the nightside is transport from the dayside. Since water vapour is predominantly found in the troposphere (below 100~hPa, see Figure~\ref{fig:tair_vmrs_intercomp}b), this transport only occurs at low latitudes as part of the equatorial jet. The edges of the gyre regions at higher latitudes act as mixing barriers and, instead, these regions are fed from the stratosphere by a dayside-to-nightside circulation mechanism \citep[][]{braam_stratospheric_2023}. The stratosphere has relatively low water vapour abundances (Figure~\ref{fig:tair_vmrs_intercomp}b), explaining the small H$_2$O(g) column over the gyres. For the 3:2 SOR, water vapour is produced during the daytime in both hemispheres, resulting in two enhanced H$_2$O column regions and a latitudinal gradient, in line with the temperature distribution in Figure~\ref{fig:pcb_32_temp}. The peak of the H$_2$O(g) column is found in the center of the daytime hemisphere (longitude 0$^{\circ}$ in Figure~\ref{fig:pcb_11_32_h2o}b and 180$^{\circ}$ in Figure~\ref{fig:pcb_11_32_h2o}c). The distribution of OH in Figure~\ref{fig:pcb_11_32_oh} (also representing the HO$_2$ distribution due to the rapid cycling between the two) is strongly affected by the distribution of water vapour and the spatial variations in its subsequent photolytic destruction. One notable difference is that stratospheric abundances of OH and HO$_2$ are much higher than water vapour (see Figure~\ref{fig:tair_vmrs_intercomp}f--g), thus providing a source for the gyre regions in the 1:1 SOR through the stratospheric dayside-to-nightside circulation mechanism.

\begin{figure}
\includegraphics[width=0.33\columnwidth]{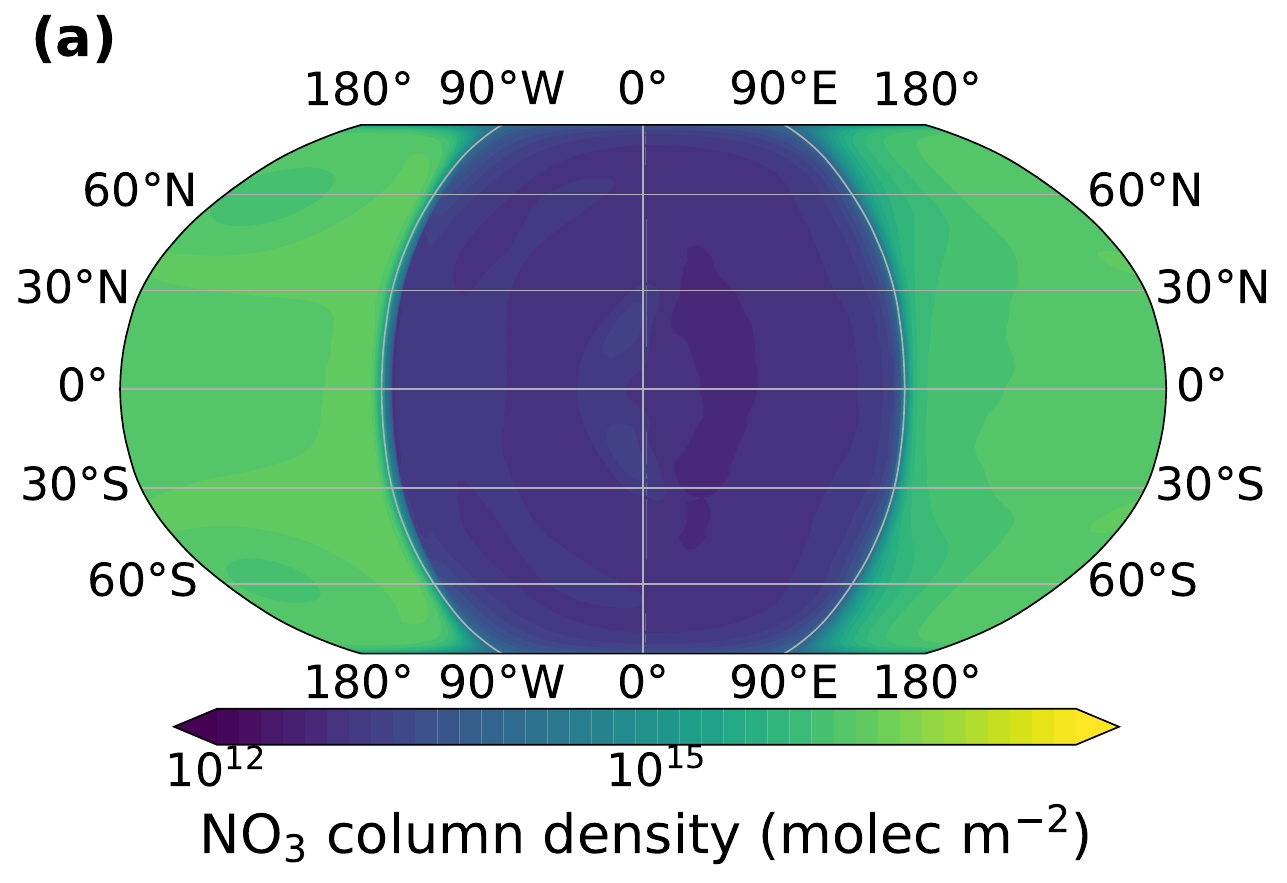}
\includegraphics[width=0.33\columnwidth]{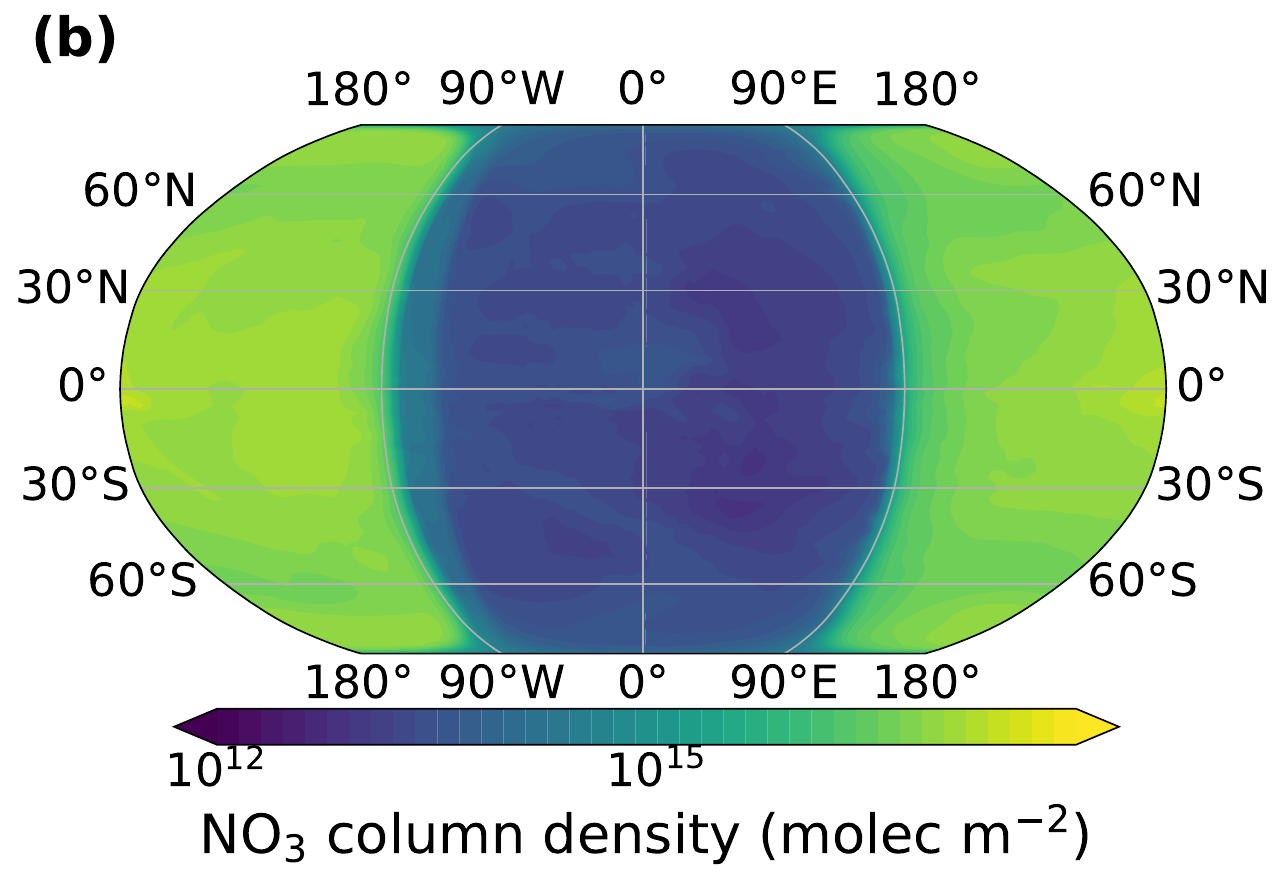}
\includegraphics[width=0.33\columnwidth]{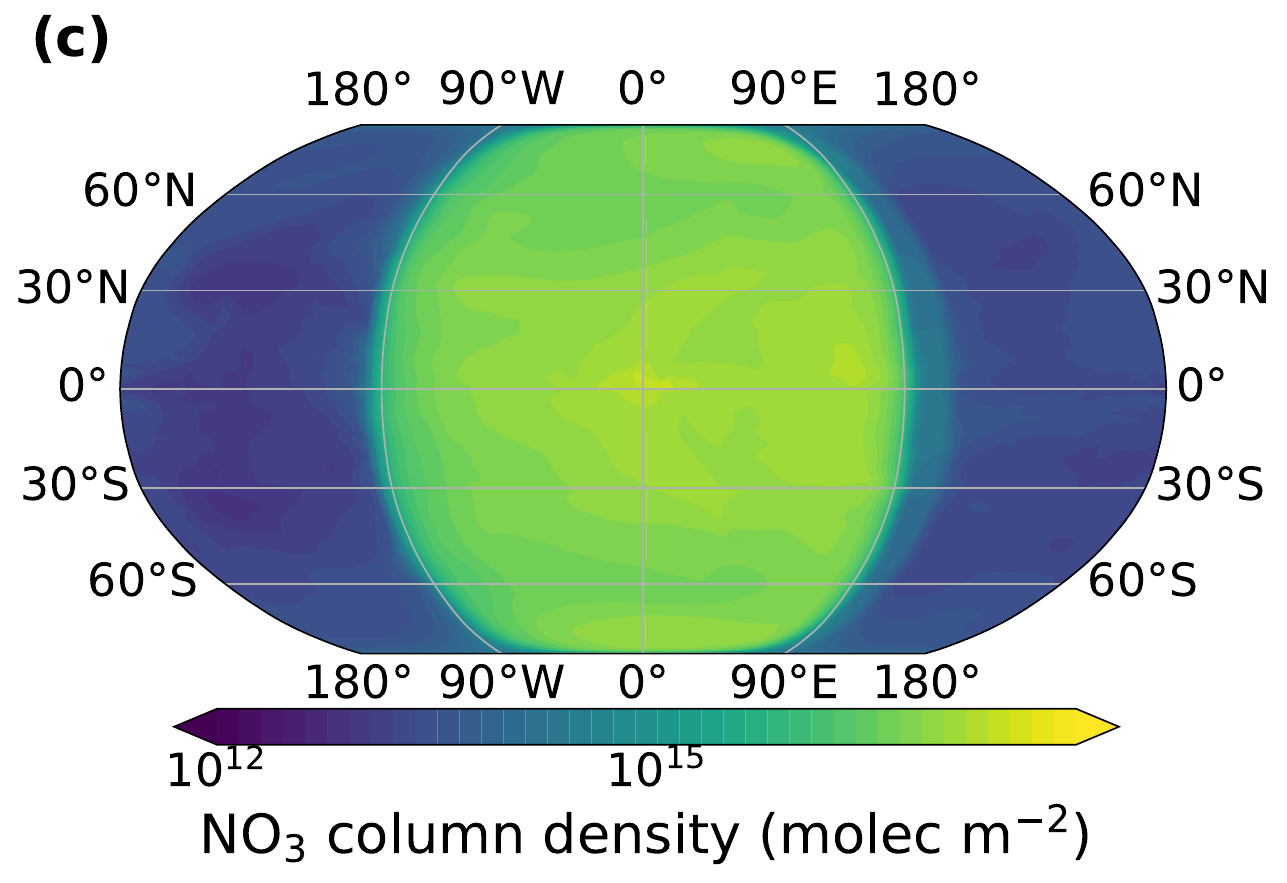}
\caption{Mean vertically integrated NO$_3$ column density for Proxima Centauri b in (a) a 1:1 SOR, (b) a 3:2 SOR with daytime on the 0$^{\circ}$ longitude hemisphere, and (c) a 3:2 SOR with daytime on the 180$^{\circ}$ longitude hemisphere.}
\label{fig:pcb_11_32_no3}
\end{figure}

As explained in Section~\ref{sec:chemical3d}, the coupled interaction between lightning flash distributions, atmospheric circulation, and photochemistry determines the distribution of NO$_3$ as shown in Figure~\ref{fig:pcb_11_32_no3}. For the 1:1 SOR, a clear hemispheric difference is shown with column densities that are three to four orders of magnitude higher on the nightside hemisphere, indicating its role as a reservoir species for NO$_\mathrm{x}$ in the absence of incoming radiation. The distribution of the NO$_3$ column density for the 3:2 SOR also confirms its role as a reservoir species in the absence of radiation. During daytime on the 0$^{\circ}$ longitude hemisphere, NO$_3$ rapidly accumulates on the nighttime hemisphere. One orbit later, daytime has shifted to the 180$^{\circ}$ longitude hemisphere, leading to the photolytic destruction of NO$_3$ here and its accumulation on the 0$^{\circ}$ longitude hemisphere. 

\begin{figure}
\includegraphics[width=0.33\columnwidth]{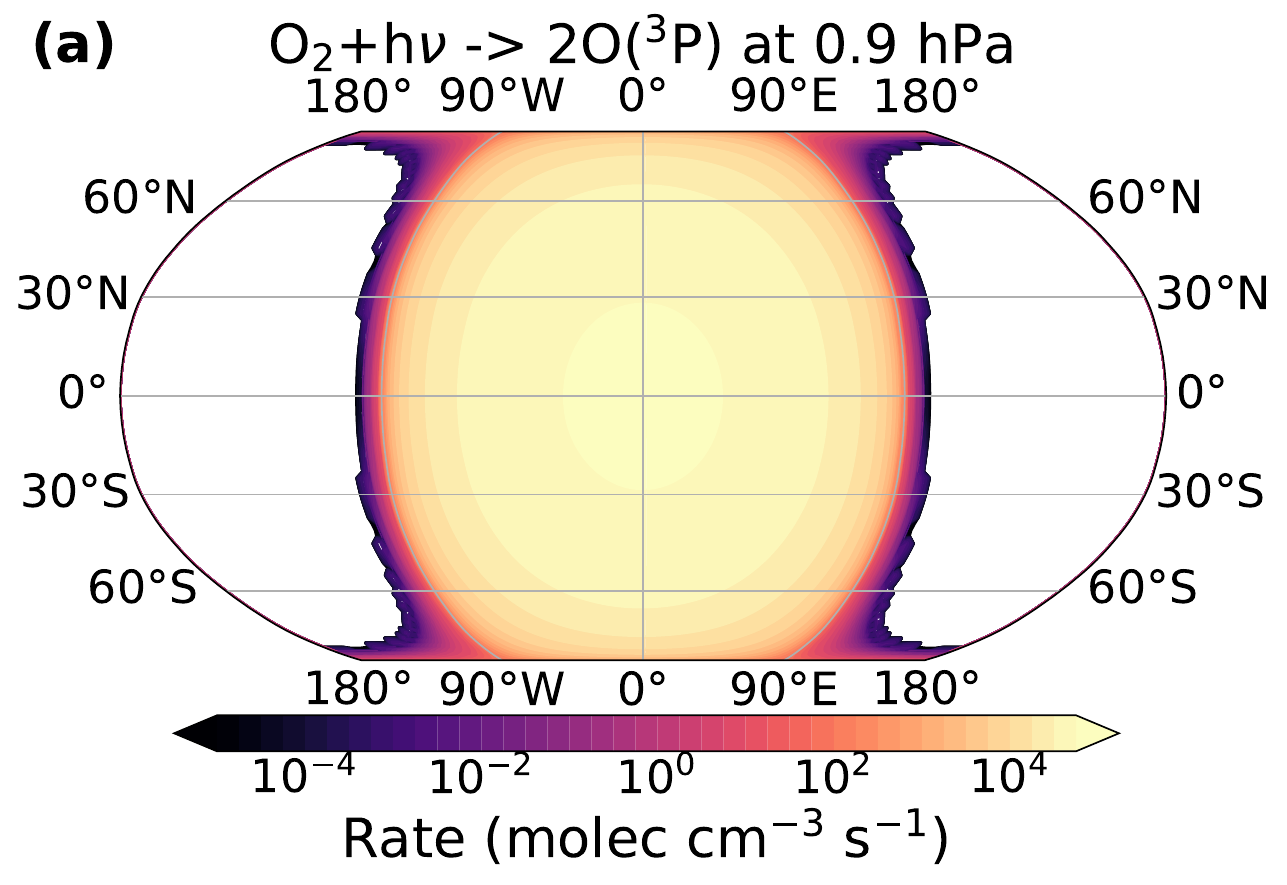}
\includegraphics[width=0.33\columnwidth]{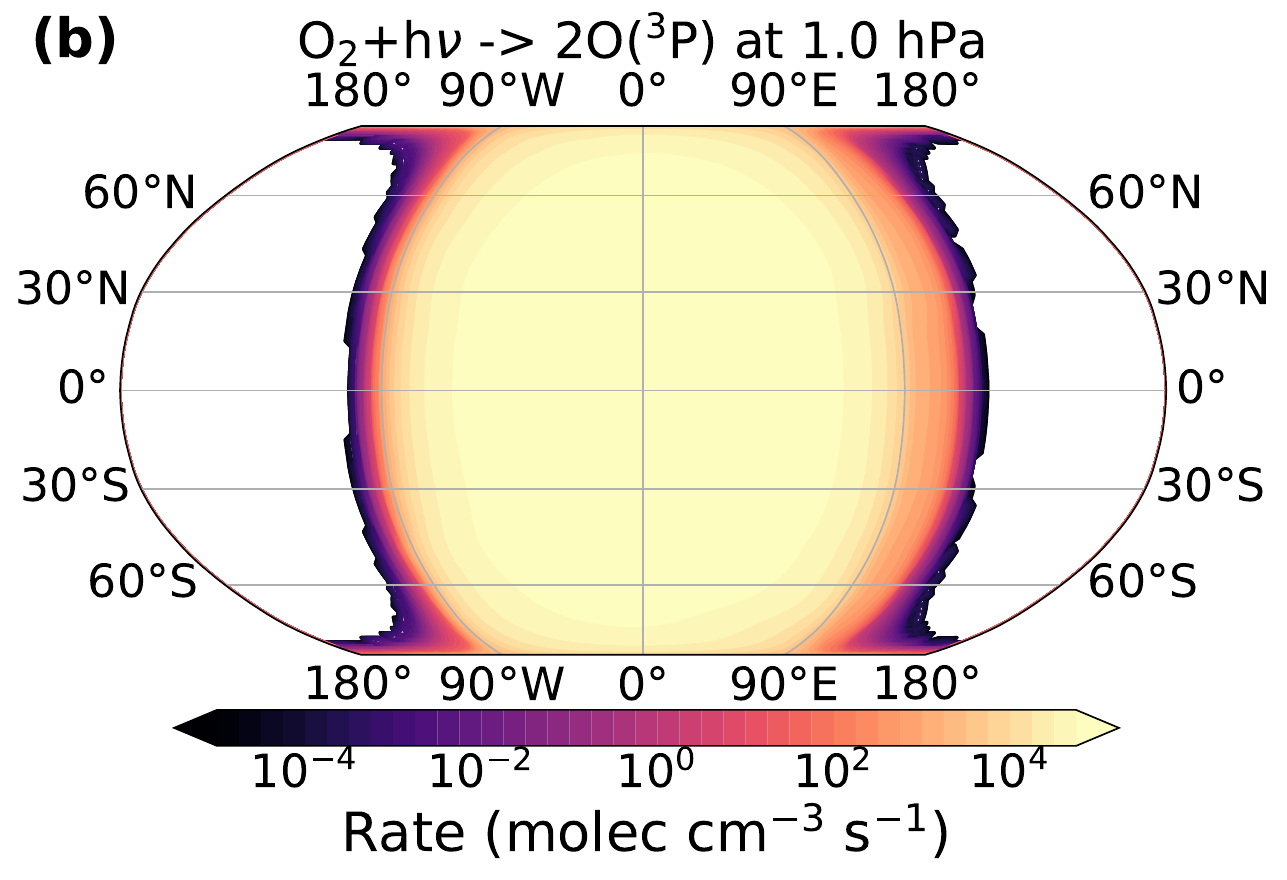}
\includegraphics[width=0.33\columnwidth]{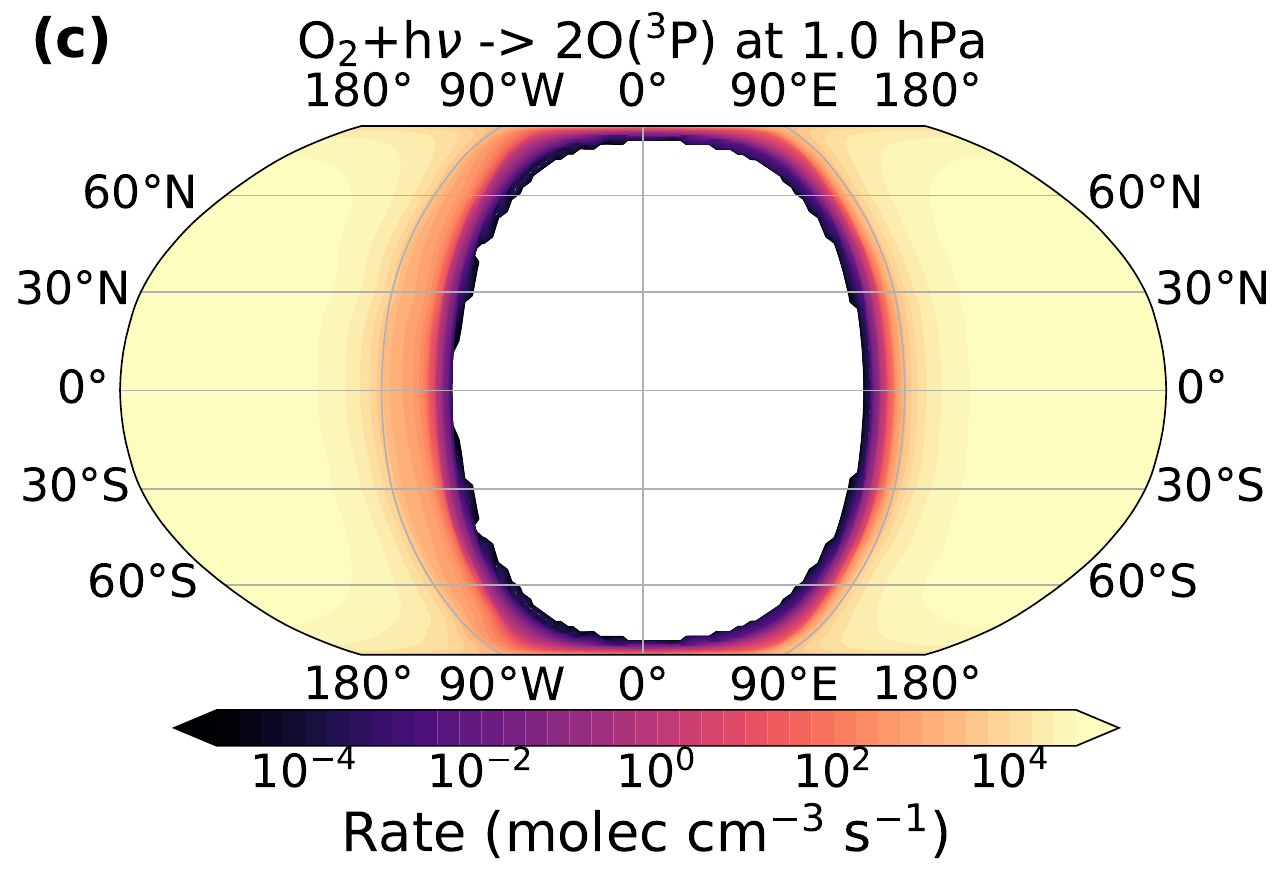}
\caption{The reaction rate corresponding to reaction R1 (\ce{O$_2$ + h$\nu$ -> O($^3$P) + O($^3$P)}) at a pressure level around ${\sim}$1~hPa for Proxima Centauri b in (a) a 1:1 SOR, (b) a 3:2 SOR with daytime on the 0$^{\circ}$ longitude hemisphere, and (c) a 3:2 SOR with daytime on the 180$^{\circ}$ longitude hemisphere. The pressure level corresponds to relatively fast reaction rates in the vertical distribution in Figure~\ref{fig:rratesz_intercomp}a.}
\label{fig:pcb_11_32_50101}
\end{figure}

Since the photolysis of molecular oxygen (also shown in Figure~\ref{fig:rratesz_intercomp}a) is a primary step for the chemistry discussed in this paper (Section~\ref{sec:chemical3d}), we present the spatial distributions of this reaction rate in Figure~\ref{fig:pcb_11_32_50101}. For the 1:1 SOR, the hemispheric contrast is evident, with photolysis mainly occurring on the dayside hemisphere. However -- as also indicated in Figure~\ref{fig:rratesz_intercomp}a -- photolysis of O$_2$ still occurs on the nightside hemisphere, although at significantly slower reaction rates. Figure~\ref{fig:pcb_11_32_50101}a shows that this photolysis is limited to the regions closest to the dayside hemisphere, indicating that these photolytic reactions are caused by radiation from the dayside hemisphere that is scattered to the nightside. The planet in a 3:2 SOR shows a similar distribution of the O$_2$ photolysis rate in Figure~\ref{fig:pcb_11_32_50101}b (daytime centred on the 0$^{\circ}$ hemisphere), albeit with a slight deviation in the exact longitude of the substellar point. The time averaging in daily simulation output causes this eastward deviation of the substellar point in Figure~\ref{fig:pcb_11_32_50101}b by a few degrees from 0$^{\circ}$ longitude. Figure~\ref{fig:pcb_11_32_50101}c shows a similar effect but with daytime centred on the 180$^{\circ}$ longitude hemisphere. Over two successive orbits for the 3:2 SOR (temporally averaging Figures~\ref{fig:pcb_11_32_50101}b and c), variations in photolysis rates are thus latitudinal as opposed to the longitudinal variations in Figure~\ref{fig:pcb_11_32_50101}a. 

\begin{figure}
\includegraphics[width=0.33\columnwidth]{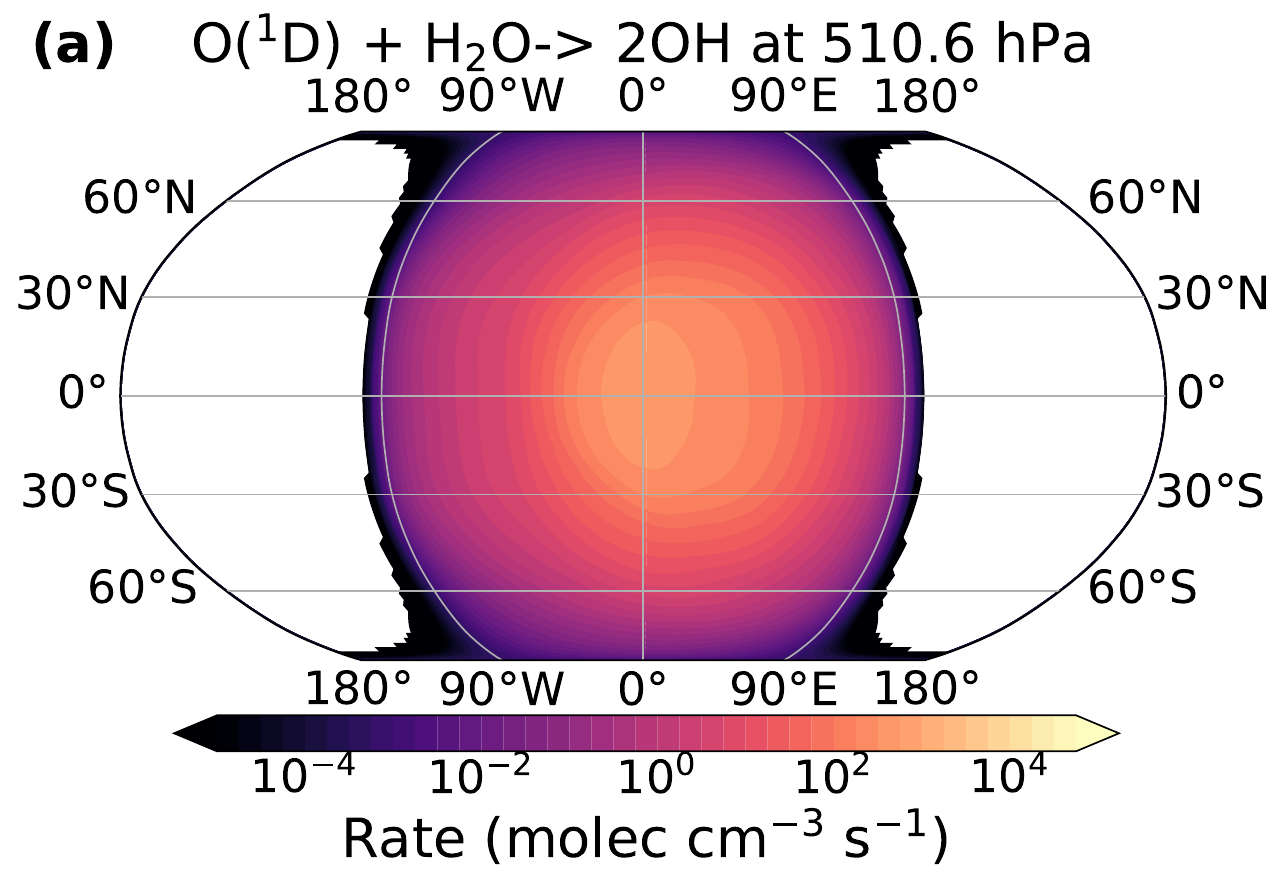}
\includegraphics[width=0.33\columnwidth]{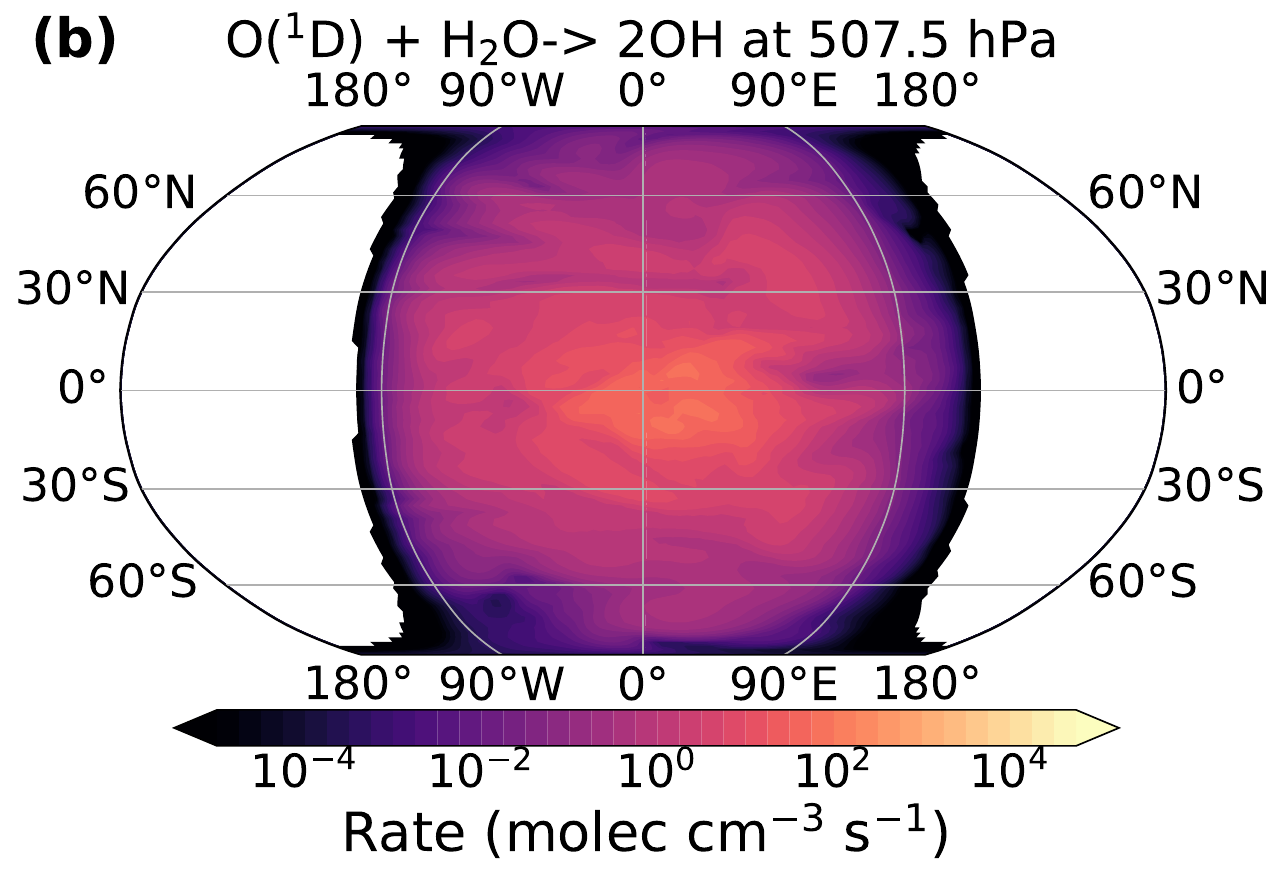}
\includegraphics[width=0.33\columnwidth]{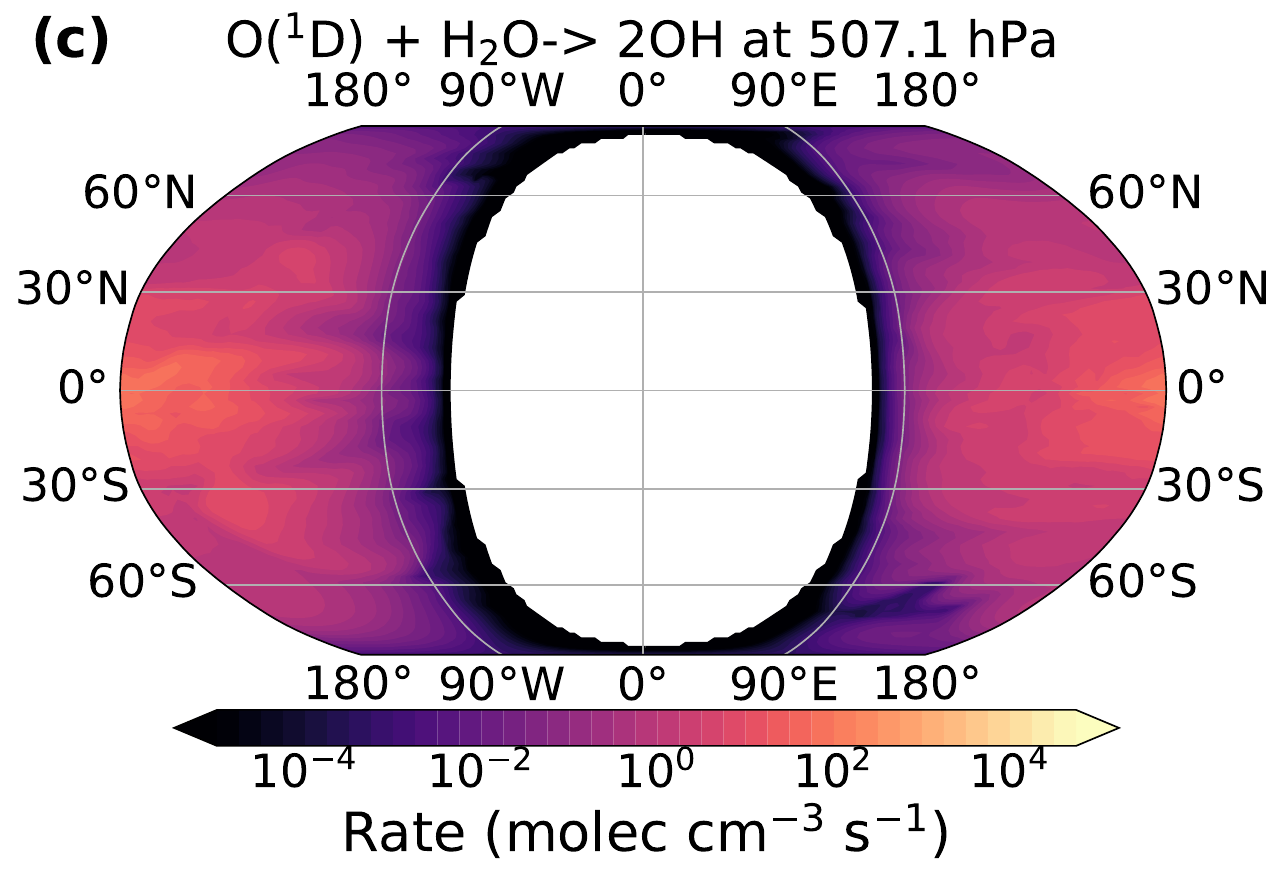}
\caption{The reaction rate corresponding to reaction R6 (\ce{H$_2$O + O($^1$D) -> 2OH}) at a pressure level of ${\sim}$500~hPa for Proxima Centauri b in (a) a 1:1 SOR, (b) a 3:2 SOR with daytime on the 0$^{\circ}$ longitude hemisphere, and (c) a 3:2 SOR with daytime on the 180$^{\circ}$ longitude hemisphere. The pressure level corresponds to relatively fast reaction rates in the vertical distribution in Figure~\ref{fig:rratesz_intercomp}c.}
\label{fig:pcb_11_32_50011}
\end{figure}

\begin{figure}
\includegraphics[width=0.33\columnwidth]{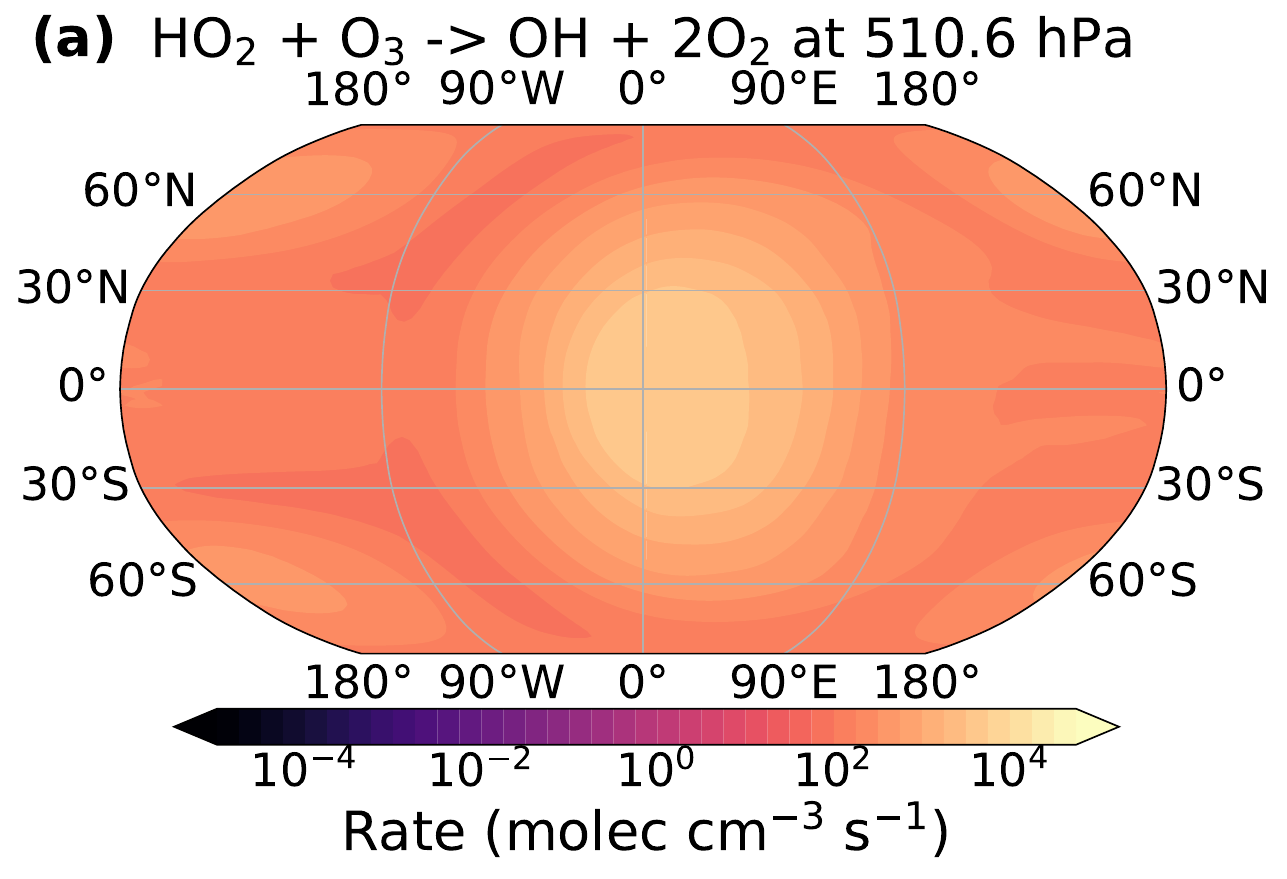}
\includegraphics[width=0.33\columnwidth]{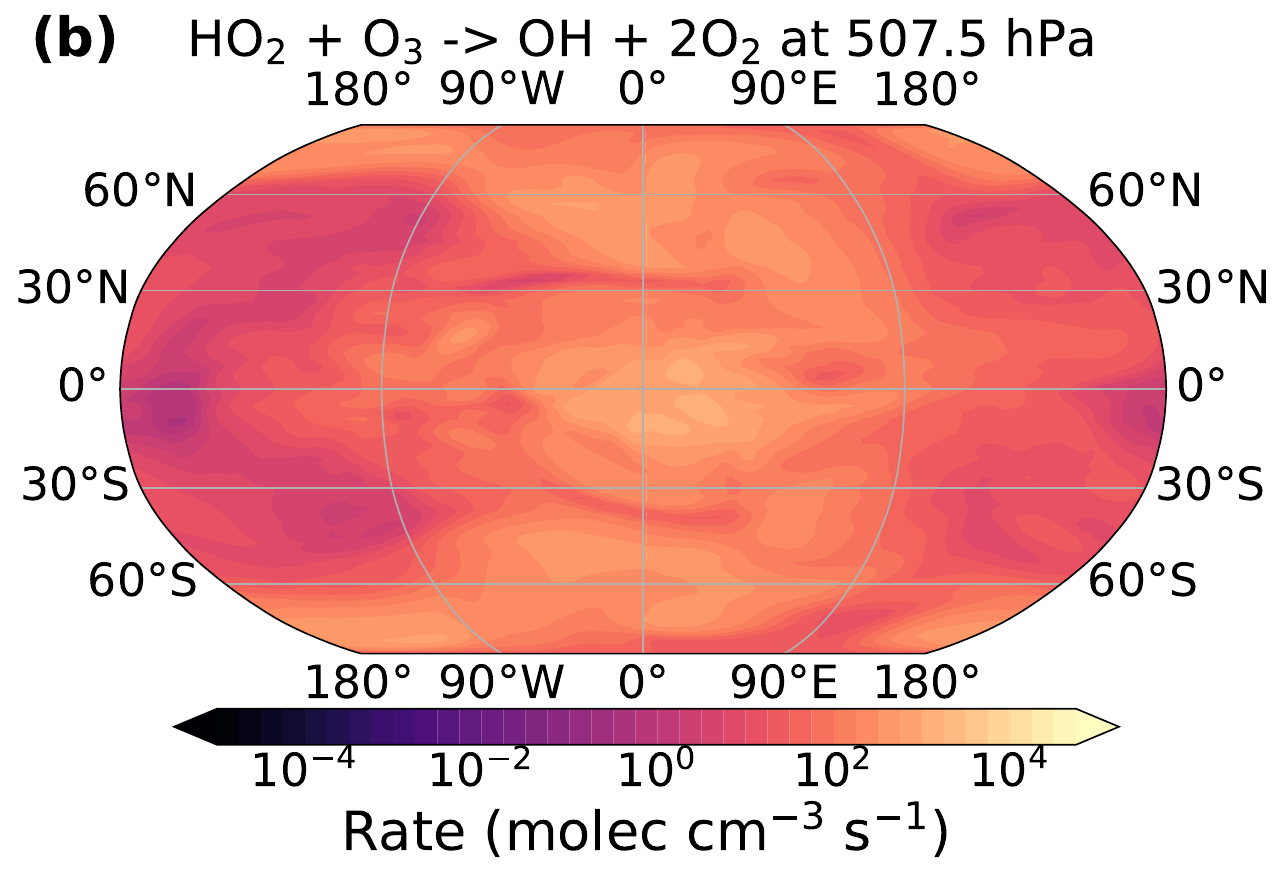}
\includegraphics[width=0.33\columnwidth]{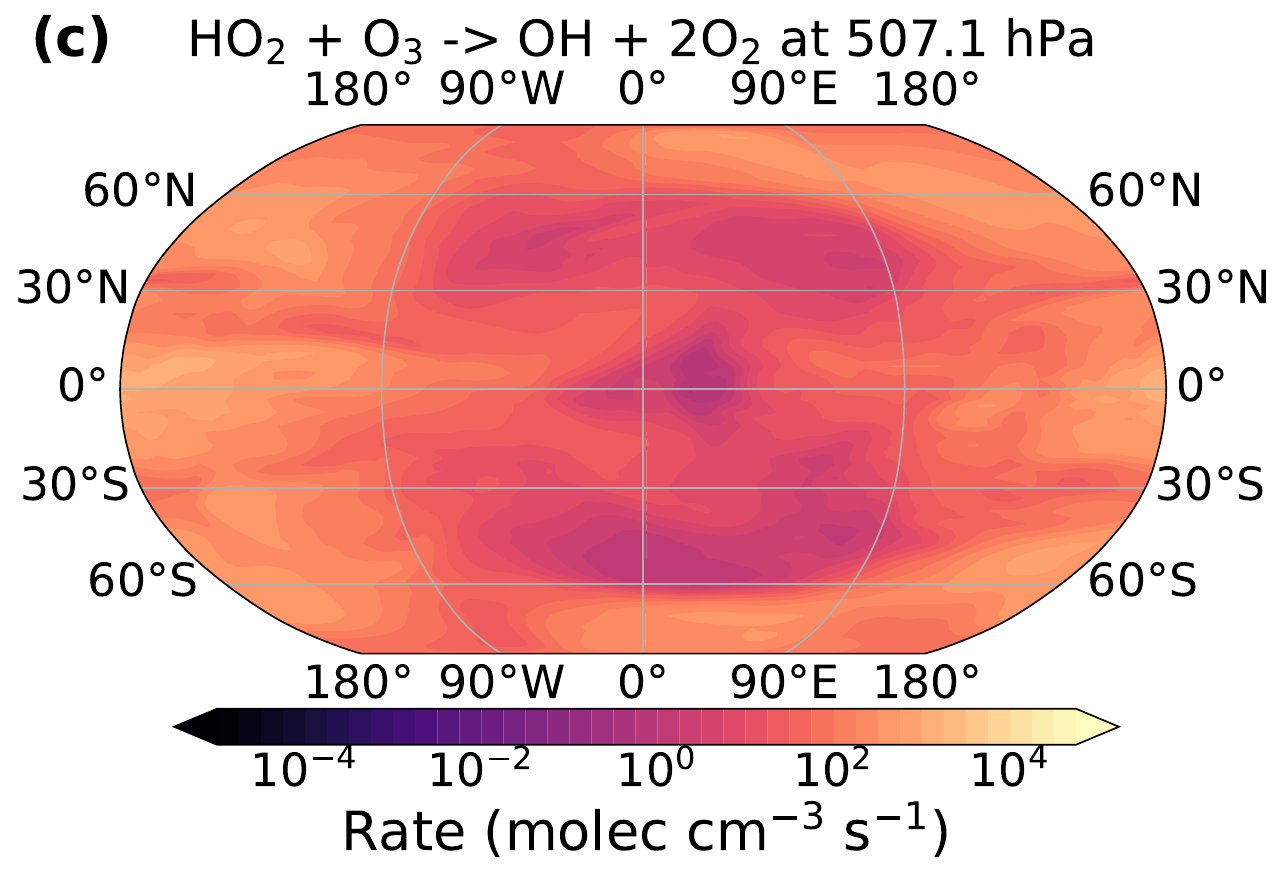}
\caption{The reaction rate corresponding to reaction R9 (\ce{HO$_2$ + O$_3$ -> OH + 2O$_2$}) at a pressure level of ${\sim}$500~hPa for Proxima Centauri b in (a) a 1:1 SOR, (b) a 3:2 SOR with daytime on the 0$^{\circ}$ longitude hemisphere, and (c) a 3:2 SOR with daytime on the 180$^{\circ}$ longitude hemisphere. The pressure level corresponds to relatively fast reaction rates in the vertical distribution in Figure~\ref{fig:rratesz_intercomp}e.}
\label{fig:pcb_11_32_50014}
\end{figure}

The reaction rate for the oxidation of water vapour and its subsequent production of OH is shown in Figure~\ref{fig:pcb_11_32_50011}, indicating its dependence on photochemistry (for the production of O($^1$D)) and the distribution of water vapour in Figure~\ref{fig:pcb_11_32_h2o}. OH is produced mainly during the daytime and the contrast between relatively smooth rates in Figure~\ref{fig:pcb_11_32_50011}a and variable rates in  Figure~\ref{fig:pcb_11_32_50011}b--c resembles the variations in the water vapor distribution in Figure~\ref{fig:pcb_11_32_h2o}. The daytime production mechanisms explains the tendency for enhanced daytime OH abundances, although OH is still transported to the nighttime (Figure~\ref{fig:pcb_11_32_oh}). Catalytic ozone destruction through Reaction R9 (\ce{HO$_2$ + O$_3$ -> OH + 2O$_2$}, see Figure~\ref{fig:pcb_11_32_50014}) thus occurs globally but more strongly during the daytime. The relatively strong reaction rates over the nightside gyres for the 1:1 SOR (Figure~\ref{fig:pcb_11_32_50014}a) are driven by the higher local ozone abundances (Figure~\ref{fig:toc_11}).

\begin{figure}
\includegraphics[width=0.33\columnwidth]{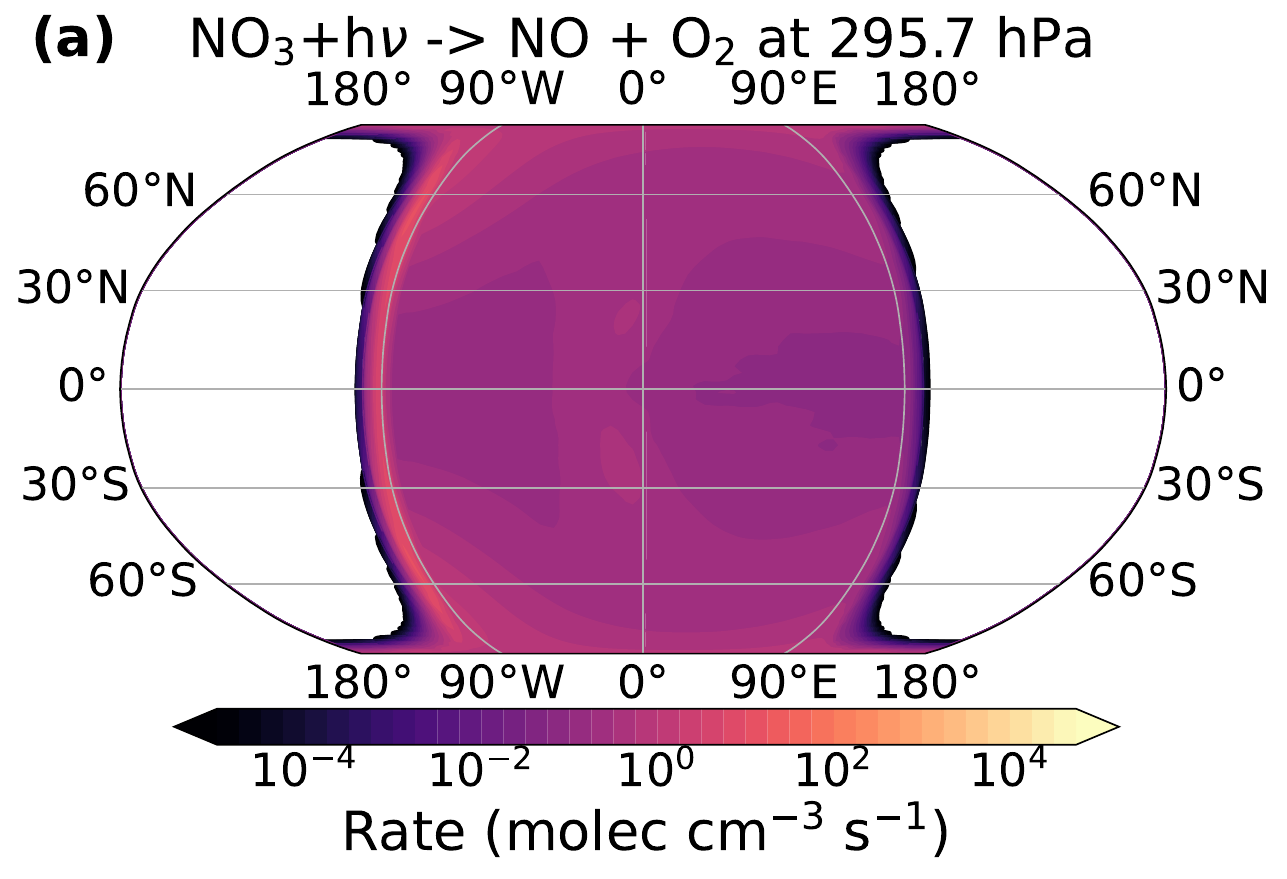}
\includegraphics[width=0.33\columnwidth]{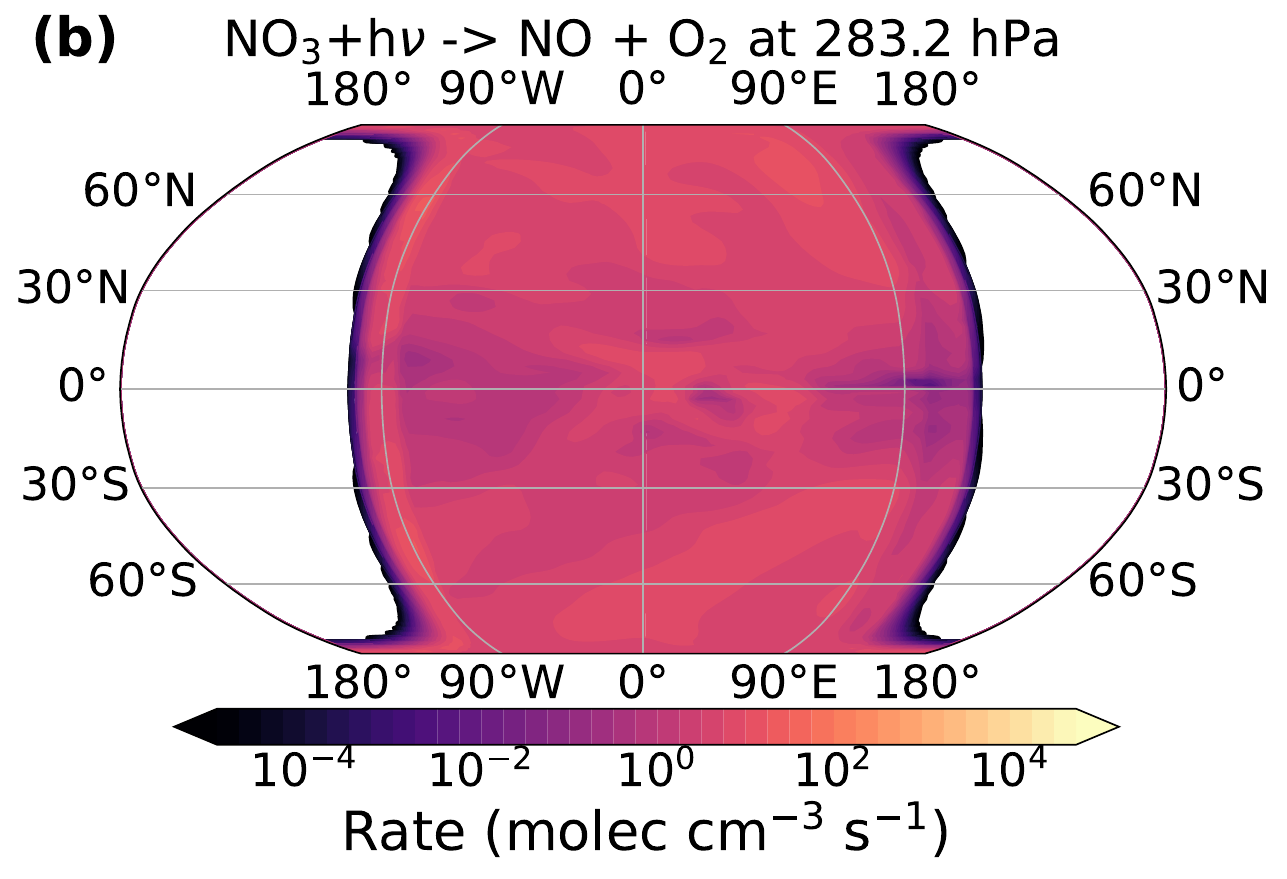}
\includegraphics[width=0.33\columnwidth]{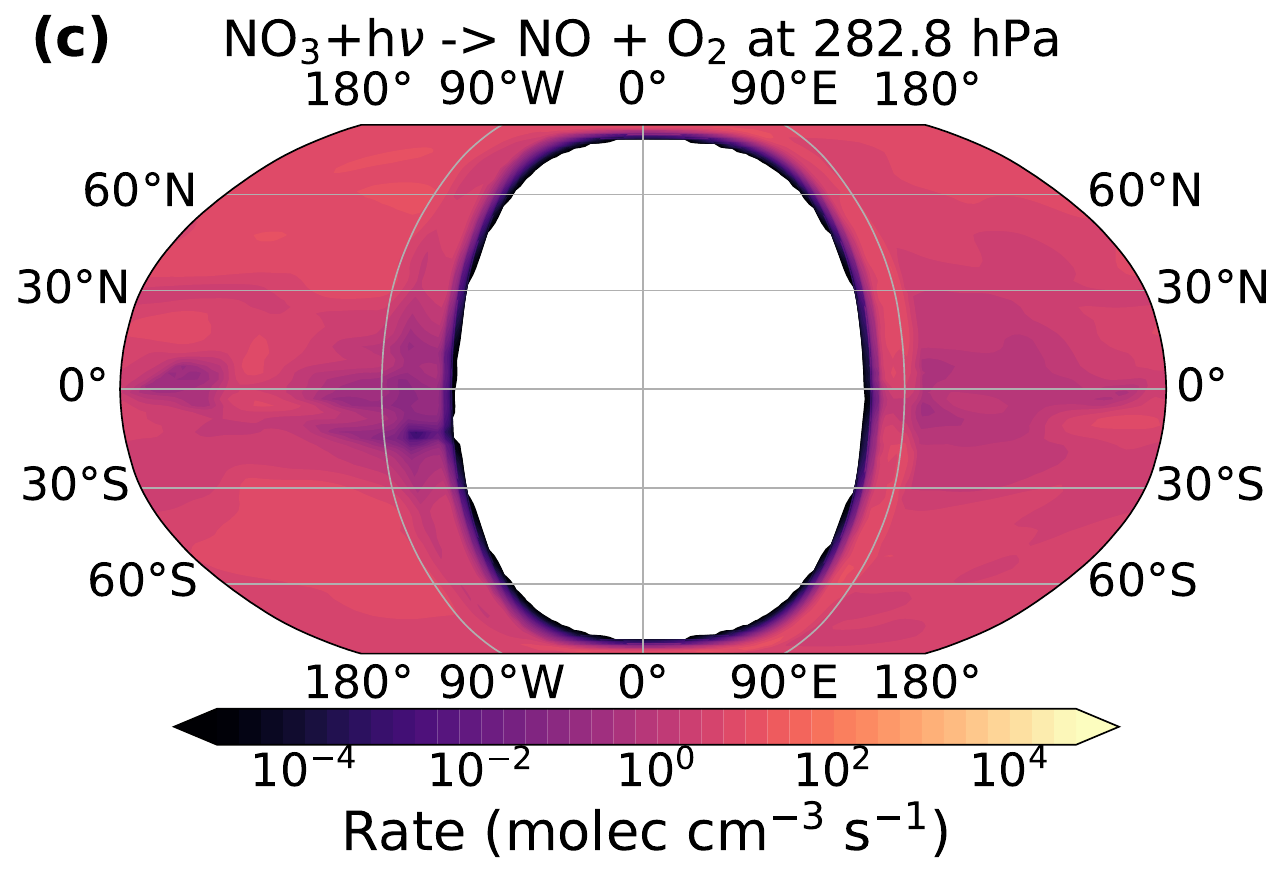}
\caption{The reaction rate corresponding to reaction R16 (\ce{NO$_3$ + h$\nu$ -> NO + O$_2$}) at a pressure level of ${\sim}$290~hPa for Proxima Centauri b in (a) a 1:1 SOR, (b) a 3:2 SOR with daytime on the 0$^{\circ}$ longitude hemisphere, and (c) a 3:2 SOR with daytime on the 180$^{\circ}$ longitude hemisphere. The pressure level corresponds to relatively fast reaction rates in the vertical distribution in Figure~\ref{fig:rratesz_intercomp}i.}
\label{fig:pcb_11_32_50121}
\end{figure}

Reaction R16 (\ce{NO$_3$ + h$\nu$ -> NO + O$_2$}) is an important step in the NO$_\mathrm{x}$ catalytic cycle (as described in Section~\ref{sec:chemical3d}) with, interestingly, nightside rates that are only marginally smaller than those on the dayside for the 1:1 SOR (Figure~\ref{fig:rratesz_intercomp}i), despite being a photolysis reaction. The spatial distributions of reaction R16 in Figure~\ref{fig:pcb_11_32_50121} show that NO$_3$ photolysis over the nightside hemisphere is limited to the regions closest to the dayside hemisphere, with similar effects seen over the nighttime hemispheres of the planet in a 3:2 SOR. Hence, incoming radiation scattered in the daytime atmosphere probably causes the nighttime photolysis rates. The nightside photolysis rates for the 1:1 SOR are strongest over the western terminator (crescent shape between 90-100$^{\circ}$~W, see Figure~\ref{fig:pcb_11_32_50121}a), corresponding to the higher NO$_3$ column density over the western as compared to the eastern terminator in Figure~\ref{fig:pcb_11_32_no3}. In turn, the distinct terminator abundances are due to the nightside production of NO$_3$ \citep[][]{braam_lightning-induced_2022} followed by eastwards atmospheric circulation. Hence, the distribution of NO$_3$ again illustrates a coupled dependence on orbital configuration, photochemistry, and atmospheric circulation.

\bibliography{bib_clean}{}
\bibliographystyle{aasjournal}



\end{document}